\documentclass[twocolumn]{aastex63}
\usepackage{threeparttable}
\usepackage{lineno}

\usepackage{amsmath}
\usepackage{subfigure}

\graphicspath{{./}{figures/}}

\begin{document}

\title{Magnetar Engines in Fast Blue Optical Transients and Their Connections with SLSNe, SNe Ic-BL, and lGRBs}

\author{Jian-Feng Liu}
\affiliation{Institute of Astrophysics, Central China Normal University, Wuhan 430079, China; \url{yuyw@ccnu.edu.cn}, \url{liuld@ccnu.edu.cn}}
\affiliation{Key Laboratory of Quark and Lepton Physics (Central China Normal University), Ministry of Education, Wuhan 430079, China}

\author[0000-0002-9195-4904]{Jin-Ping Zhu}
\affiliation{Department of Astronomy, School of Physics, Peking University, Beijing 100871, China}
%\affiliation{Monash Centre for Astrophysics, School of Physics and Astronomy, Monash University, Clayton, VIC 3800, Australia}

\author[0000-0002-8708-0597]{Liang-Duan Liu}
\affiliation{Institute of Astrophysics, Central China Normal University, Wuhan 430079, China; \url{yuyw@ccnu.edu.cn}, \url{liuld@ccnu.edu.cn}}
\affiliation{Key Laboratory of Quark and Lepton Physics (Central China Normal University), Ministry of Education, Wuhan 430079, China}

\author[0000-0002-1067-1911]{Yun-Wei Yu}
\affiliation{Institute of Astrophysics, Central China Normal University, Wuhan 430079, China; \url{yuyw@ccnu.edu.cn}, \url{liuld@ccnu.edu.cn}}
\affiliation{Key Laboratory of Quark and Lepton Physics (Central China Normal University), Ministry of Education, Wuhan 430079, China}

\author[0000-0002-9725-2524]{Bing Zhang}
\affiliation{Nevada Center for Astrophysics, University of Nevada, Las Vegas, NV 89154, USA}
\affiliation{Department of Physics and Astronomy, University of Nevada, Las Vegas, NV 89154, USA}

\begin{abstract}

We fit the multi-band lightcurves of 40 fast blue optical transients (FBOTs) with the magnetar engine model. The mass of the FBOT ejecta, the initial spin period and polar magnetic field of the FBOT magnetars are respectively constrained to $M_{\rm{ej}}=0.11^{+0.22}_{-0.09}\,M_\odot$, $P_{\rm{i}}=9.1^{+9.3}_{-4.4}\,{\rm{ms}}$, and $B_{\rm p}=11^{+18}_{-7}\times10^{14}\,{\rm{G}}$. The wide distribution of the value of $B_{\rm p}$ spreads the parameter ranges of the magnetars from superluminous supernovae (SLSNe) to broad-line Type Ic supernovae (SNe Ic-BL; some are observed to be associated with long-duration gamma-ray bursts), which are also suggested to be driven by magnetars. Combining FBOTs with the other transients, we find a strong universal anti-correlation as $P_{\rm{i}}\propto{M_{\rm{ej}}^{-0.41}}$, indicating them could share a common origin. To be specific, it is suspected that all of these transients originate from collapse of extreme-stripped stars in close binary systems, but with different progenitor masses. As a result, FBOTs distinct themselves by their small ejecta masses with an upper limit of ${\sim}1\,M_\odot$, which leads to an observational separation in the rise time of the lightcurves $\sim10\,{\rm d}$. In addition, the FBOTs together with SLSNe can be separated from SNe Ic-BL by an empirical line in the $M_{\rm peak}-t_{\rm rise}$ plane corresponding to an energy requirement of a mass of $^{56}$Ni of $\sim0.3M_{\rm ej}$, where $M_{\rm peak}$ is the peak absolute magnitude of the transients and $t_{\rm rise}$ is the rise time.

\end{abstract}


\section{Introduction} \label{sec:intro}

In the past decade, several unique, fast-evolving and luminous transients have been discovered, thanks to the improved cadence and technology of wide-field surveys. These transients are usually quite blue ($g - r\lesssim -0.2$) and luminous (an absolute magnitude of $-16\lesssim M_{\rm peak}\lesssim-23$) at peak and their lightcurves show fast rise and decline with a duration shorter than about ten days. They are, hence, named as fast blue optical transients \citep[FBOTs; e.g.,][]{Drout2014,Inserra2019}. Since \cite{Drout2014} reported a sample of FBOTs identified from a search within the Pan-STARRS1 Medium Deep Survey (PS1-MDS) data, the observations of $\sim100$ FBOT candidates have been presented \citep[e.g.,][]{Arcavi2016,Whitesides2017,Pursiainen2018,Tampo2020,Ho2019,Ho2020,Ho2021}. 
The event rate density of FBOTs is $\sim1-10\%$ of that of local core-collapse supernovae \citep[SNe;][]{Drout2014,Pursiainen2018,Ho2021}.

The progenitor and energy source of FBOTs are still very unclear. Two different classes of models have been proposed in literature to explain the observational proprieties of FBOTs. The first class broadly contains binary neutron star (BNS), binary white dwarf (BWD) or NS--WD mergers \citep[e.g.,][]{Yu2013,Yu2015,Yu2019b,Zenati2019};  accretion-induced collapse (AIC) of a WD \citep[e.g.,][]{Kasliwal2010,Brooks2017,Yu2015,Yu2019a}; SN explosions of ultra-stripped progenitor stars \citep[e.g.,][]{Tauris2013,Tauris2015,Tauris2017,Suwa2015,Hotokezaka2017,De2018,Sawada2022} including electron capture SNe \citep[e.g.,][]{Moriya2016,Mor2022}; common envelope jets SNe \citep[][]{Soker2019,Soker2022}; and tidal disruption of a star by a NS or a black hole \citep[e.g.,][]{Liu2018,Perley2019,Kremer2021,Metzger2022}. The common feature of this class of models is that the fast evolution of FBOTs is attributed to a small ejecta mass, and the luminous brightness of FBOTs is attributed to additional energy injection from a central engine sources besides the radioactive decay power by $^{56}{\rm Ni}$ \citep[e.g.,][]{Drout2014,Pursiainen2018}. The extra energy source could be a  spinning-down NS or an accreting black hole (i.e., in the tidal disruption models).  The second class of models invoke shock breakouts from a dense stellar wind \citep[e.g.,][]{Chevalier2011,Ginzburg2012,Drout2014}; interaction between the ejecta from a massive star and a dense circumstellar material \citep[CSM; e.g.,][]{Rest2018,Fox2019,Leung2020,Xiang2021,Pellegrino2022}; and jet-cocoon interaction and emission \citep{Gottlieb2022}. FBOTs in this class of models are attributed to the breakout of the accumulated energy in the shock. 
%}\textbf{It is also possible that interaction between FBOT ejecta and dense hydrogen-poor CSM convert the ejecta kinetic energy into radiation energy to power a fraction light curves of FBOTs \citep[e.g.,][]{Wang2019,Xiang2021,Pellegrino2022}.  Ejecta-CSM interaction and magnetar engine spin-down can be hard to distinguish through photometry alone. 
By fitting the bolometeric light curves of FBOTs with the CSM interaction plus $^{56}$Ni decay model, \cite{Xiang2021} and \cite{Pellegrino2022} found that, in order to account for the rapid and luminous light curves, the mass loss rates of the progenitor should be up to $\sim 1 M_{\odot}$ yr$^{-1}$, which is however inconsistent with the limits obtained from the radio observations of FBOTs.

Recent studies revealed that the hosts of FBOTs are exclusively star-forming galaxies \citep{Drout2014,Pursiainen2018,Pellegrino2022}, whose star-formation rates and metallicities are consistent with those of extreme stripped-envelope explosions \citep{Wiseman2020} including hydrogen-poor Type Ic superluminous SNe \citep[SLSNe; e.g.,][]{Lunnan2014,Chen2016}, broad-lined Type Ic SNe \citep[SNe Ic-BL; e.g.,][]{Arcavi2010}, and long-duration gamma-ray bursts \citep[lGRBs; e.g.,][]{Kruhler2015,Perley2016}. Furthermore, it is worth noticing that these extreme stripped-envelope explosions are widely believed to harbor a long-lived millisecond magnetar \citep{Dai1998,Wheeler2000,Zhang2001,Yu2010,Kasen2010,Woosley2010,Piro2011,Inserra2013,Zhang2018}, which can lose its rotational energy via  spin-down processes to provide an additional energy injection for the explosion. Therefore, in view of the similarity between the host galaxies of FBOTs and those of SLSNe, SNe Ic-BL, and lGRBs, it would be reasonable to suspect that the FBOTs are also powered by a magnetar engine. Indeed, the existence of such an engine can provide a good explanation to the lightcurves of some FBOTs \citep{Yu2015,Hotokezaka2017,Rest2018,Margutti2019,Wang2019,Sawada2022}.

In additional, the benchmark FBOT event, AT2018cow, from a nearby luminosity distance $\approx60\,{\rm Mpc}$ \citep{Prentice2018,Perley2019} provided an opportunity for the broad-band observations from radio to $\gamma$-rays \citep{RiveraSandoval2018,Ho2019,Margutti2019,Huang2019}. In particular, the radio observation revealed a dense magnetized environment of AT2018cow, which plausibly supports the existence of a newly formed magnetar \citep{Mohan2020}.

Based the above considerations, in this paper, we collect a large number of FBOTs from the literature and fit their light curves within the framework of the magnetar engine model. The obtained parameters are further compared with those of SLSNe and SNe Ic-BL associated/unassociated with lGRBs. Previously, \cite{Yu2017} had suggested a united scenario to connect SLSNe and SNe Ic-BL. Therefore, in this paper, we will investigate whether such a united understanding can be extended to the FBOT phenomena, which could provide a key rule to the physical origin of the FBOTs. 

\section{Lightcurve Modeling}

\subsection{Sample Selection}

The criteria for our sample selections are as follows: (1) reported rise time above half-maximum $t_{\rm rise}\lesssim 10\,{\rm d}$, (2) spectroscopic redshift measurement from its host galaxy spectral features; (3) published lightcurves observed in at least two filters; (4) at least some data are available close to the peak. 

Following the above criteria, we collect 40 FBOTs from the literature: 7 events from the PS1-MDS \citep{Drout2014,Inserra2019}; 25 events reported in the Dark Energy Survey Supernova Program \citep[DES-SN;][]{Pursiainen2018}; 3 events collected from the Supernova Legacy Survey \citep[SNLS;][]{Arcavi2016}; 2 events discovered by the Hyper Suprime-Cam SSP (HSC–SSP) Transient Survey \citep{Tampo2020}; and 3 well studied samples, i.e., PTF10iam, AT2018cow, and Koala, discovered by the Palomar Transient Factory \citep[PTF;][]{Arcavi2016}, the Asteroid Terrestrial-impact Last Alert System (ATLAS) Survey \citep{Prentice2018,Perley2019}, and the Zwicky Transient Facility One-d Cadence (ZTF-1DC) Survey \citep{Ho2020}, respectively.

Our FBOT sample contains 22 robust cases whose rises were recorded in survey projects. For these events, sufficient data on the rise and decline phases of the lightcurve pose a strict constraint on the model parameters, especially for the ejecta mass. The sample also includes 18 events without any detection during the rise phase of the lightcurve. Due to the lack of observational epoch before the peak, the model parameters are related to the rise time we set. 
% Table \ref{tab:ParameterResults} summarizes the detailed observational information for the FBOTs collected in our sample.

\subsection{Magnetar Engine Model} \label{sec:MagnetarPoweredModel}

As usual, the spin-down luminosity of a magnetar can be generally expressed according to the luminosity of magnetic dipole radiation as

\begin{equation}
L_{\rm sd}(t) = L_{\rm sd,i}\left(1 + \frac{t}{t_{\rm sd}}\right)^{-2},
\end{equation}
where $L_{\rm sd,i} =  10^{47}\,{\rm erg}\,{\rm s}^{-1}\,P_{\rm i,-3}^{-4}B_{\rm p,14}^2$ is the initial value of the luminosity, $t_{\rm sd} \simeq 2\times10^5\,{\rm s}\,P_{\rm i,-3}^2B_{\rm p,14}^{-2}$ is the spin-down timescale, and $P_{\rm i}$ and $B_{\rm p}$ are the initial spin period and polar magnetic strength of the magnetar, respectively. The total rotational energy of the magnetar can be written as $E_{\rm rot} = L_{\rm sd,i}t_{\rm sd}= 2\times10^{52}\,{\rm erg}\,P_{\rm i,-3}^{-2}$. Here the conventional notation $Q_x = Q/10^x$ is adopted in cgs units.

We adopt the common analytic solution derived by \cite{Arnett1982} to calculate the bolometric luminosity of an FBOT powered by a magnetar as:

\begin{equation}
\begin{split}
L_{\rm rad}(t) &= e^{-(t/t_{\rm diff})^2}(1-e^{-At^{-2}})\times \\
&\int^t_02L_{\rm sd}(t')\frac{t'}{t_{\rm diff}}e^{(t'/t_{\rm diff})^2}\frac{dt'}{t_{\rm diff}},
\end{split}
\end{equation}
where $t_{\rm diff}$ is the photon diffusion timescale of the FBOT ejecta and $A$ is the leakage parameter. For an ejecta of a mass $M_{\rm ej}$ and velocity $v_{\rm ej}$, the diffusion timescale is given by $t_{\rm diff} = (3\kappa M_{\rm ej}/4\pi v_{\rm ej}c)^{1/2}$, where $\kappa$ is the optical opacity. Here the dynamical evolution of the ejecta is
ignored. The kinetic energy of the ejecta is assumed to
be directly determined by the rotational energy of the
magnetar so that the ejecta velocity can be estimated
as $v_{\rm ej} \simeq \sqrt{2E_{\rm rot} / M_{\rm ej}}$. This assumption is viable as long
as $t_{\rm sd}\lesssim t_{\rm diff}$ and the initial value of the kinetic energy is not much higher than $10^{50}$ erg \footnote{he explosion energy of a typical SNe Ib/c is around $10^{51}$ erg \citep{Ugliano2012,Ertl2016}. However, for FBOTs, if they mostly originate from ultra-stripped SNe as suggested below, then it could be natural to expect a relatively weak explosion. In this case, it would be simple and reasonable to assume that the final kinetic energy of the FBOT ejecta is determined by the injected energy (i.e., the rotational energy of the magnetar). }.
%\textbf{If the initial kinetic energy of the ejecta, $E_{\rm ej}$ is much larger than the rotational energy of the magnetar, that is  $E_{\rm ej } \gg E_{\rm rot}$, the characteristic velocity of the ejecta is not affected by the energy injection from the magnetar. On the contrary, if the magnetar input energy overwhelms the initial kinetic ejecta energy, that is  $E_{\rm ej } \ll E_{\rm rot}$, the ejecta is swept up into a thin shell with a characteristic velocity $v_{\rm ej} \simeq \sqrt{2E_{\rm rot} / M_{\rm ej}}$. Since the ejecta mass of the FBOTs is relatively small, the corresponding initial kinetic energy is also relatively small \citep[e.g.,][]{Tauris2015}, so the latter case is more likely. In our calculations, we assume that the final kinetic energy of the ejecta is derived from the rotational energy of the magnetar. We will further verify the rationality of this assumption in Section \ref{sec:shape}.} 
By considering that the energy injected into the ejecta could be in the form of high-energy photons, we write the leakage parameter of the ejecta as $A = 3\kappa_\gamma M_{\rm ej}/4\pi v_{\rm ej}^2$, where $\kappa_{\gamma}$ is the opacity for gamma-rays. The frequency-dependence of $\kappa_{\gamma}$ is ignored for simplicity.

% Here the dynamical evolution of the ejecta is ignored. The kinetic energy of the ejecta is assumed to be directly determined by the rotational energy of the magnetar so that the ejecta velocity can be estimated as $v_{\rm ej} \approx \sqrt{2E_{\rm rot} / M_{\rm ej}}$. This assumption is viable as long as $t_{\rm sd}\lesssim t_{\rm diff}$. 

Finally, in order to calculate the monochromatic luminosity of the FBOT emission, we define an photosphere temperature as

\begin{equation}
T_{\rm ph}(t) = \max\left[ \left( \frac{L_{\rm rad}}{4\pi\sigma_{\rm SB}v_{\rm ph}^2t^2} \right)^{1/4} , T_{\rm floor} \right]
\end{equation}
with the Stefan-Boltzmann constant $\sigma_{\rm SB}$, floor temperature $T_{\rm floor}$, and photospheric velocity $v_{\rm ph}\simeq v_{\rm ej}$ (which is a standard approximation in the literature).

\begin{deluxetable}{cccl}
\tablecaption{Fitting Parameters and Priors \label{tab:PriorsParameters}}
\tablecolumns{4}
\tablewidth{0pt}
\tablehead{
\colhead{Parameter} &
\colhead{Min} &
\colhead{Max} &
\colhead{Prior}
}
\startdata
$M_{\rm ej}/M_{\odot}$ & $0.001$ & $20$ & Log-flat\\
$P_{\rm i}/{\rm ms}$ & $0.5$ & $50$ & Log-flat \\
$B_{\rm p}/{\rm G}$ & $10^{13}$ & $10^{17}$ & Log-flat \\
$\kappa/{\rm cm}^{2}\,{\rm g}^{-1}$ & $0.01$ & $0.2$ & Flat\\
$\kappa_\gamma/{\rm cm}^{2}\,{\rm g}^{-1}$ & $10^{-2}$ & $10^2$ & Log-flat \\
$T_{\rm floor}/10^{3}\,{\rm K}$ & $3$ & $25$ & Flat \\
$t_{\rm shift}/{\rm d}$ & $0$ & $20$ & Flat \\
$A_{V}/{\rm mag}$ & $0$ & $0.5$ & Flat \\
\enddata
\end{deluxetable}

\subsection{Lightcurve Fitting}

By adopting a Markov Chain Monte Carlo method, we use the magnetar engine model described in Section \ref{sec:MagnetarPoweredModel} with the \texttt{emcee}  package \citep{Foreman-Mackey2013} to fit the multi-band lightcurves of the collected FBOTs. For the Milky Way
extinction, we take values from the dust maps of \citep{Schlafly2011} and fix $R_{V}=3.1$.
Because the extinction of the host galaxy is unknown, we set $A_{V}$ as a free parameter, with a uniform distribution prior between $0$ and $0.5$ magnitudes. There are 8 free parameters: ejecta mass $M_{\rm ej}$, initial spin period $P_{\rm i}$, magnetic field strength $B_{\rm p}$, opacity $\kappa$, opacity to high-energy photons $\kappa_\gamma$, floor temperature $T_{\rm floor}$, host extinction $A_V$, and the time of explosion relative to the first observed data point $t_{\rm shift}$. \cite{Ho2021} recently reported 22 FBOTs with spectroscopic observations, most of which were classified as Type Ib/Ibn/IIb SNe or hybrid IIn/Ibn SNe. Furthermore, a fraction of FBOTs were found to be Type Ic SNe \citep[e.g.,][]{Drout2013,De2018}. Although a major fraction of FBOTs lack spectroscopic classifications, we assume that these collected FBOTs could plausibly contain a large amount of helium, carbon or oxygen. Thus, the prior of $\kappa$ for FBOTs is preferably set in a range of $0.05-0.2\,{\rm cm}^2{\rm g}^{-1}$, which is suitable for scattering in ionized helium, carbon or oxygen. For those events without any detection before the peak, we note that the upper limit of the prior for $t_{\rm shift}$ is defined as the time between the pre-explosion non-detection and the first detection. The priors of these fitting parameters are listed in Table \ref{tab:PriorsParameters}.

For each lightcurve fitting, we run the code in parallel using 12 nodes with at least 10,000 iterations where the first 100 iterations are used to burn in the ensemble. We list the fitting results of the derived model parameters in Table \ref{tab:ParameterResults}, while the detailed fittings to the multi-band lightcurves for each event are shown in the Appendix \ref{sec:SampleAndFittingResults}. Generally, the FBOT lightcurves can be well fitted by the magnetar engine model with high quality. For example, we present the posteriors of the fitting parameters for DES16C3gin in Figure \ref{fig:DES16C3gin_corner_fit}.

\section{Results and Discussions}

\subsection{Properties of FBOTs and their possible origins}\label{FBOTresults}

\begin{figure*}
\centering\includegraphics[width = 0.75\linewidth, trim= 0 0 20 0,clip]{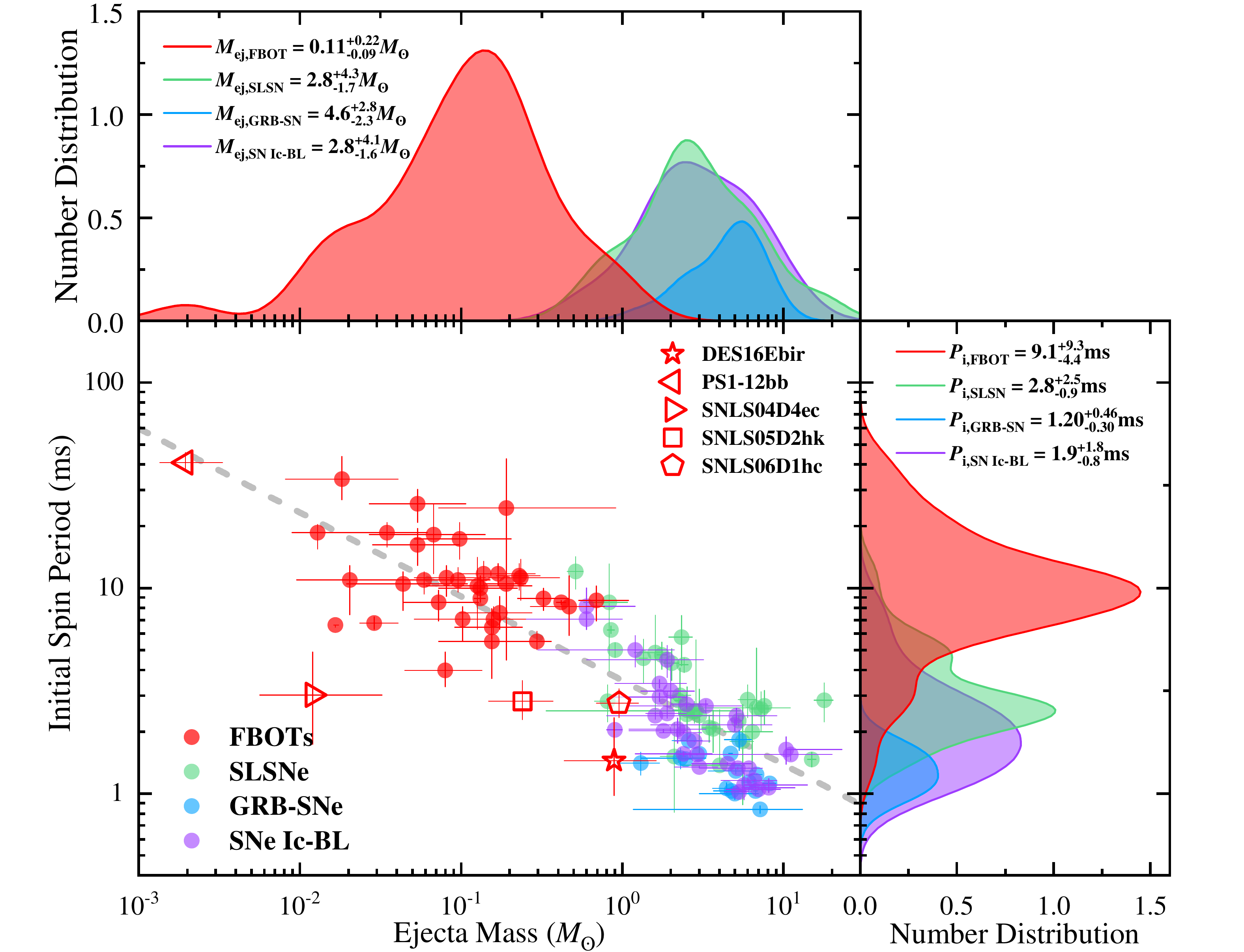}
\caption{Relationship between ejecta masses of FBOTs and initial spin periods of FBOT magnetars (red points). The green, blue, and violet points correspond to the cases of SLSNe \citep{Yu2017}, GRB-SNe \citep{Lv2018}, and SNe Ic-BL \citep{Lyman2016,Taddia2019}, respectively. The best-fitting log-linear relation $P_{\rm i}\propto M_{\rm ej}^{-0.41}$ is shown by the dashed line. The top and right panels display the number distributions of ejecta masses and initial NS periods, derived by the method of the kernel density estimation, for these four types of explosions.}
\label{fig:EjectaMassAndPeriod}
\end{figure*}

\begin{figure*}
\centering\includegraphics[width = 0.75\linewidth, trim= 0 0 20 0,clip]{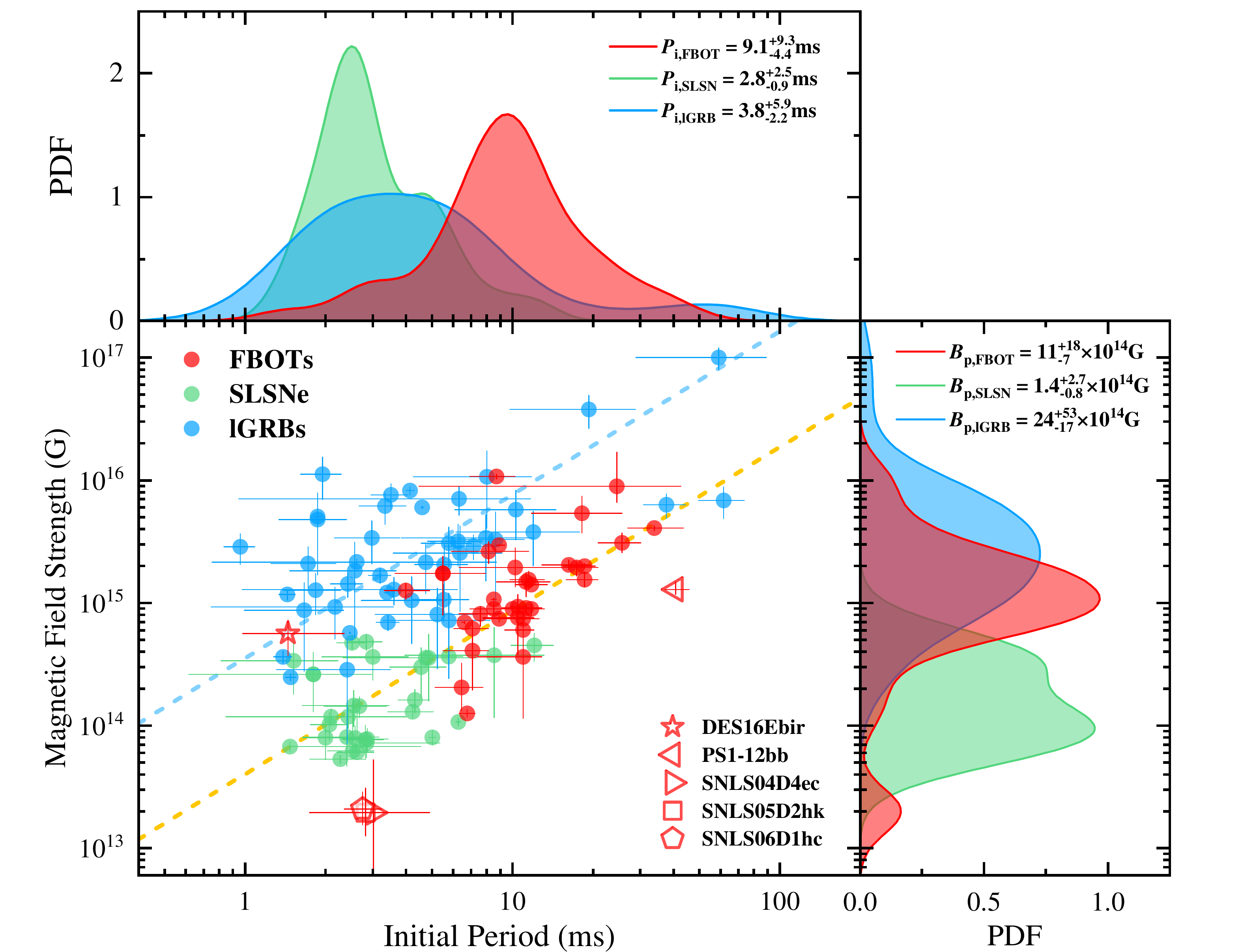}
\caption{Magnetic filed strengths of FBOT magnetars against the initial spin periods (red points). Magnetar parameters for SLSN (green points) and lGRB (blue points) are collected from \cite{Yu2017} and \cite{Lv2014}, respectively. Orange and blue dashed lines represent the parallel fitting $P_{\rm i}-B_{\rm p}$ anti-correlations of the magnetars for the FBOTs plus SLSNe and lGRBs, respectively. The probability density functions of the magnetic filed strength and initial spin period distributions for these three types of explosions are displayed in the top and right panels, respectively.}
\label{fig:PeriodAndMagneticFieldStrength}
\end{figure*}

The most important outputs of the lightcurve modeling are the ejecta mass $M_{\rm ej}$, initial spin period $P_{\rm i}$, and magnetic field strength $B_{\rm p}$. We plot $M_{\rm ej}$ vs. $P_{\rm i}$ and $P_{\rm i}$ vs. $B_{\rm p}$ in Figures \ref{fig:EjectaMassAndPeriod} and \ref{fig:PeriodAndMagneticFieldStrength}, respectively. The ejecta masses for most of FBOTs we collected are in the range of $\sim0.002-1\,M_\odot$ with a median value of $\sim0.11\,M_\odot$. For the magnetars, the initial spin periods are centered at $\sim9.1^{+9.3}_{-4.4}\,{\rm ms}$. The magnetic field strengths, in which the median value is $\sim 25\,B_{\rm c}$, have a wide distribution mostly between $\sim B_{\rm c}$ and $\sim200\,B_{\rm c}$. Here, $B_{\rm c} = m_e^2c^3/(q\hbar) = 4.4\times10^{13}\,{\rm G}$ represents the Landau critical magnetic field defined by electron mass $m_e$, electron charge $q$ and reduced Planck constant $\hbar$.

Corresponding to our sample selection criterion as $t_{\rm rise}\lesssim10\,{\rm d}$, the upper limit of the ejecta masses of FBOTs can be set to be around $1M_{\odot}$, which hints that FBOTs could have the following types of origins with the formation of a rapidly rotating magnetar: (I) BNS mergers to produce massive NS remnants \citep[i.e., the mergernova model;][]{Yu2013}, (II) mergers of a NS and a WD \citep{Zenati2019}, (III) AICs of WDs, including both single- and double-degenerate cases\footnote{Binary NS or WD mergers and WD AICs can occur in active galactic nucleus accretion disks to drive bright magnetar-powered explosions \citep{zhu2021b,zhu2021a}.} \citep{Yu2019a,Yu2019b}, and (IV) SN explosions of ultra-stripped stars \citep[i.e., ultra-stripped SNe; e.g.,][]{Tauris2015,Hotokezaka2017,Sawada2022}. For Case I, since the derived masses here are generally higher than the masses that can be produced by the BNS mergers \citep[e.g., ][]{Radice2018}, the mergernova model could be ruled out for most FBOTs. Nevertheless, the model could still account for some special sources such as PS1-12bb, which have the fastest evolution and relatively low luminosities that are consistent with the prediction of the mergernova model. Furthermore, if a larger opacity is taken into account which can be caused by lanthanides, then more samples could be classified to the mergernova candidates as their ejecta masses become smaller than $\lesssim 0.01\,M_{\odot}$. In any case, the relatively low event rate of the BNS mergers definitely makes them can only account for a very small faction of the observed FBOTs.  Alternatively, in comparison, the relatively wide range of the ejecta masses of FBOTs most favors the ultra-stripped SN model in close binaries, although it is unclear how newborn NSs formed via this channel can have an initial spin period as high as $P_{\rm i} \sim 2-40\,{\rm ms}$. We infer that the compact companion in a close binary can increase the angular momentum of the ultra-stripped star by the tidal torque, possibly resulting in the rapid rotation of the NS. In addition, for the FBOTs with ejecta masses around $\sim0.1M_{\odot}$, the WD-related models still cannot be ruled out, which have some advantages in explaining the multi-wavelength features of some FBOTs.

\subsection{Connection with SLSNe and SNe Ic-BL}
%\textbf{Some observational results suggest a possible connections between SNe Ib/c and SLSNe. For example, \cite{Milisavljevic2013} found links between the late-time emission properties of a subset of energetic, slow-evolving H-poor SN and SLSNe.} 
It has been widely suggested that SLSNe and SNe Ic-BL associated/unassociated with lGRBs, at least a good fraction of them, are also driven by millisecond magnetars \citep[e.g.,][]{Kasen2010,Lv2014,Mazzali2014,Metzger2015,Kashiyama2016,Yu2017,Liu2017,Nicholl2017,Lv2018}. Therefore, it is necessary and interesting to investigate the possible connection and differences between FBOTs and these explosion phenomena, as did in \citep[e.g.,][]{Milisavljevic2013,Yu2017,Margutti2019,Pian2020} for connecting normal SNe Ib/c, SLSNe, and lGRBs. For comparison, we display FBOTs along with SLSNe, GRB-SNe and normal SNe Ic-BL in the $M_{\rm ej}$ vs. $P_{\rm i}$ space  in Figure \ref{fig:EjectaMassAndPeriod}, and in the $P_{\rm i}$ vs. $B_{\rm p}$ space in Figure \ref{fig:PeriodAndMagneticFieldStrength}. 
%\textbf{The multidimensional numerical simulations of neutrino-driven SNe suggest that the upper limit of the initial kinetic energy of ejecta provided by neutrinos $\sim 2 \times 10^{51}$ erg  can hardly explain more energetic explosions\citep{Ugliano2012,Ertl2016}. This upper limit is smaller than the inferred kinetic energy ($E_{\rm ej } \sim 10^{52}$ erg) of some GRB-SNe and normal SNe Ic-BL. 
The $P_{\rm i}$ values for GRB-SNe and normal SNe Ic-BL are collected from \cite{Lyman2016,Lv2018,Taddia2019} by assuming that the kinetic energy of the ejecta is derived from the rotational energy of the magnetar. 

As shown in Figure \ref{fig:EjectaMassAndPeriod}, the combination of the four different types of transients shows a clear universal correlation between ejecta mass and initial spin period as

\begin{equation}
\label{equ:M_ej_P_i}
   P_{\rm i}\propto M_{\rm ej}^{-0.41},
\end{equation}
with a Pearson correlation coefficient $\rho = -0.84$, which is well consistent with the results discovered by \cite{Yu2017} for the SLSN sample only. This anti-correlation strongly indicates that these explosion phenomena may share a common origin. It can be also found that the clearest criterion defining FBOTs could be their small ejecta masses, which generally corresponds to relatively large initial spin periods because of the strong anti-correlation. Therefore, on the one hand, FBOTs very likely originate from stellar collapses, just having a progenitor much lighter and much more stripped than those of SLSNe and SNe Ic-BL. On the other hand, the $M_{\rm ej}-P_{\rm i}$ anti-correlation indicates that more massive progenitors have larger angular momenta. In Section \ref{FBOTresults}, we suspect that the FBOT progenitors could be ultra-stripped stars in close binary systems, which can be spun up by their compact companions. Following this consideration, the $M_{\rm ej}-P_{\rm i}$ anti-correlation could be a natural result of the interaction between the progenitor and the compact companion \citep[see also][Hu et al. 2022, in Preparation]{Blandchard2020,Fuller2022}. If this hypothesis is true, then it is expected that the progenitors of SLSNe and SNe Ic-BL can also be substantially influenced by a compact companion.

\cite{Yu2017} have found that the primary difference between SLSNe and lGRBs could be the magnetic field strengths of their magnetar engines. Specifically, SLSNe have $B_{\rm c}\lesssim B_{\rm p}\lesssim10B_{\rm c}$, while lGRBs have $ B_{\rm p}\gtrsim 10B_{\rm c}$. Therefore, in \cite{Yu2017}, it was suspected that the ultra-high magnetic fields can play a crucial role in launching a relativistic jet to produce GRB emission. Here, it is however found that the surface magnetic fields of more than half FBOTs can be higher than $10B_{\rm c}$, but no GRB has been detected to be associated with FBOTs. One possibility is that the GRB emission associating these FBOTs is highly beamed and the emission beam largely deviates from the line of sight. This, however, is disfavored by the difference between the event rate density of FBOTs and lGRBs. Therefore, a more promising explanation is that these FBOT magnetars intrinsically cannot produce GRB emission, even though their magnetic fields satisfy $B_{\rm p}>10B_{\rm c}$. The probable reason is that the FBOT magnetars rotate too slowly to provide sufficiently large energy for a relativistic jet. Additionally, in view of the small masses of the FBOT ejecta, the possible fallback accretion onto the magnetar is also potentially weak and thus cannot help to launch the jet.

Finally, in view of the significant similarity between FBOTs and SLSNe, it would be reasonable to regard them as an unified phenomenon with different  progenitor masses. From this view, the separation between FBOTs and SLSNe is empirical but does not imply fundamentally different physics. For example, the FBOTs, DES16E1bir ($M_{\rm ej} \sim 0.9\,M_\odot$) and SNLS06D1hc ($M_{\rm ej} \sim 1\,M_\odot$), in our sample could in fact be classified into SLSNe. In any case, by combining with the FBOT and SLSN data, we can find a weak correlation between $P_{\rm i}$ and $B_{\rm p}$, as presented in Figure \ref{fig:PeriodAndMagneticFieldStrength}. Such a correlation could also exist in the lGRB data, but with a shift in $B_{\rm p}$. This indicates that GRB magnetars have magnetic fields statistically higher than those of SLSNe and FBOTs for the same initial spin period $P_{\rm i}$.

\subsection{Shape of Lightcurves} \label{sec:shape}

\begin{figure*}
    \centering
	\includegraphics[width = 0.49\linewidth , trim = 75 30 35 60, clip]{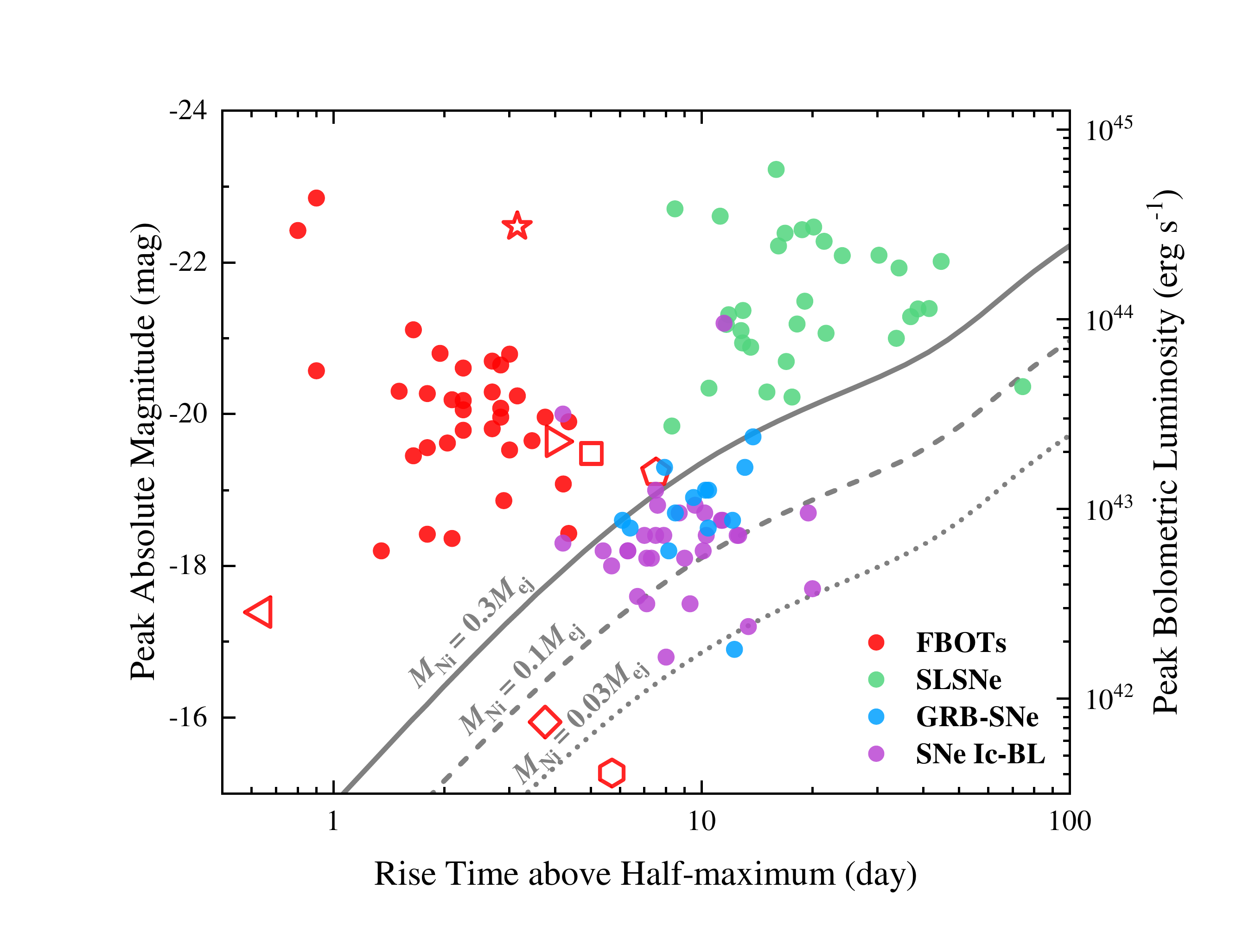}
	\includegraphics[width = 0.49\linewidth , trim = 75 30 35 60, clip]{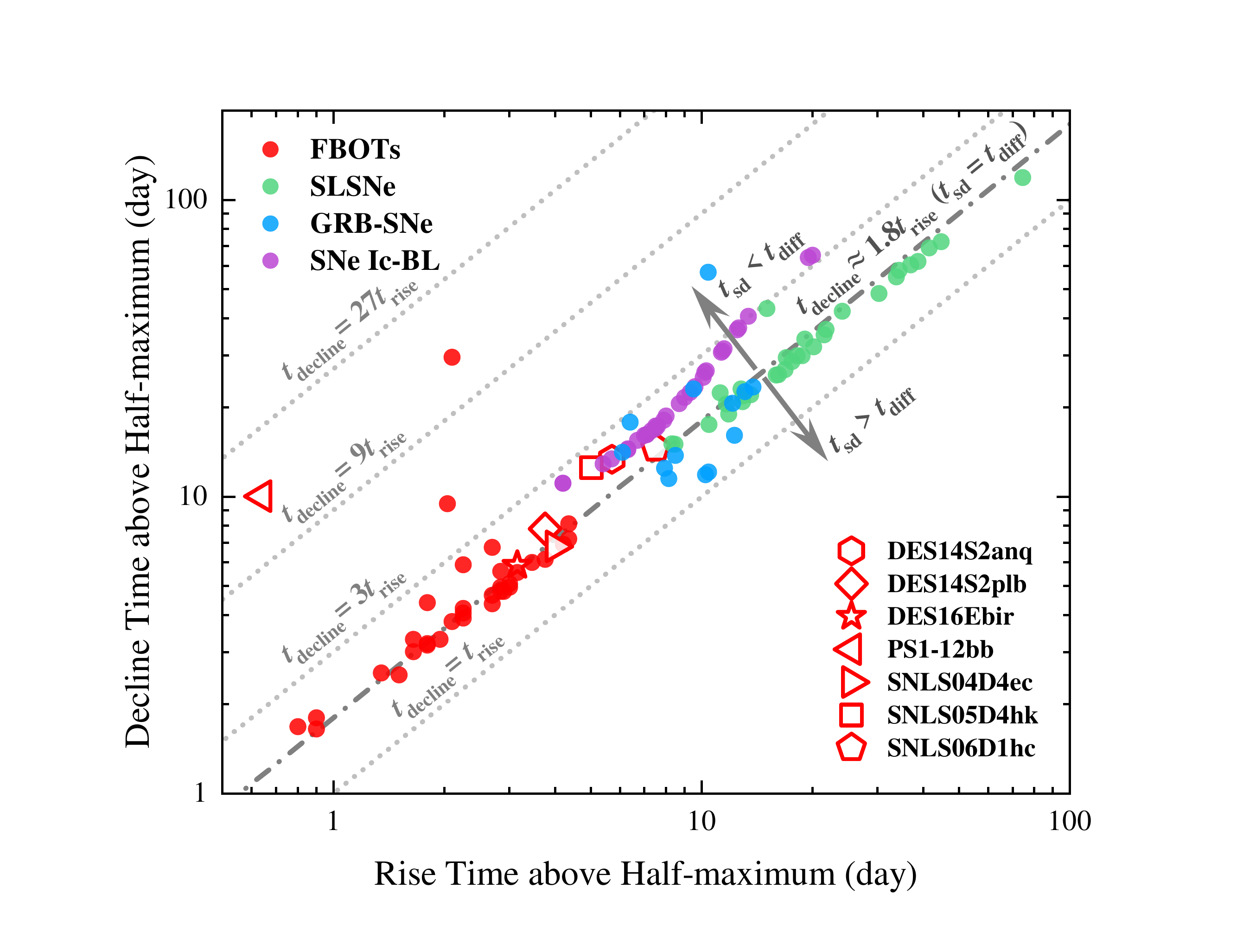}	
    \caption{Left panel: peak absolute magnitudes against the rise times for FBOTs (red points), SLSNe (green points), GRB-SNe (blue points) and SNe Ic-BL (purple points). The solid, dashed and dotted lines are the conditions of $M_{\rm Ni} = 0.3M_{\rm ej}$, $0.1M_{\rm ej}$ and $0.03M_{\rm ej}$, respectively. Right panel: decline times against the rise times for these four types of explosions. The dashed-dotted line, i.e., $t_{\rm decl}\approx 1.8t_{\rm rise}$, corresponds to $t_{\rm sd} = t_{\rm diff}$. The dotted lines represent the different relationships between these two timescales as labeled. }\label{fig:RiseTime}
\end{figure*}

According to the parameter values constrained from the fittings, we can calculate the peak absolute magnitude (or peak luminosity), the rise and decline timescales above half-maximum luminosity of the FBOTs, and their $1\sigma$ uncertainties, which are also listed in Table \ref{tab:ParameterResults}.
%The corner plot of the posterior probability distributions of $M_{\rm peak}, t_{\rm rise}$, and $t_{\rm decl}$ of DES16C3gin as an example are presented in Figure \ref{fig:MT_corner_DES16C3gin_v2}.}
These parameters determine the basic shape of the lightcurves of the transients and can be measured directly from observational data, which therefore are very useful for classifying the transients. As presented in the left panel of Figure \ref{fig:RiseTime}, it seems reasonable to set the boundary between FBOTs and SLSNe at $t_{\rm rise}\sim 10\,{\rm d}$ where the data are relatively sparse. So we adopt it as a sample selection criterion. Strictly speaking, it cannot be ruled out that the ambiguous gap between FBOTs and SLSNe could just be a result of selection effects and the distribution between these two phenomenological types of explosion phenomena could in fact be intrinsically continuous.

Generally speaking, FBOTs together with SLSNe can be separated from SNe Ic-BL including GRB-SNe by the separation line 

\begin{equation}
\begin{split}
M_{\text {peak }}=\left\{\begin{array}{l}
a_{1}\tilde{t}_{\rm rise}-b_{1} \log_{10}\tilde{t}_{\rm rise}-c_{1},{~\rm for~}\tilde{t}_{\rm rise} \lesssim 30, \\
a_{2}\tilde{t}_{\rm rise}-b_{2} \log _{10}\tilde{t}_{\rm rise}-c_{2},{~\rm for~}\tilde{t}_{\rm rise} \gtrsim 30,
\end{array}\right.\label{maglim}
\end{split}
\end{equation}
which corresponds to a nickel mass $M_{\rm Ni}=0.3M_{\rm ej}$, where $\tilde{t}_{\rm rise}=t_{\rm rise}/\rm d$ and the numerical coefficients read $a_1=0.083$, $b_1 =5.3$, $c_1=14.94$, $a_2=0.0089$, $b_2=5.2$, and $c_2=12.63$. This separation line is plotted using the $M_{\rm ej}-E_{\rm K}$ relationship derived from the $M_{\rm ej}-P_{\rm i}$ relationship, i.e., Equation (\ref{equ:M_ej_P_i}), by assuming that all the kinetic energy of the ejecta comes from the rotational energy of the magnetar. It is commonly believed that the mass of the $^{56}$Ni synthesized during core-collapse SNe is very difficult to reach a few tens percent of the total mass of the SN ejecta \citep[e.g.,][]{Suwa2015,Saito2022}.  Based on radiation transport calculations, \cite{Ertl2020}  found that current models employing standard assumptions of the explosions and nucleosynthesis predicts radioactive decay powered  light curves that are less luminous than commonly observed SNe Ib and Ic. 
%This means magnetars may be a promising alternative for normal SNe Ib and Ic.} 
So, both FBOTs and SLSNe of magnitudes above Equation (\ref{maglim}) definitely cannot be primarily powered by radioactive decay of $^{56}$Ni and an engine power is required. Nevertheless, some outliers still exist in our sample, e.g., DES14S2anq and DES14S2plb. In comparison, the peak luminosity of SNe Ic-BL is relatively lower, which reduces the energy requirement and, in principle, makes the radioactive power model available. Nevertheless, by considering of the continuous transition between different phenomena, it could still be nature to suggest that the emission of a fraction of SNe Ic-BL including GRB-SNe is also partly powered by the magnetar engine, although the majority of the spin-down energy of the magnetar has been converted to the kinetic energy of the SN ejecta \citep[e.g.,][]{Lin2021,Zhang2022}.

As analyzed in \cite{Yu2015,Yu2017}, the emission fraction of the spin-down energy is primarily determined by the relationship between the timescales of $t_{\rm sd}$ and $t_{\rm diff}$, which can be basically reflected by the ratio of the lightcurve rise to the decline times. As shown in the right panel of Figure \ref{fig:RiseTime}, the FBOT and SLSN data can be well fitted by the line of $t_{\rm decl}\approx1.8t_{\rm rise}$, which corresponds to $t_{\rm sd}=t_{\rm diff}$. This is a reason why we can regard FBOTs and SLSNe as an unified phenomenon. In comparison, the SNe Ic-BL data are obviously in the $t_{\rm sd}<t_{\rm diff}$ region, as expected. For GRB-SNe, although they are usually classified as SNe Ic-BL, their distribution in the $t_{\rm rise}-t_{\rm decl}$ plane is actually more diffusive than normal SNe Ic-BL.

\section{Conclusion}

In this paper, we perform a systematic analysis of multi-band lightcurves of 40 FBOTs using the magnetar engine model and most samples are fitted with high quality. Then, the explosion and magnetar parameters are well constrained. It is found that the median values with $1\sigma$ deviation of ejecta mass and initial spin period of the FBOT magnetars are $M_{\rm{ej}}=0.11^{+0.22}_{-0.09}\,M_\odot$ and $P_{\rm{i}}=9.1^{+9.3}_{-4.4}\,{\rm{ms}}$. The magnetic field strengths $B_{\rm p}$ are mostly between $\sim B_{\rm c}$ and $\sim200B_{\rm c}$ with a median value of $\sim 25B_{\rm c}$. Here, please keep in mind that the value of $M_{\rm{ej}}$ is somewhat dependent on the adoption of the ejecta velocity. If FBOT explosions can be initially as explosive as and even more explosive than normal SNe Ib/c, the value of $M_{\rm{ej}}$ would be systematically increased by a factor of $\sim2$ so that an appropriate diffusion timescale can be kept.

In view of that the star-formation rates and metallicities of the FBOT hosts are consistent with those of SLSNe and SNe Ic-BL including GRB-SNe \citep{Wiseman2020}, we compare the derived parameters of the FBOTs with the other types of extreme stripped-envelope explosions which are potentially driven by magnetar engines too. Consequently, we find a strong continuous anti-correlation between $M_{\rm{ej}}$ and $P_{\rm{i}}$ for FBOTs, SLSNe, GRB-SNe and SNe Ic-BL as $P_{\rm{i}}\propto{M_{\rm{ej}}^{-0.41}}$. A clear criterion to define FBOTs is their small ejecta masses, with an upper limit of ${\sim}1\,M_\odot$, which is around the lower limit of the masses of other explosion phenomena. Furthermore, the magnetic field strengths of the FBOT magnetars span from $\sim B_{\rm c}\lesssim B_{\rm p}\lesssim 10B_{\rm c}$ for SLSN magnetars to $B_{\rm p}\gtrsim 10B_{\rm c}$ for lGRB magnetars. These connections indicate that most FBOTs may share a common origin with SLSNe, lGRBs and normal SNe Ic-BL.
Since the progenitors of FBOTs likely have low masses, we suspect that most FBOTs originate from the collapse of ultra-stripped stars in close binary systems. However, mergernovae and WD-related models are still not ruled out, which could give nature explanations for some special outlier samples.

With the distributions of $t_{\rm rise}$ vs. $M_{\rm peak}$ for these different types of explosions, we know that the FBOT and SLSN data can be separated by a criterion of $t_{\rm rise}\sim10\,{\rm d}$, while  FBOTs together with SLSNe can be separated from GRB-SNe and normal SNe Ic-BL by the line corresponding to $M_{\rm Ni}=0.3M_{\rm ej}$. These criteria can be used to classify FBOTs, SLSNe and SNe Ic-BL in observation.

\acknowledgments

We thank the anonymous reviewer for helpful comments and feedback. We thank Sheng Yang, Ying Qin, Rui-Chong Hu, and Noam Soker for helpful comments, H-J. L$\Ddot{\rm u}$, M. R. Drout and M. Pursiainen for sharing their data. This work is supported by the National SKA program of China (2020SKA0120300), the National Key R\&D Program of China (2021YFA0718500), and the National Natural Science Foundation of China (Grant No. 11833003).

\appendix

\section{FBOT Samples and Fitting Results} \label{sec:SampleAndFittingResults}

The observed data and fitting lightcurves for the FBOTs collected in our sample are presented in Figure \ref{fig:FittingResults}.

\begin{figure}
\centering
\includegraphics[width=5.8cm]{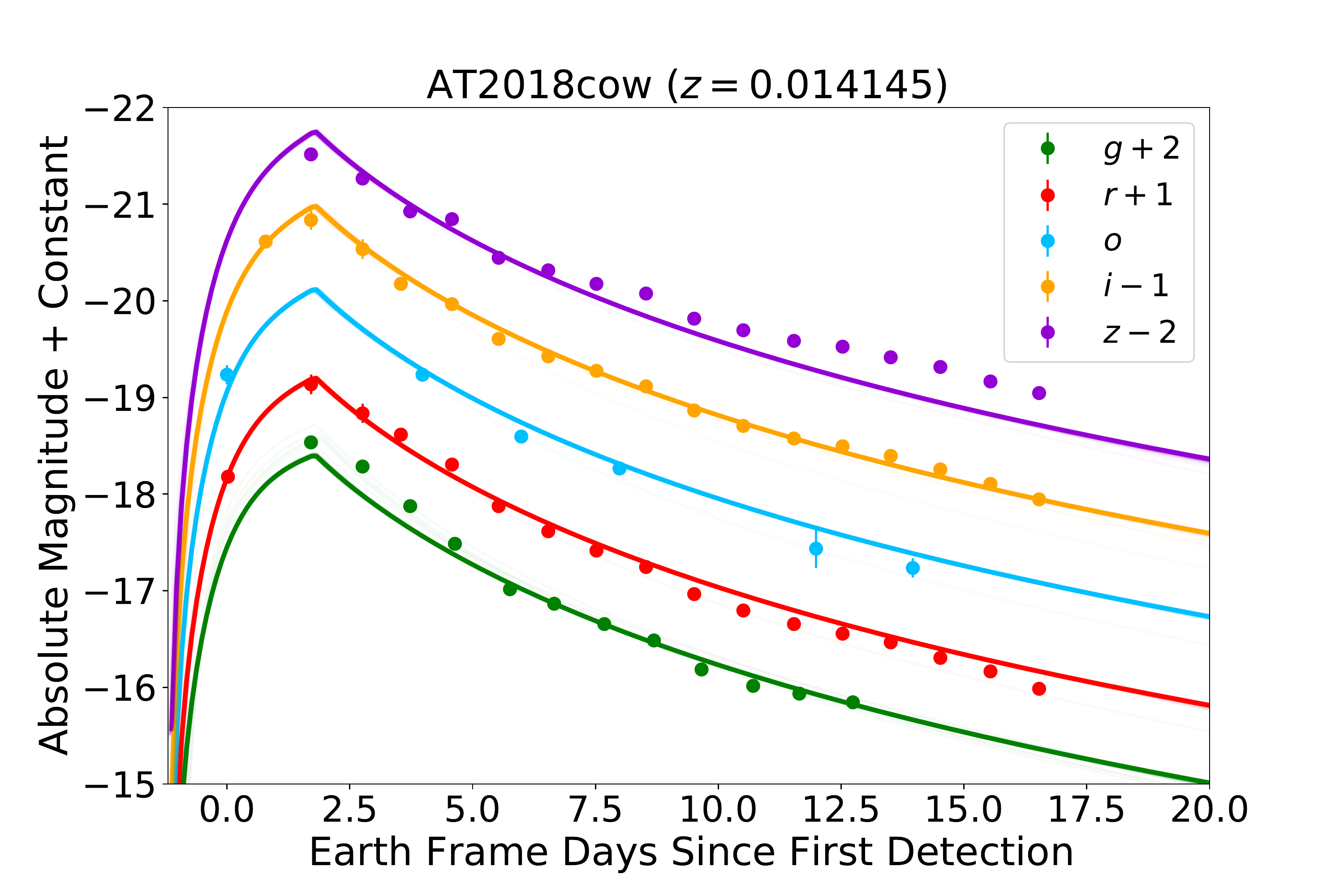}
\includegraphics[width=5.8cm]{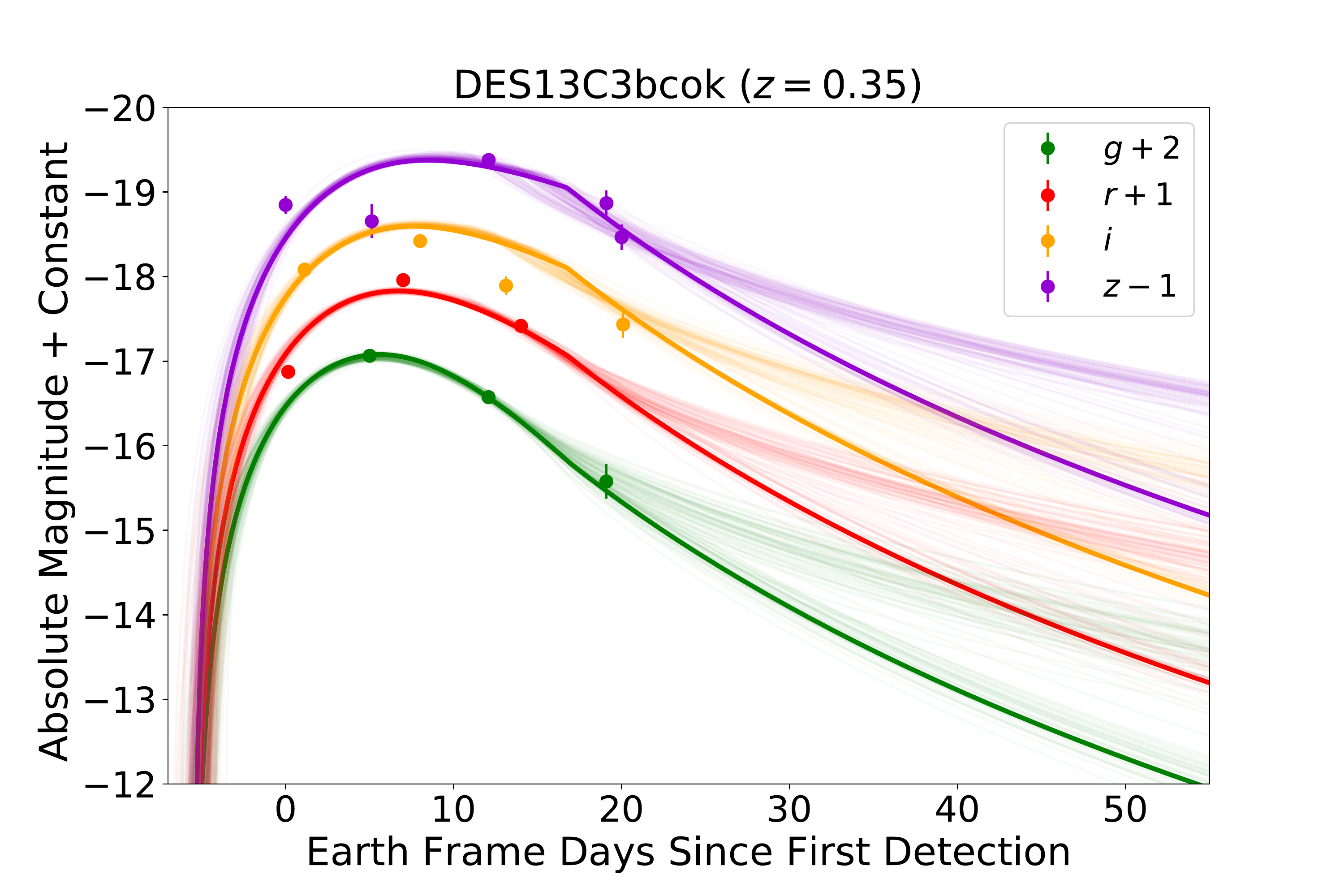}
\includegraphics[width=5.8cm]{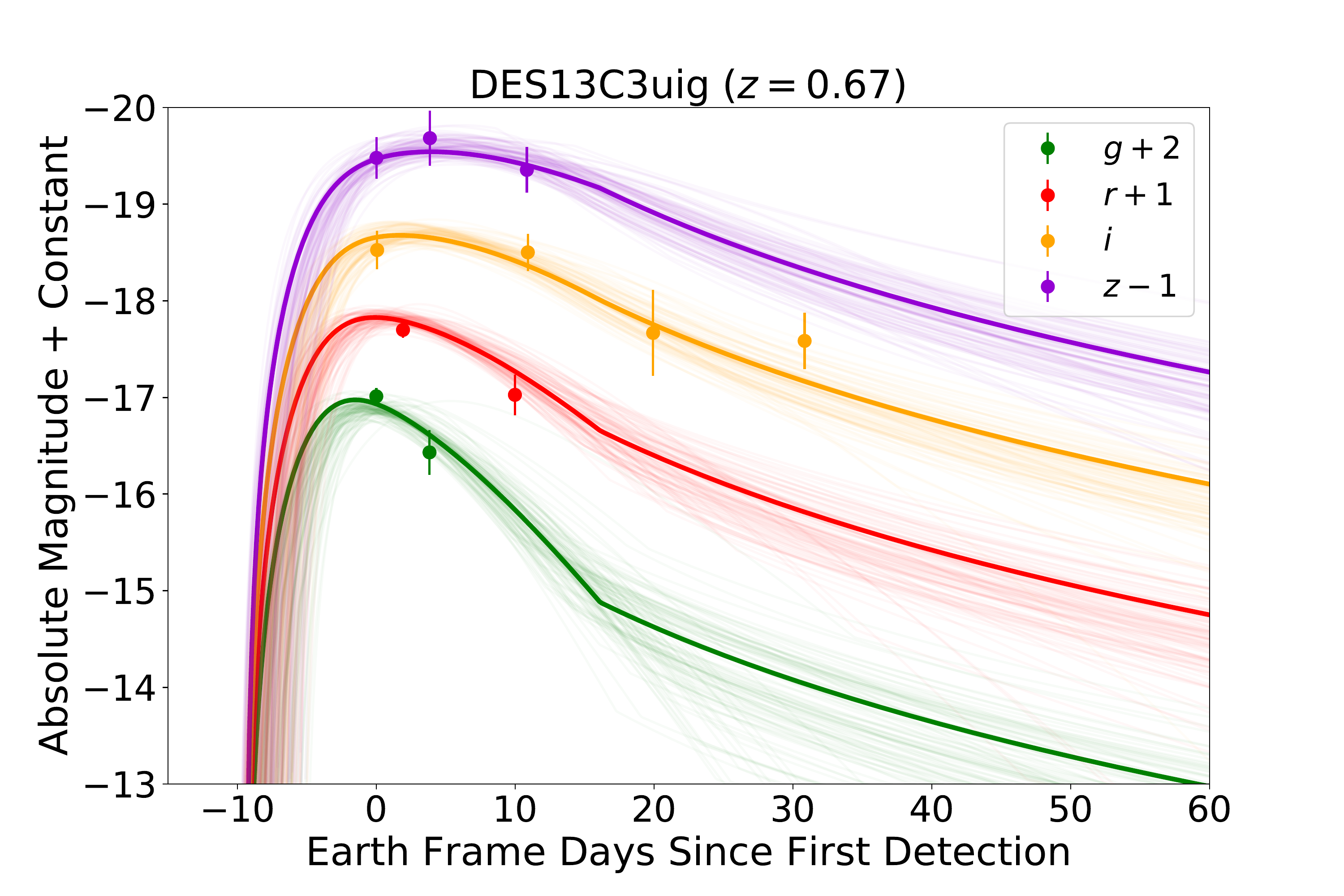}

\includegraphics[width=5.8cm]{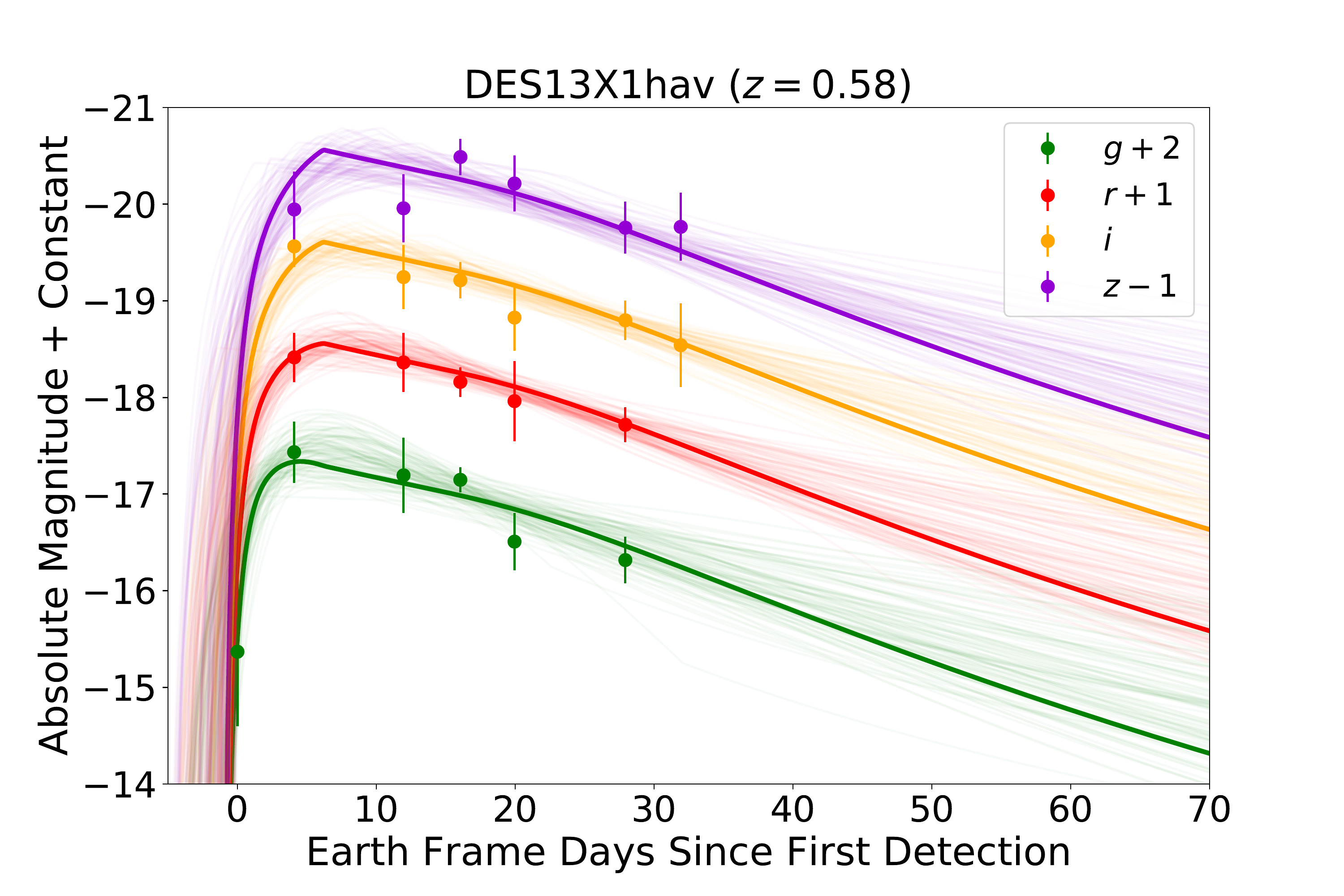}
\includegraphics[width=5.8cm]{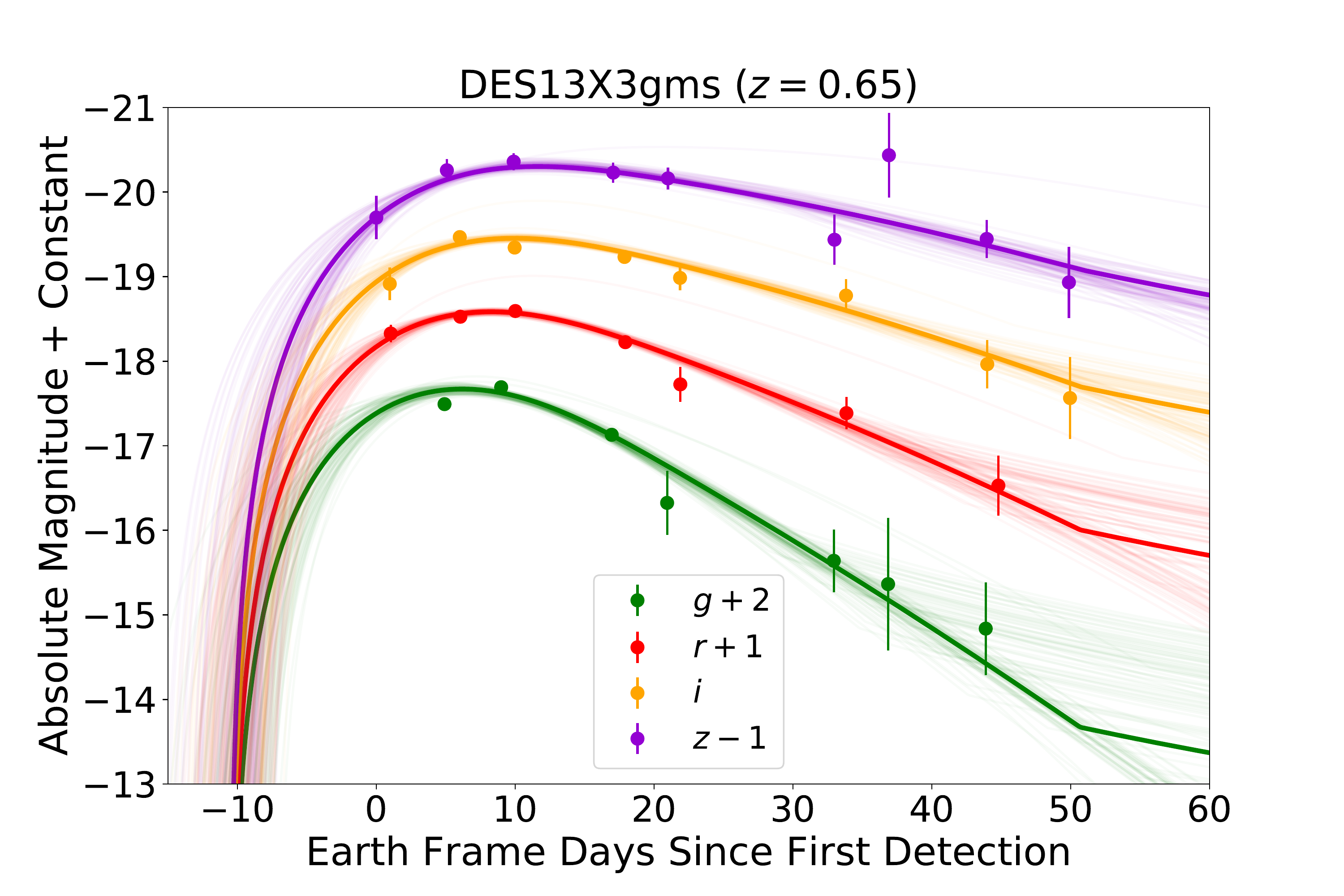}
\includegraphics[width=5.8cm]{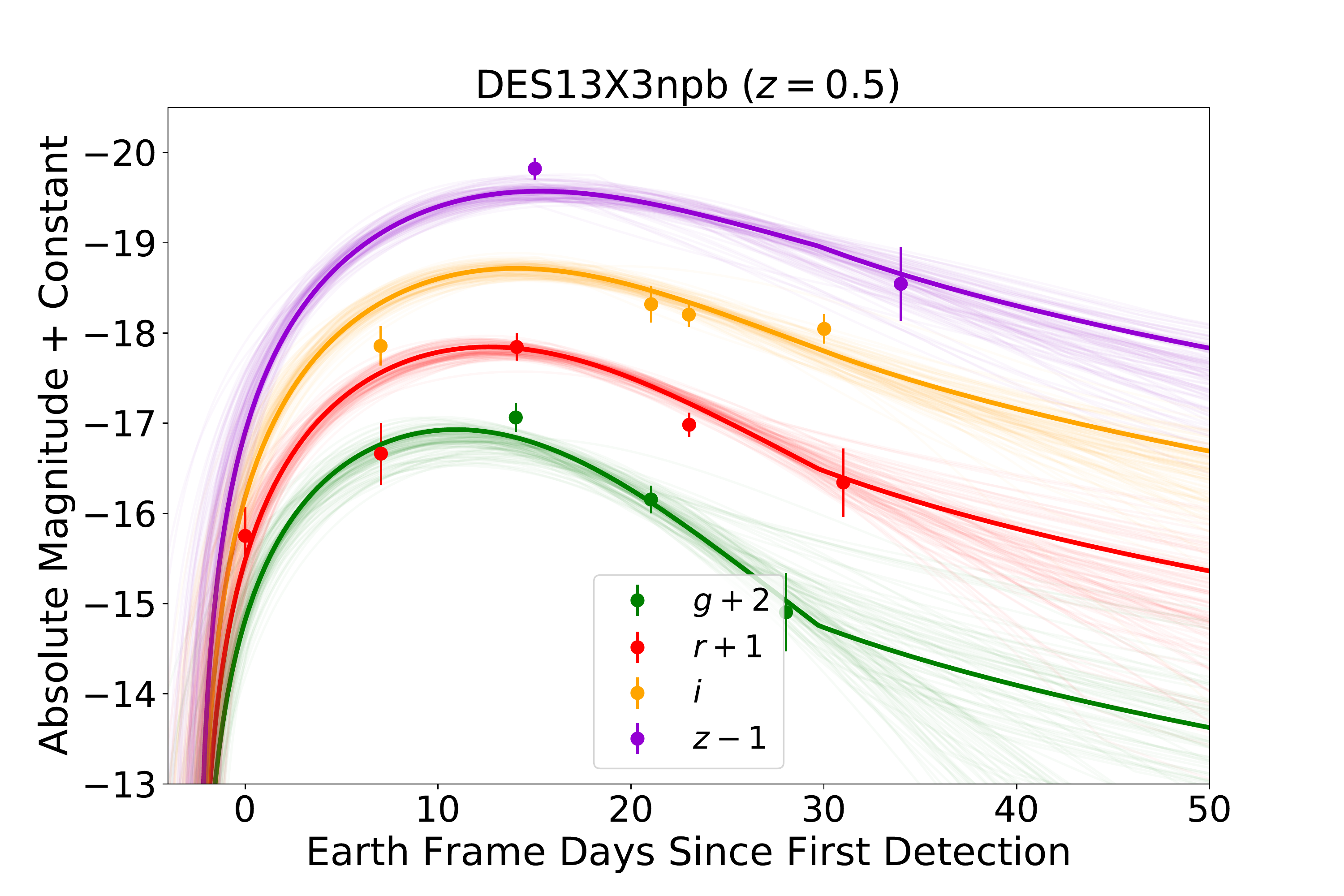}

\includegraphics[width=5.8cm]{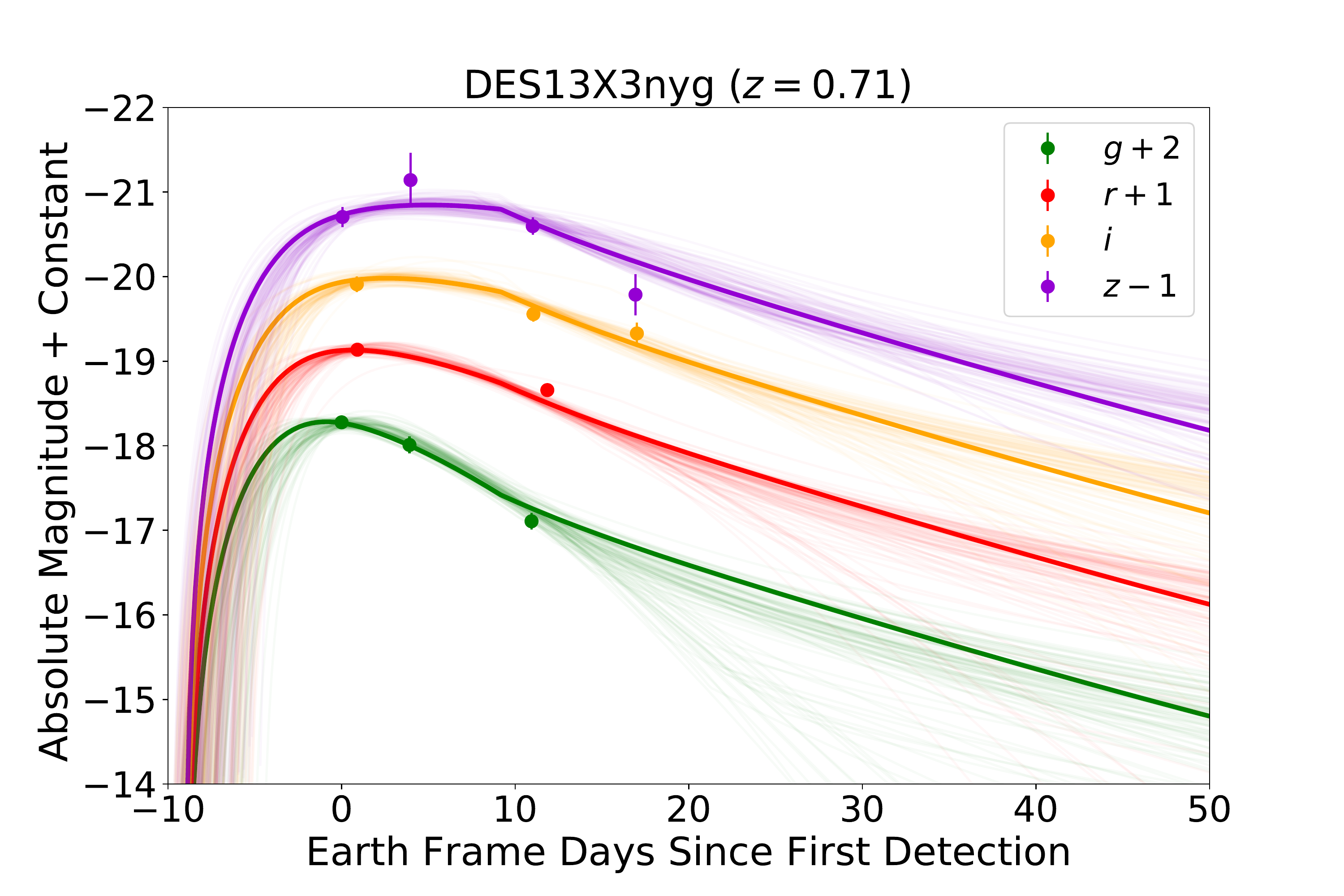}
\includegraphics[width=5.8cm]{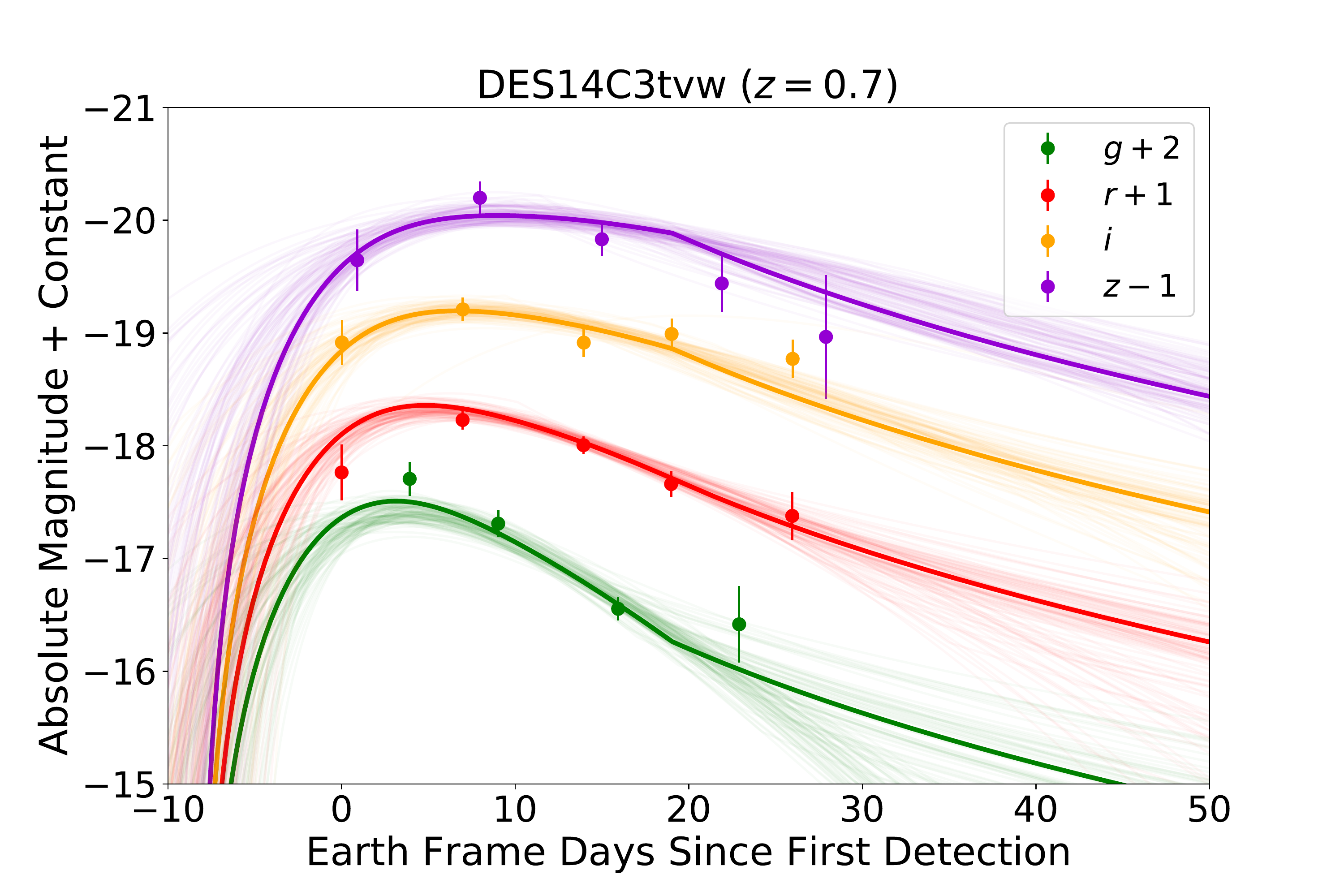}
\includegraphics[width=5.8cm]{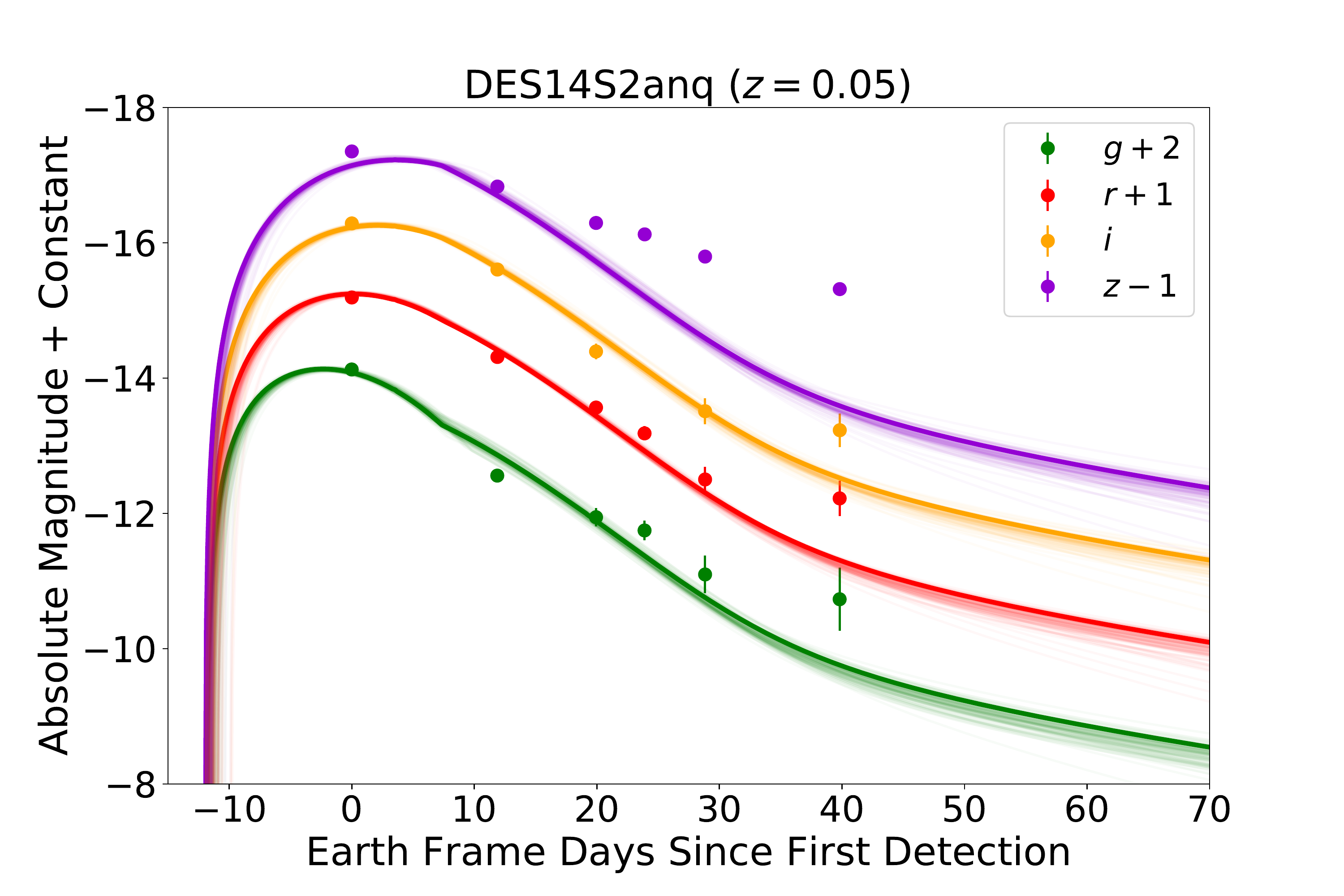}

\includegraphics[width=5.8cm]{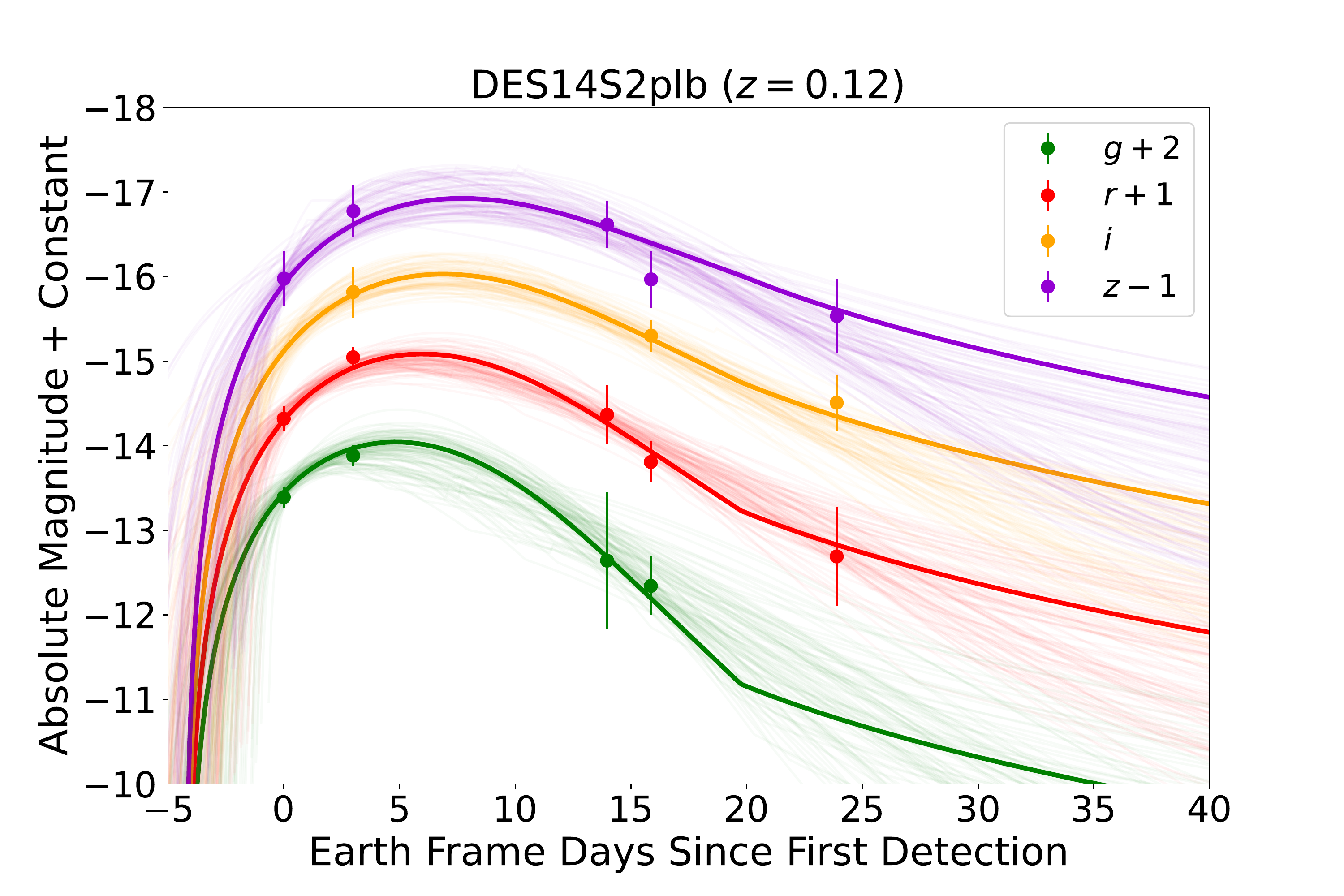}
\includegraphics[width=5.8cm]{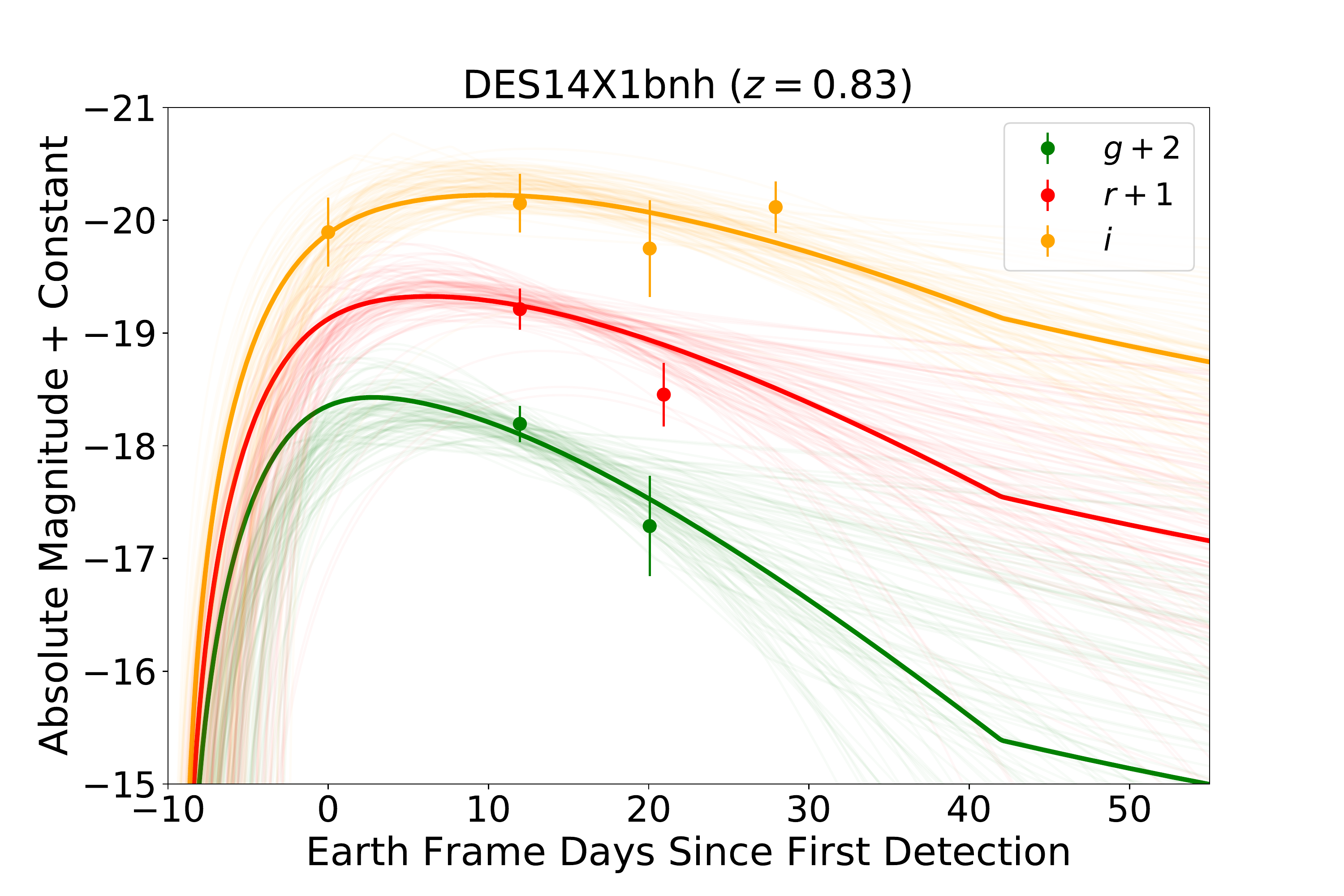}
\includegraphics[width=5.8cm]{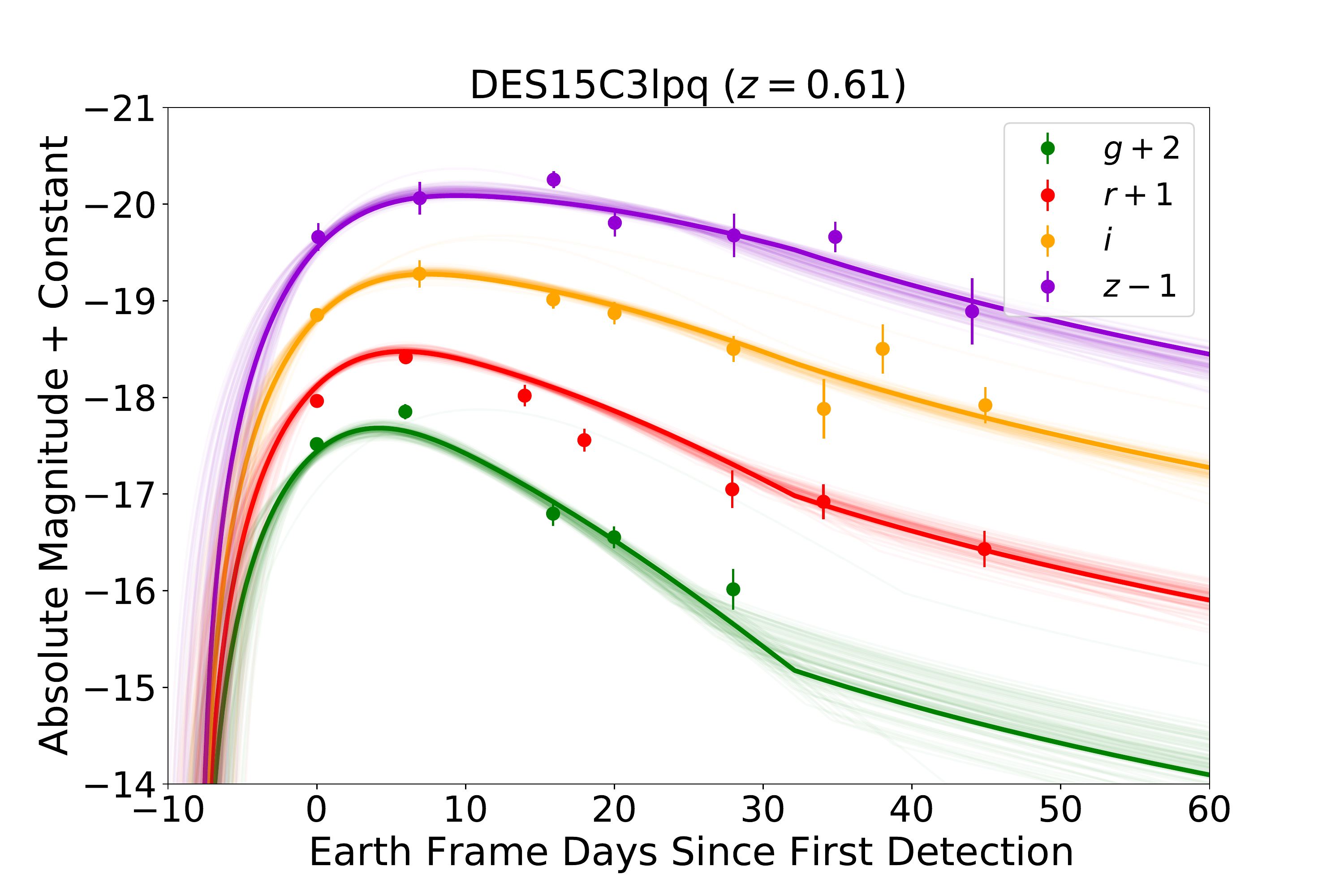}

\includegraphics[width=5.8cm]{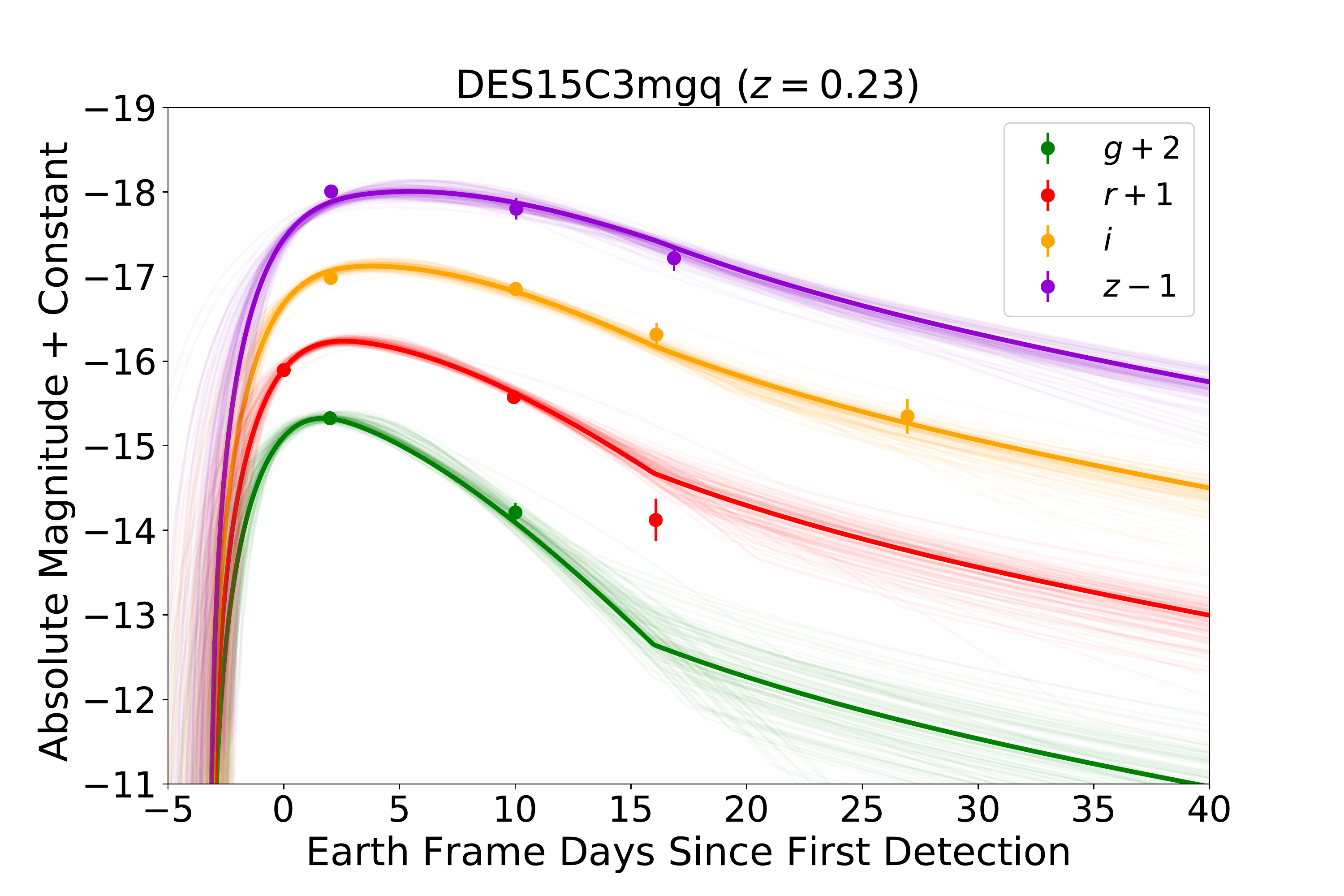}
\includegraphics[width=5.8cm]{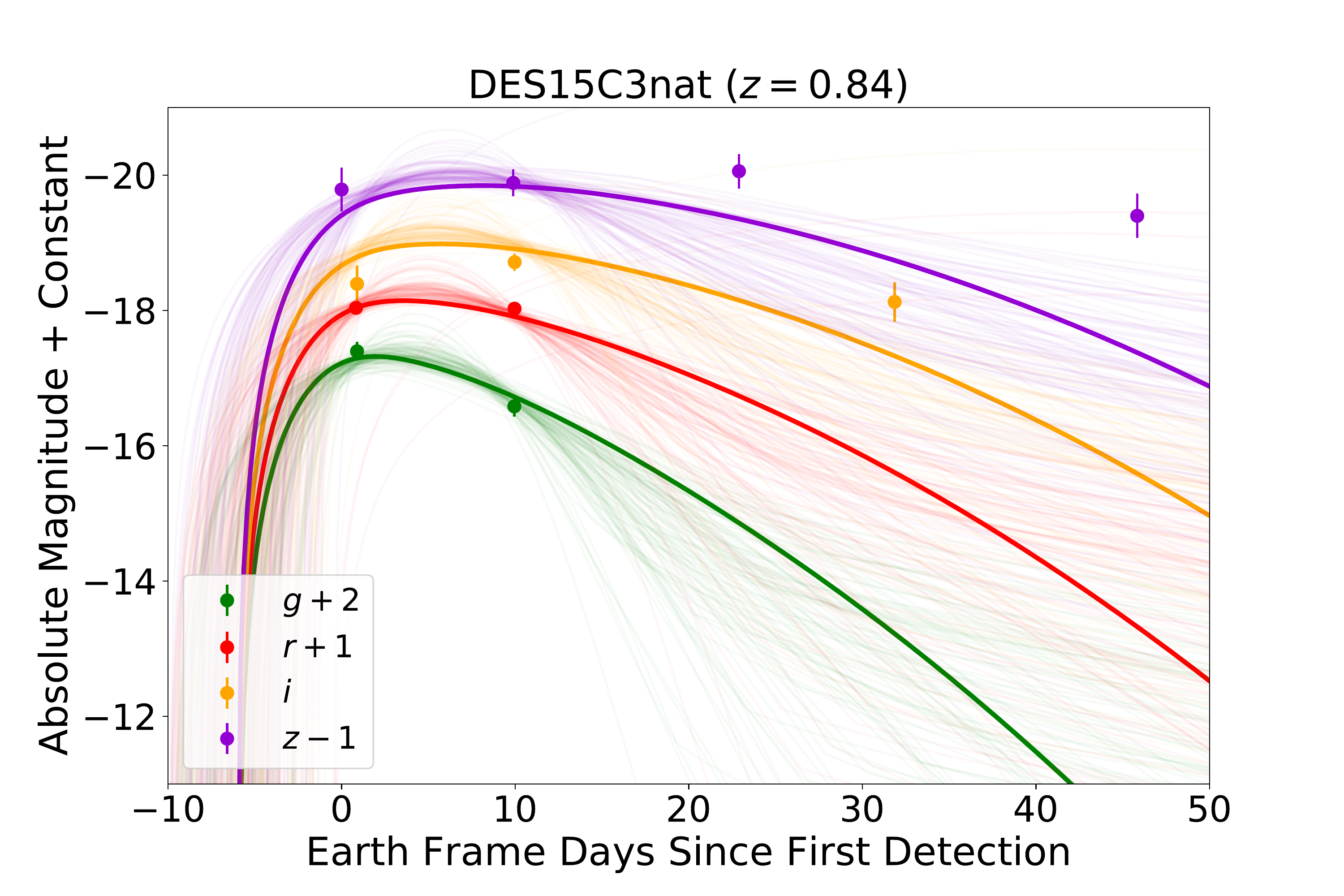}
\includegraphics[width=5.8cm]{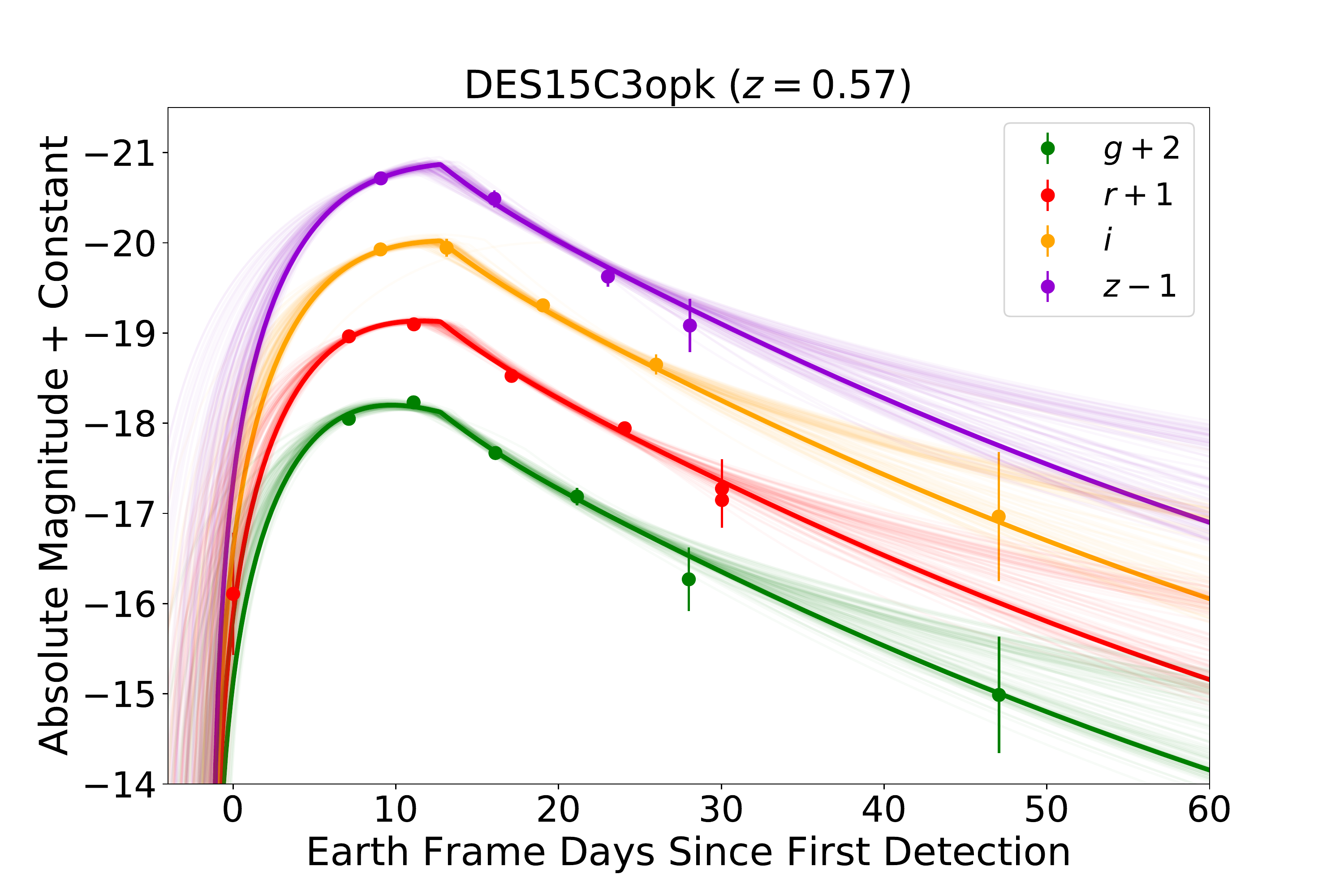}

\includegraphics[width=5.8cm]{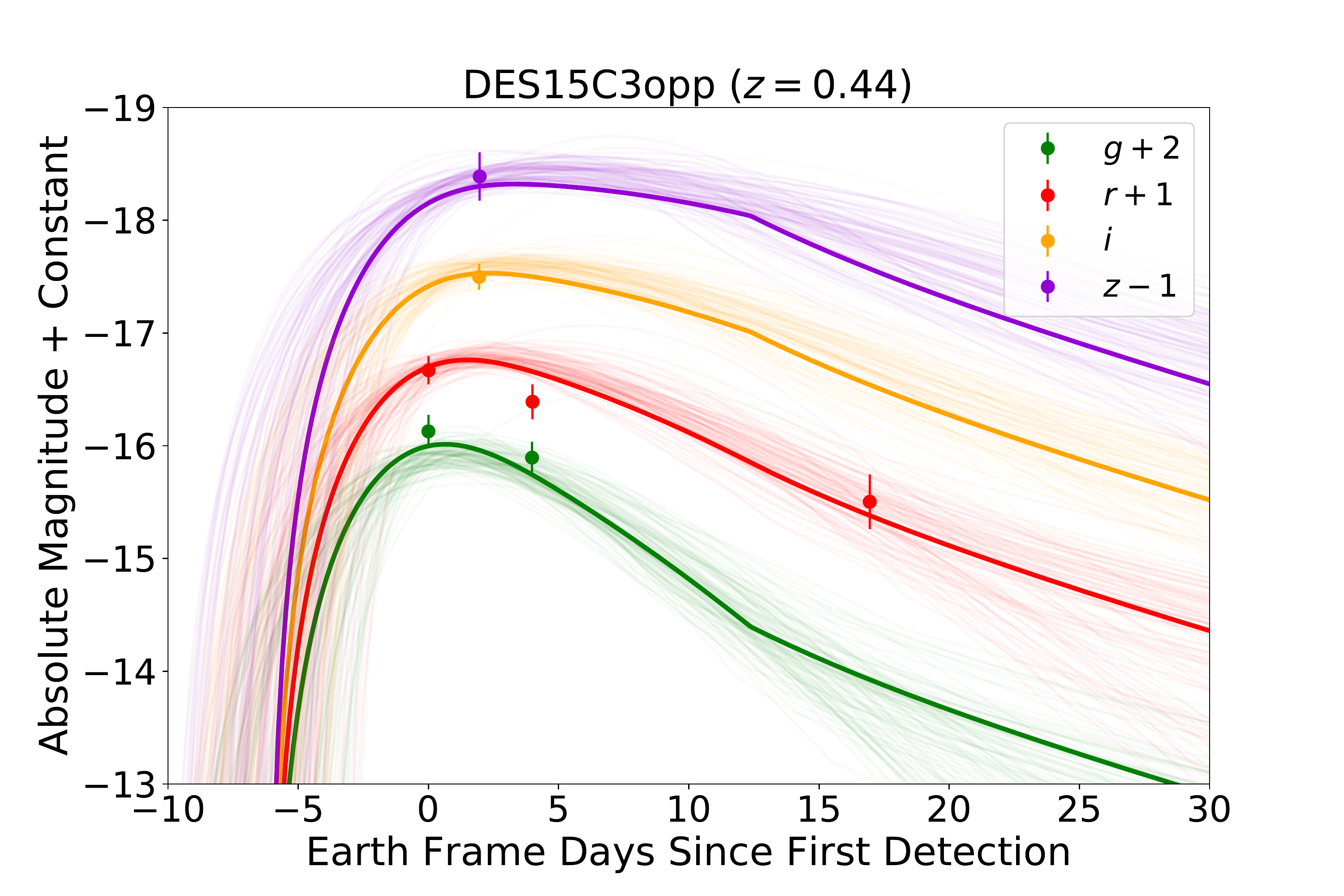}
\includegraphics[width=5.8cm]{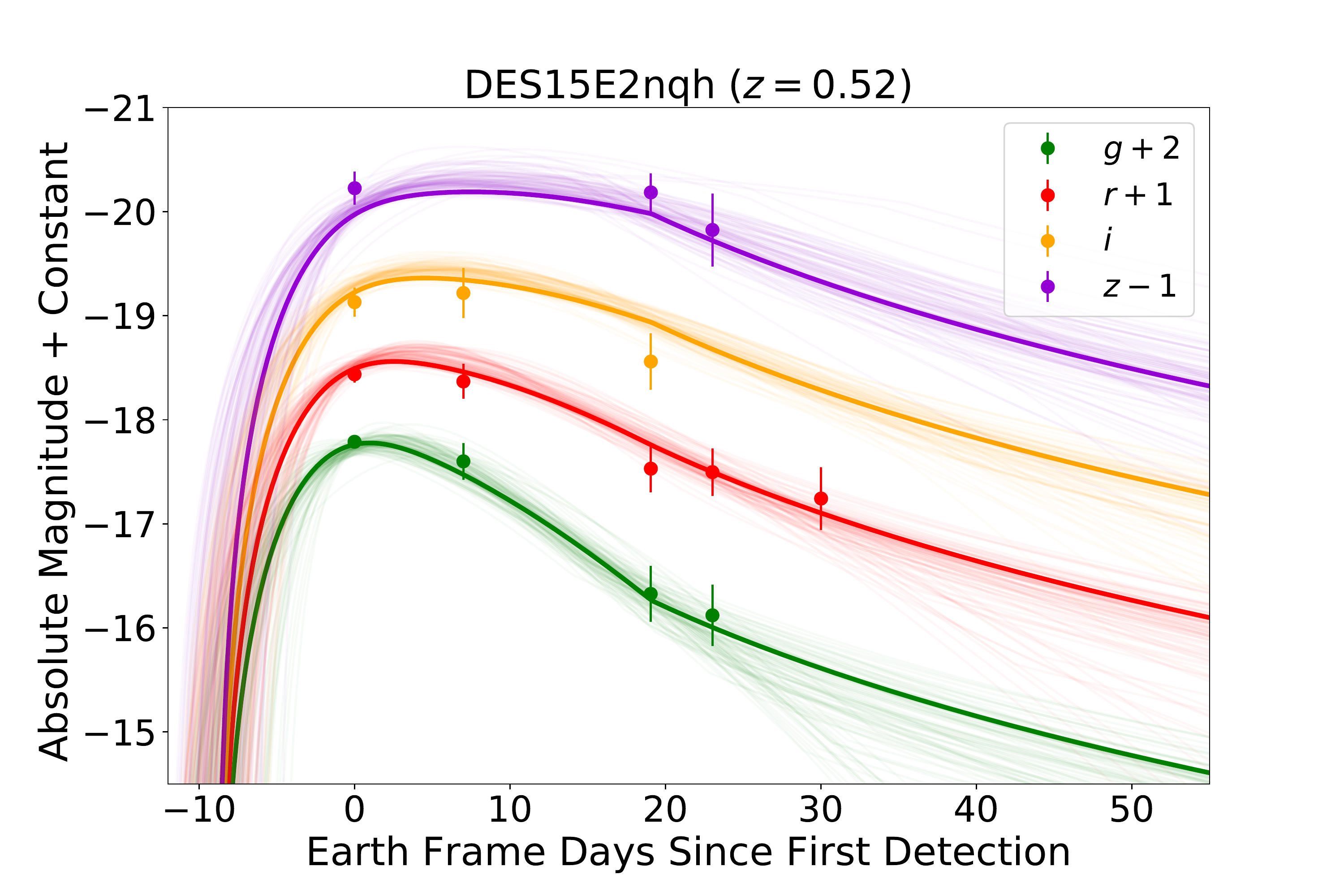}
\includegraphics[width=5.8cm]{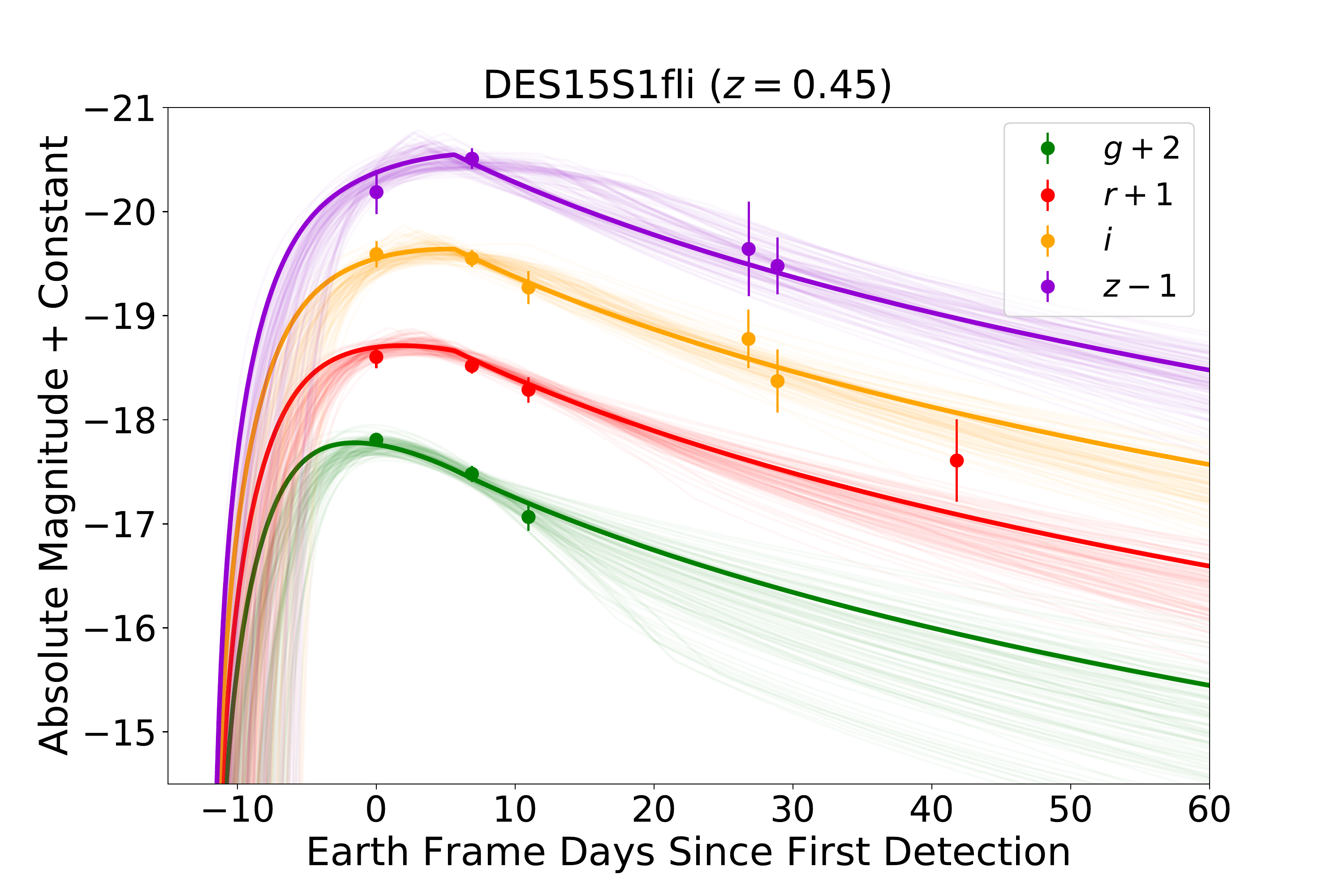}

\caption{Multi-band observations and fittings for the FBOTs by the magnetar-powered model.}
\label{fig:FittingResults}
\end{figure}

\addtocounter{figure}{-1}
\begin{figure}
\includegraphics[width=5.8cm]{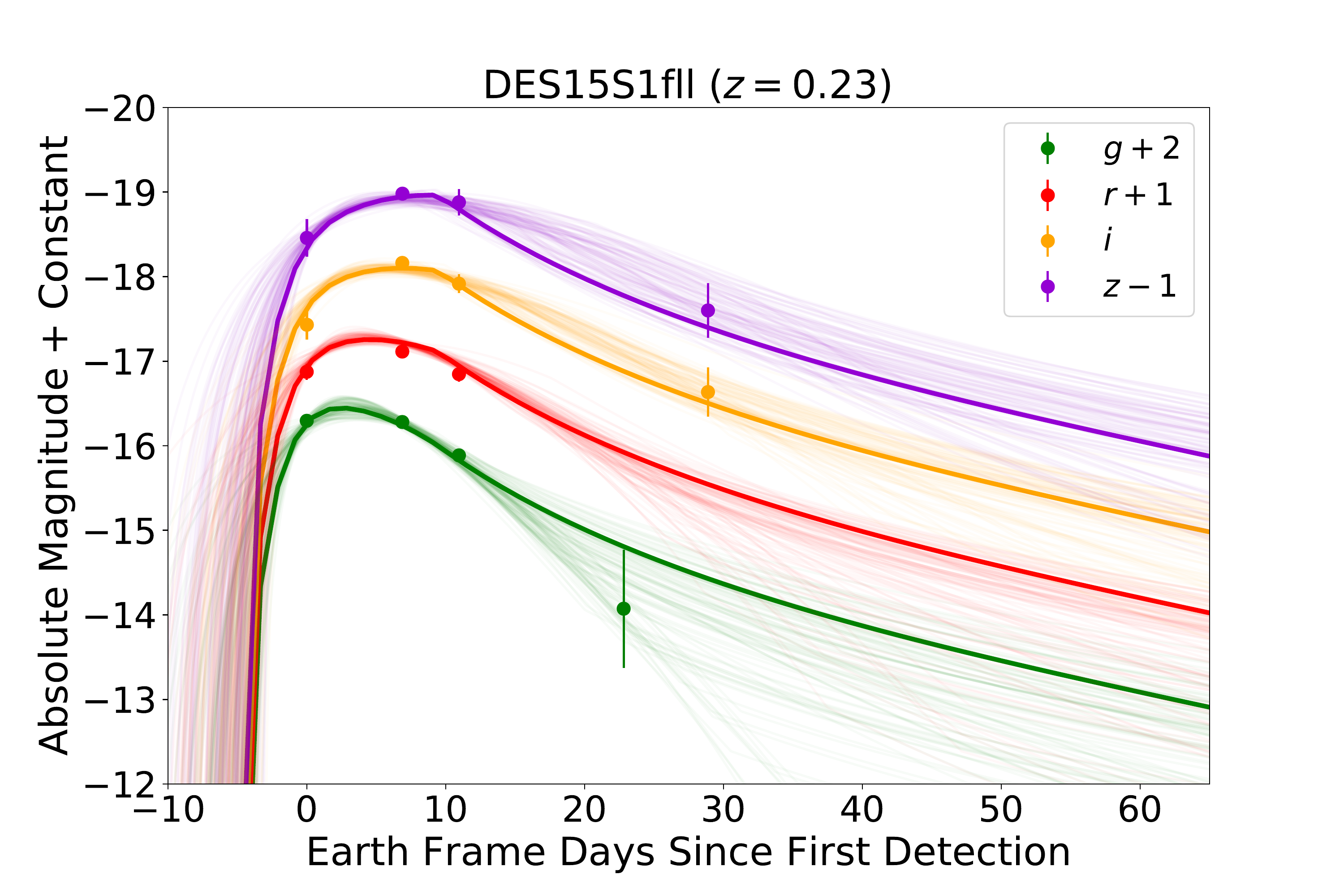}
\includegraphics[width=5.8cm]{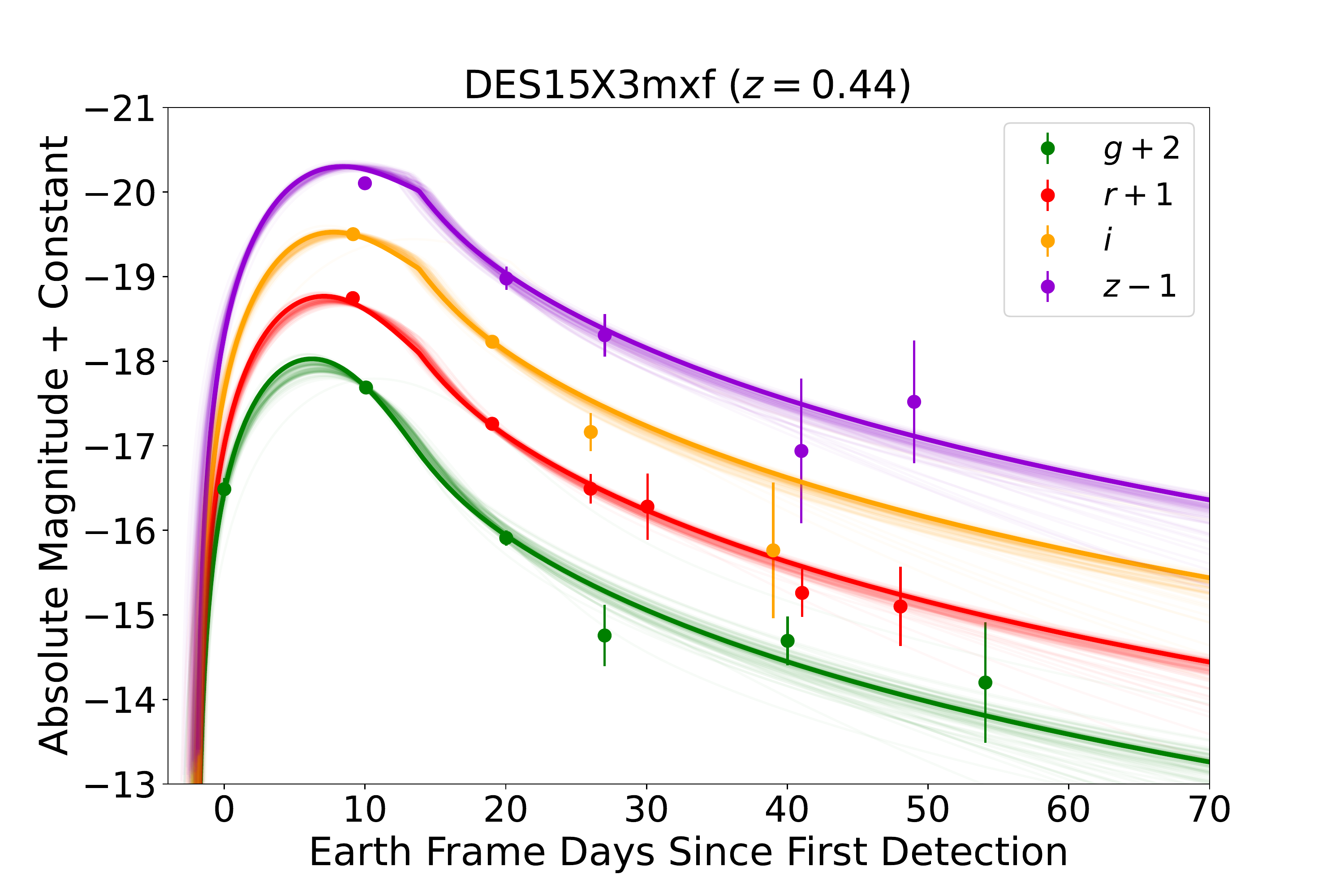}
\includegraphics[width=5.8cm]{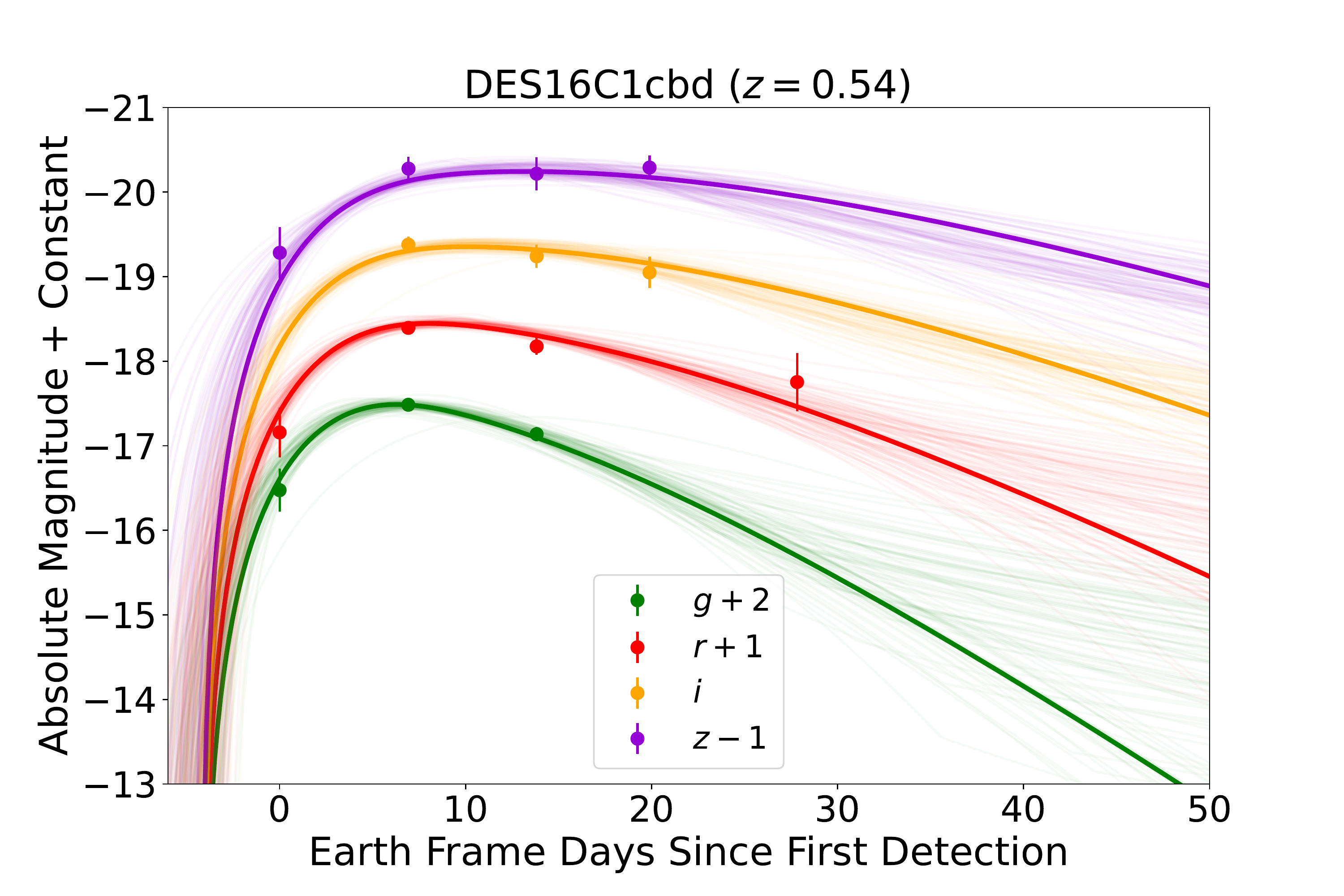}

\includegraphics[width=5.8cm]{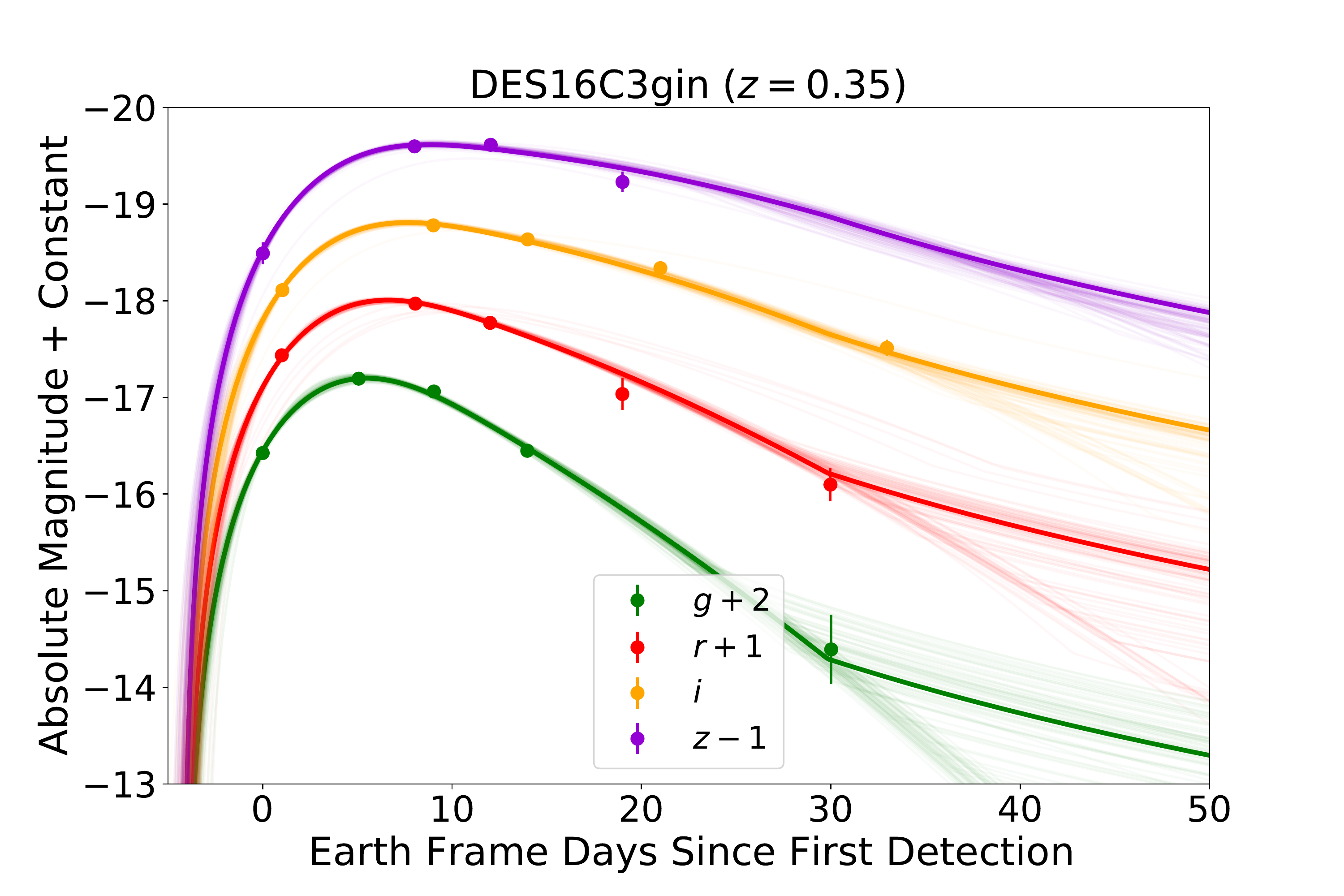}
\includegraphics[width=5.8cm]{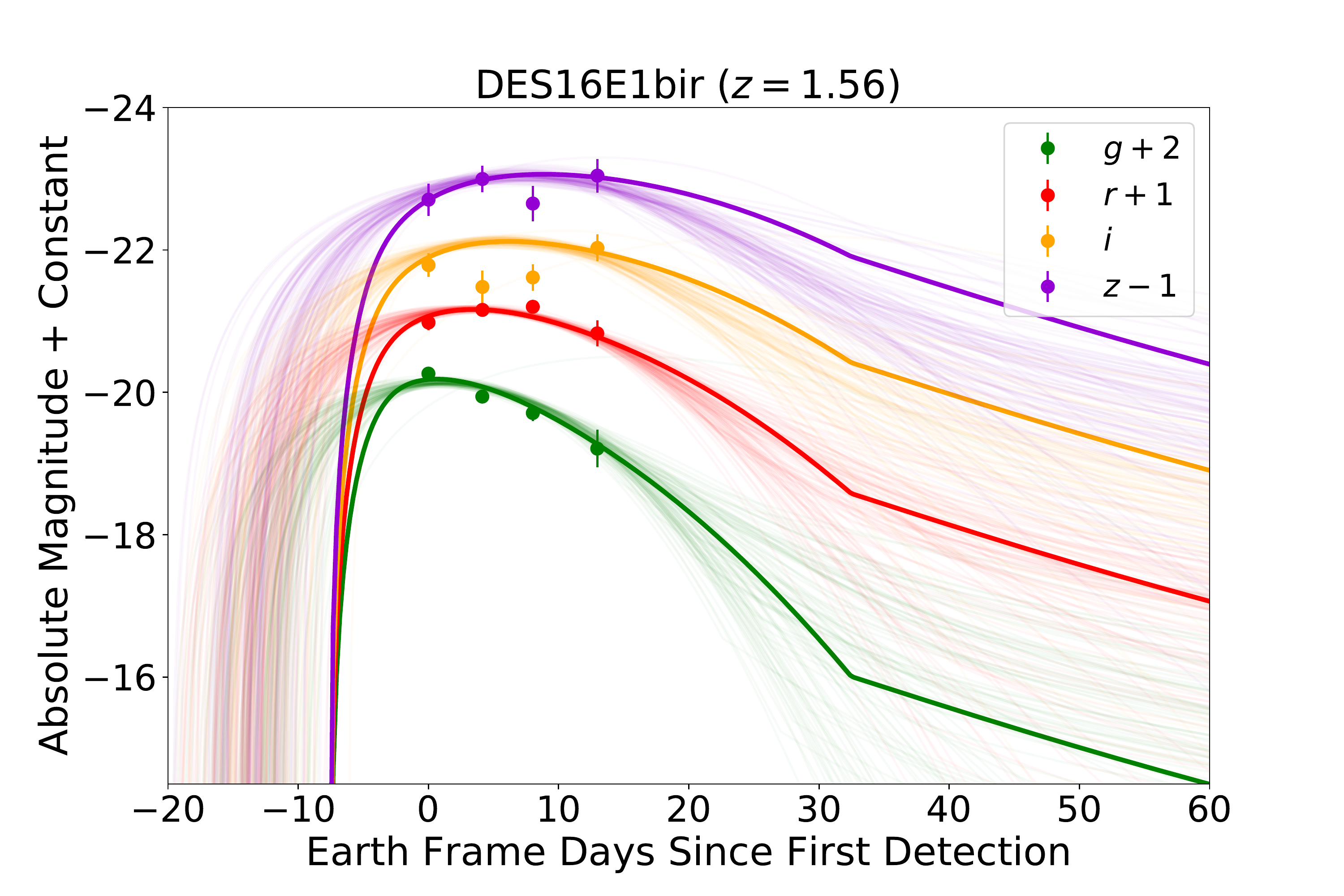}
\includegraphics[width=5.8cm]{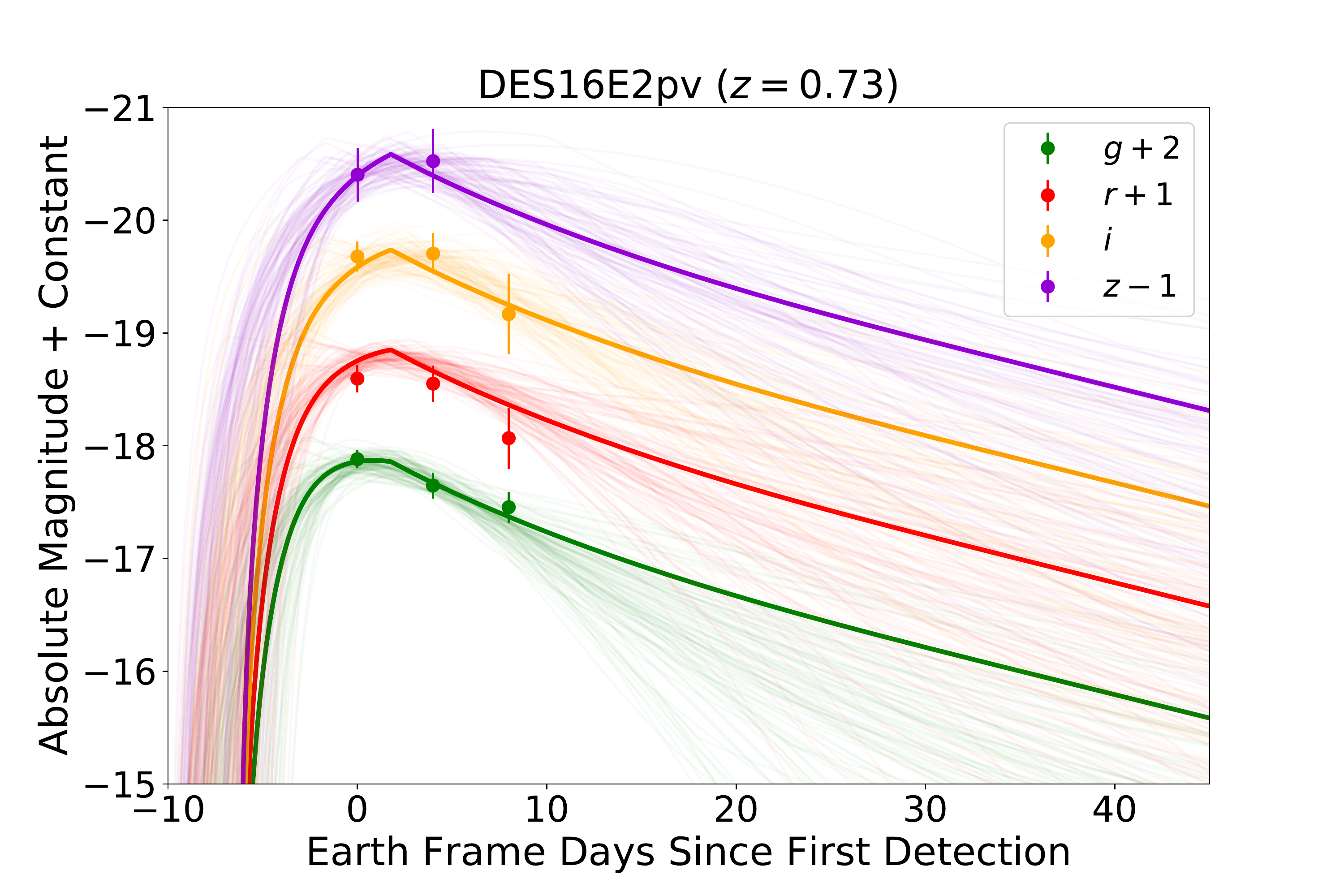}

\includegraphics[width=5.8cm]{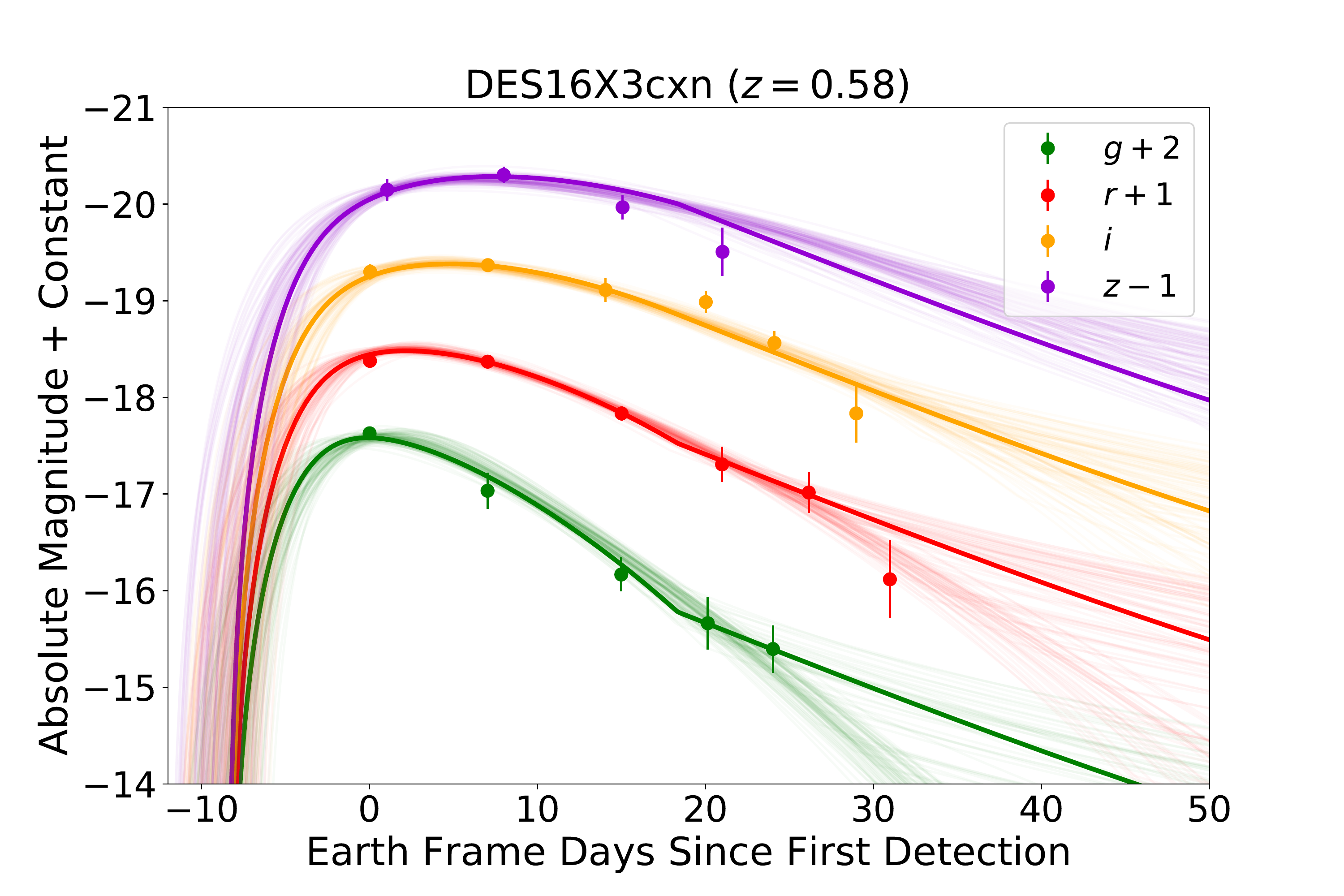}
\includegraphics[width=5.8cm]{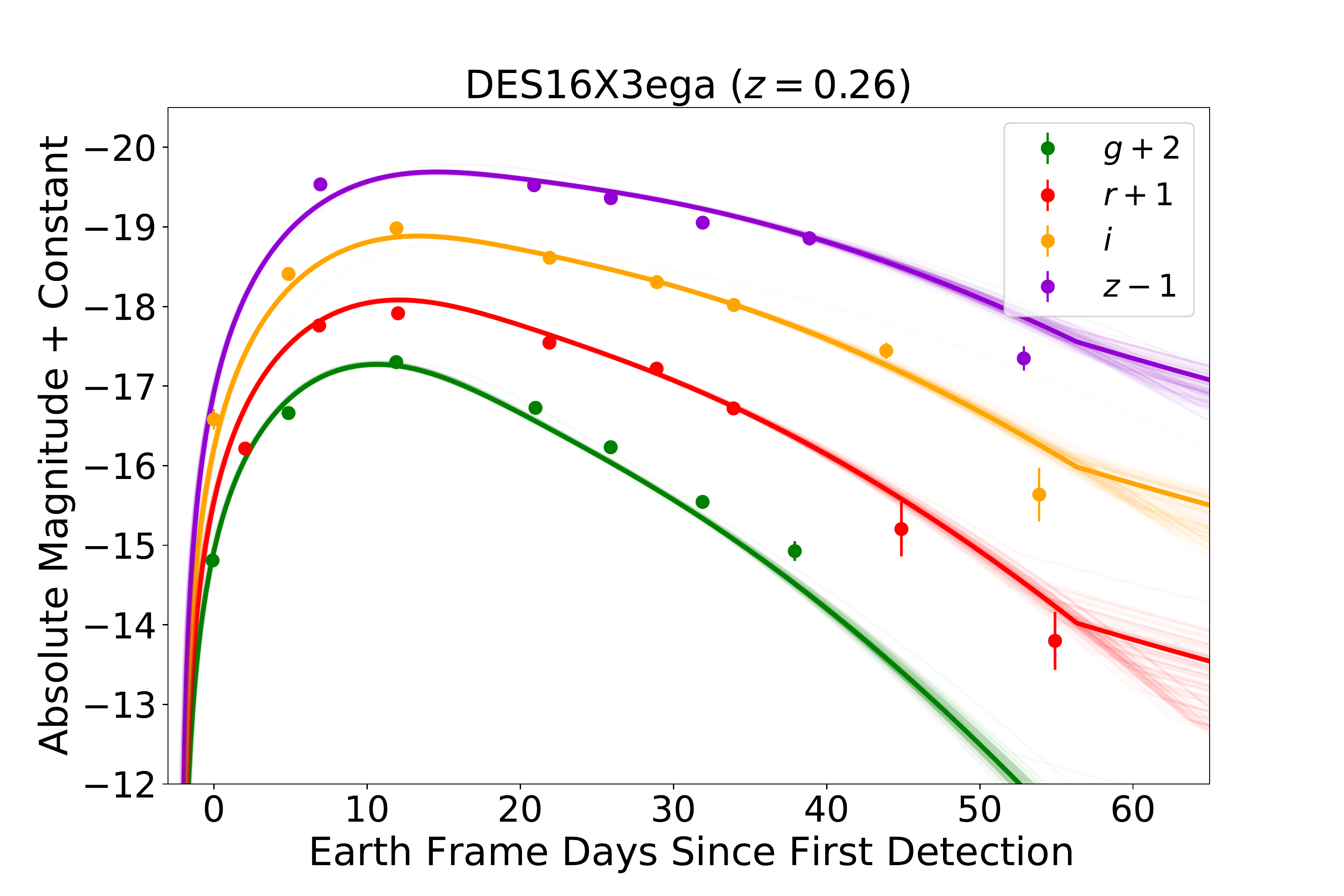}
\includegraphics[width=5.8cm]{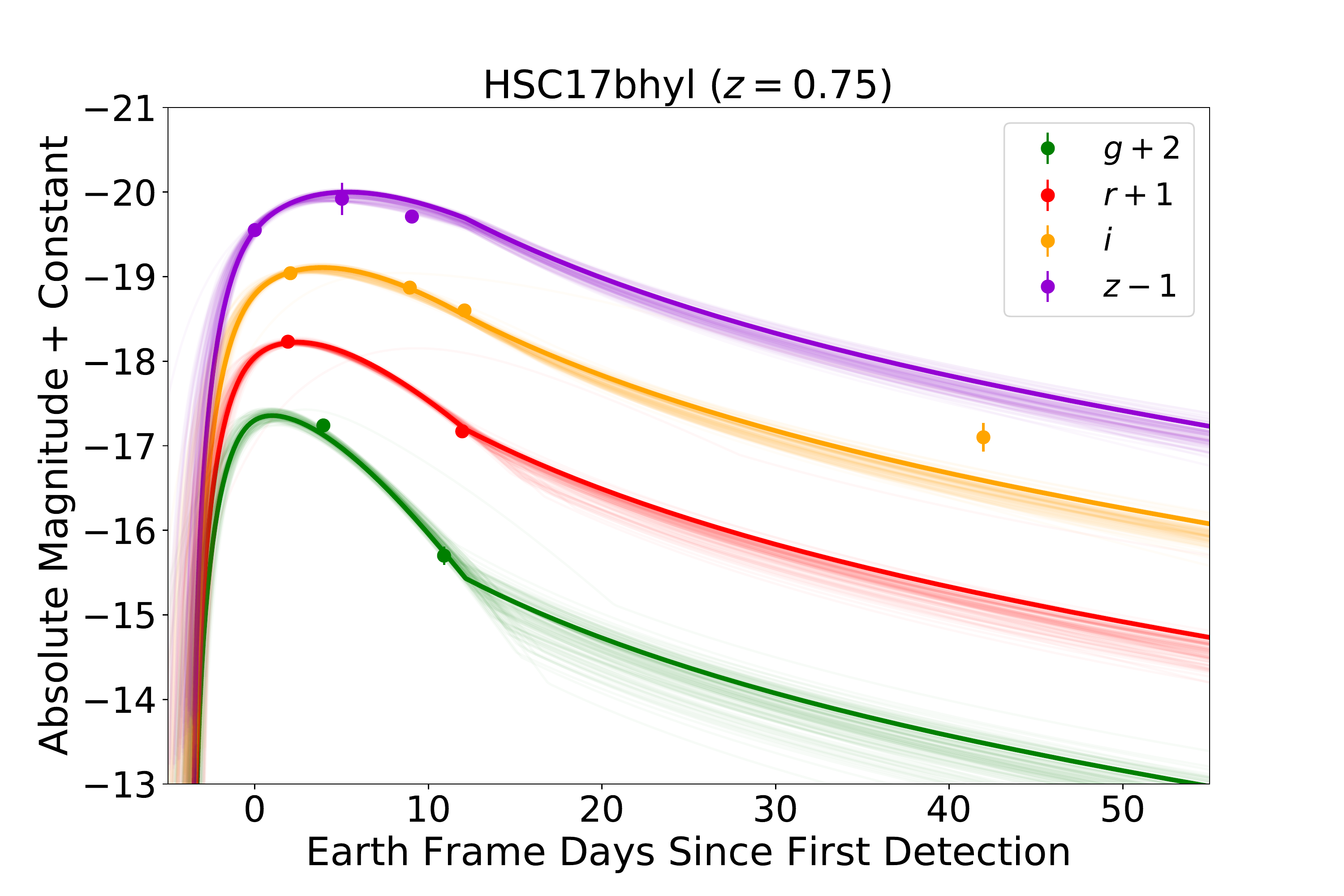}

\includegraphics[width=5.8cm]{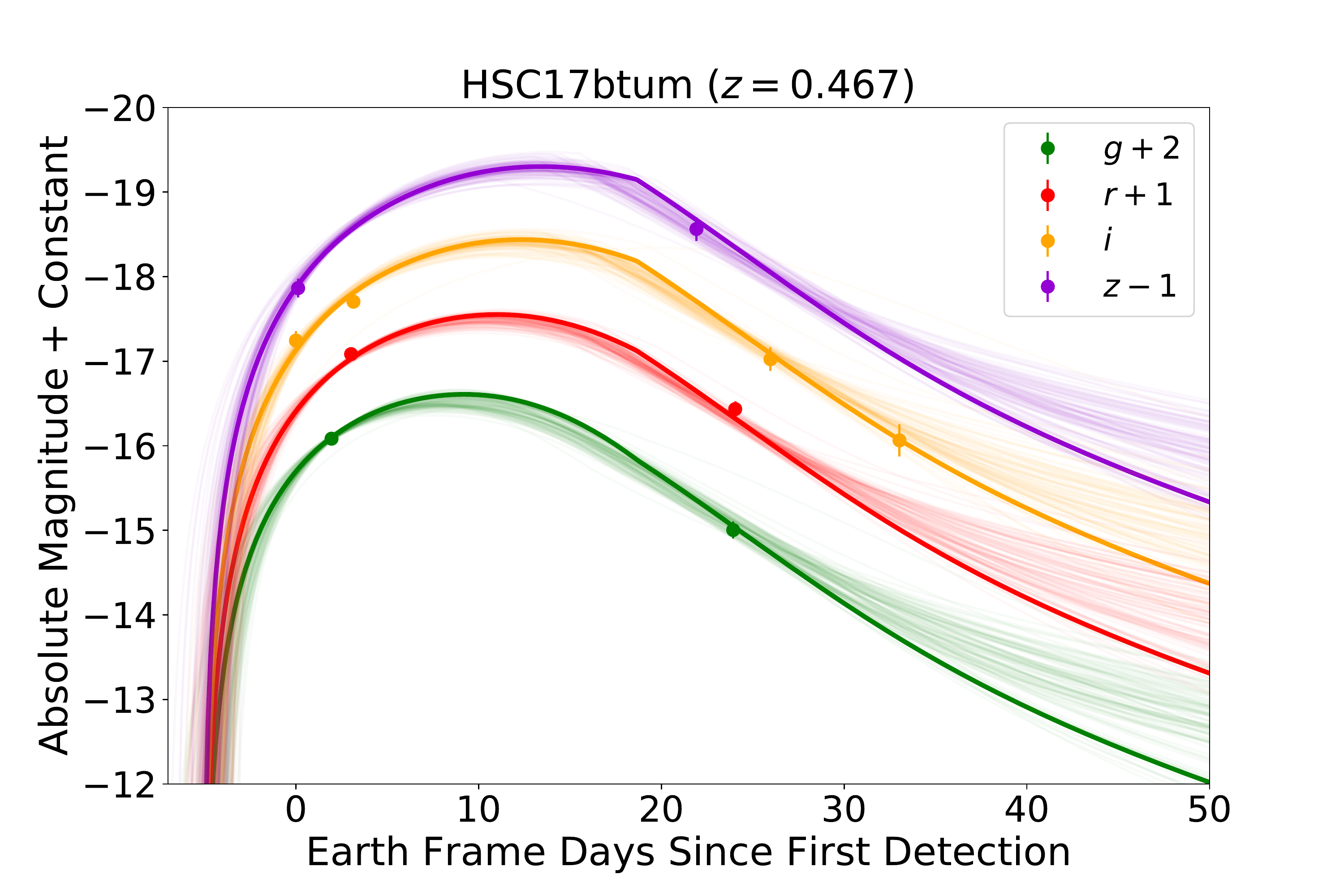}
\includegraphics[width=5.8cm]{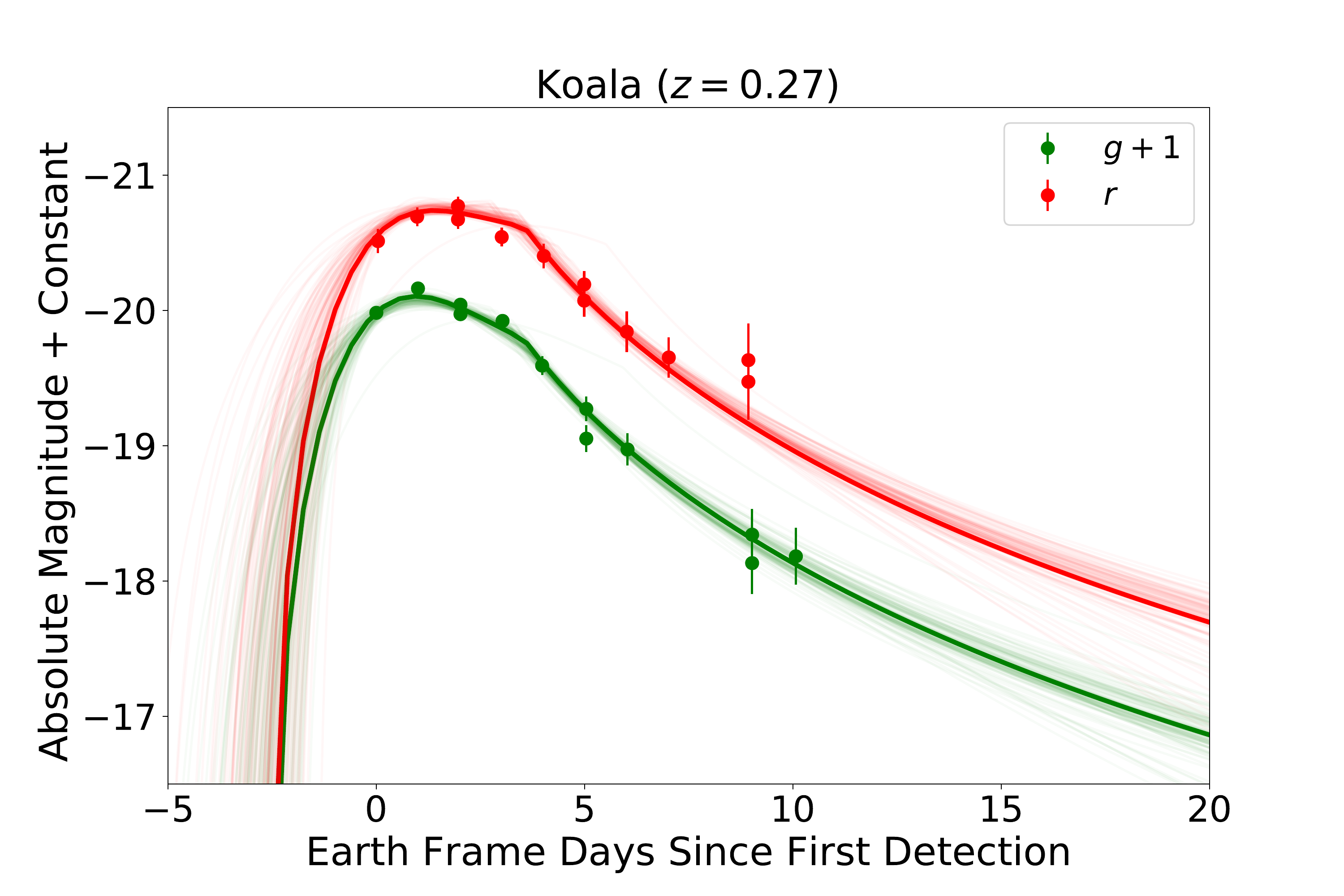}
\includegraphics[width=5.8cm]{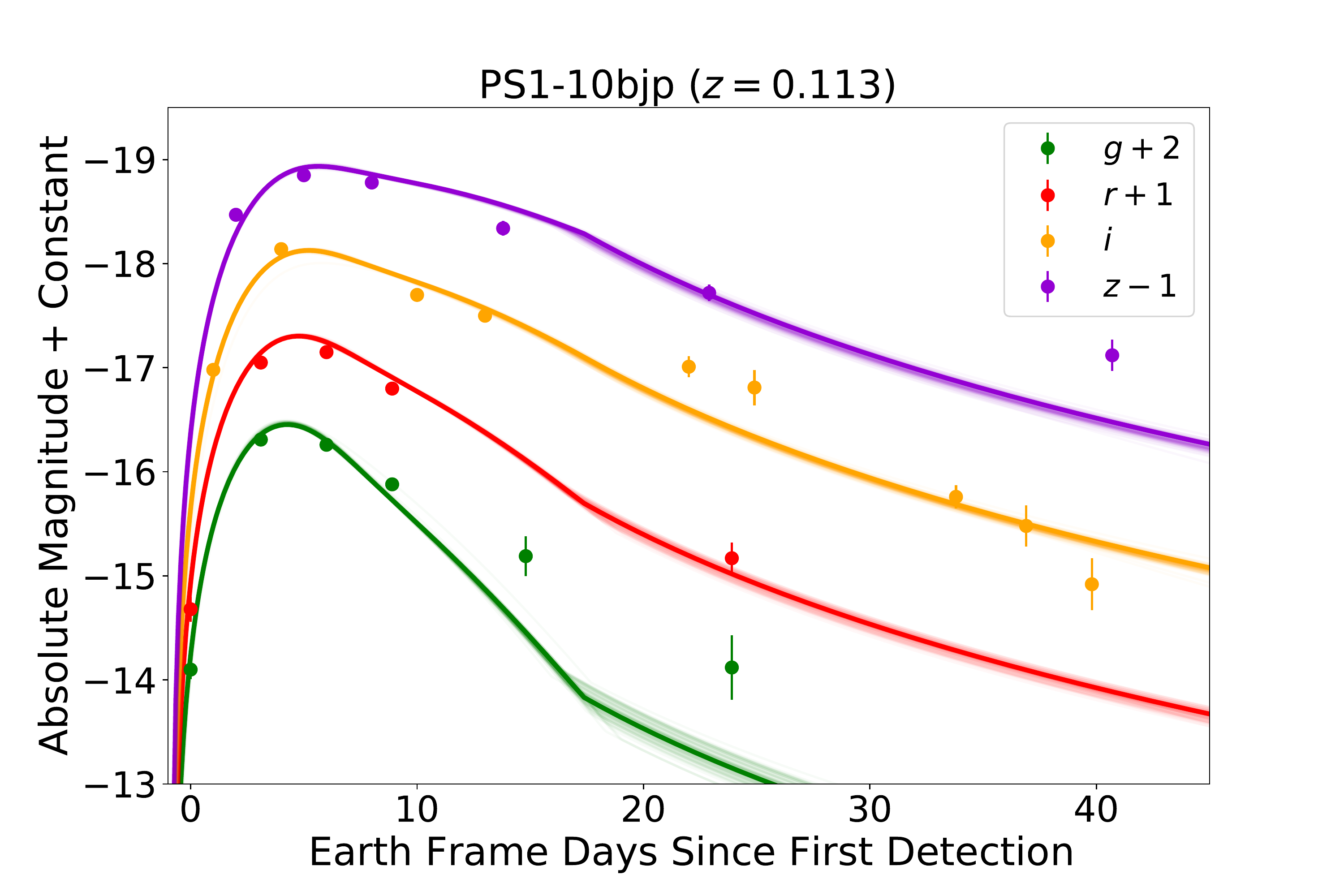}

\includegraphics[width=5.8cm]{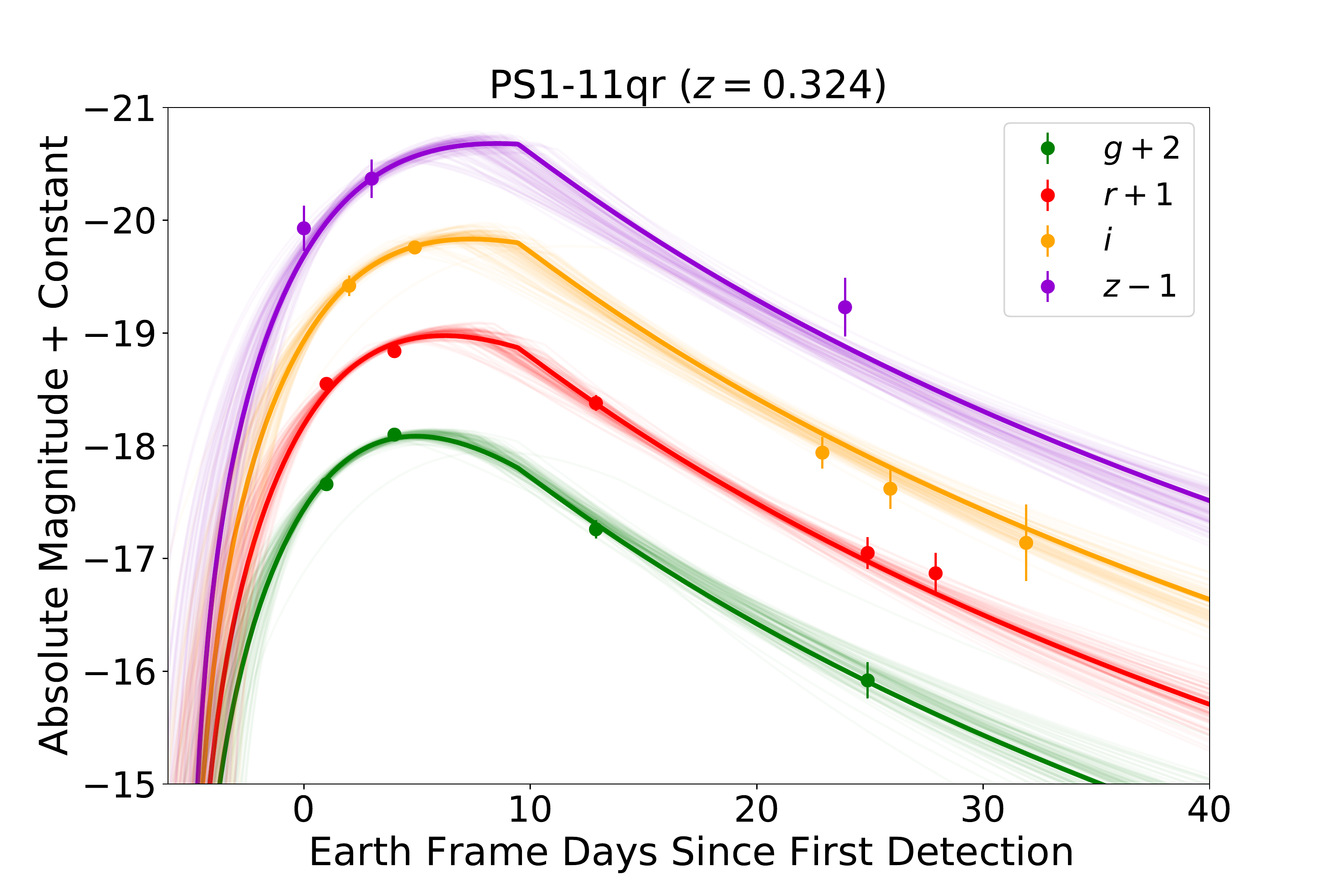}
\includegraphics[width=5.8cm]{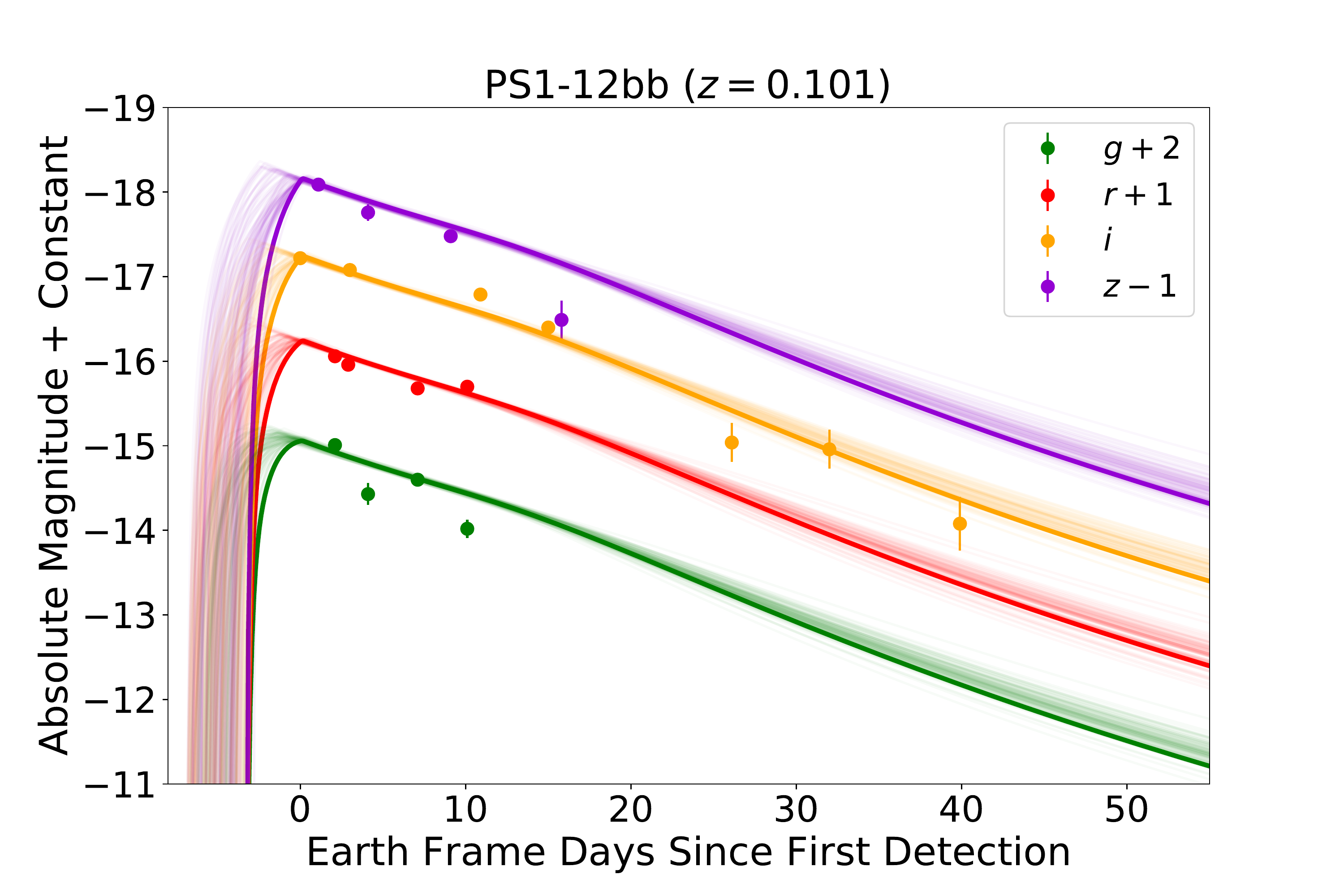}
\includegraphics[width=5.8cm]{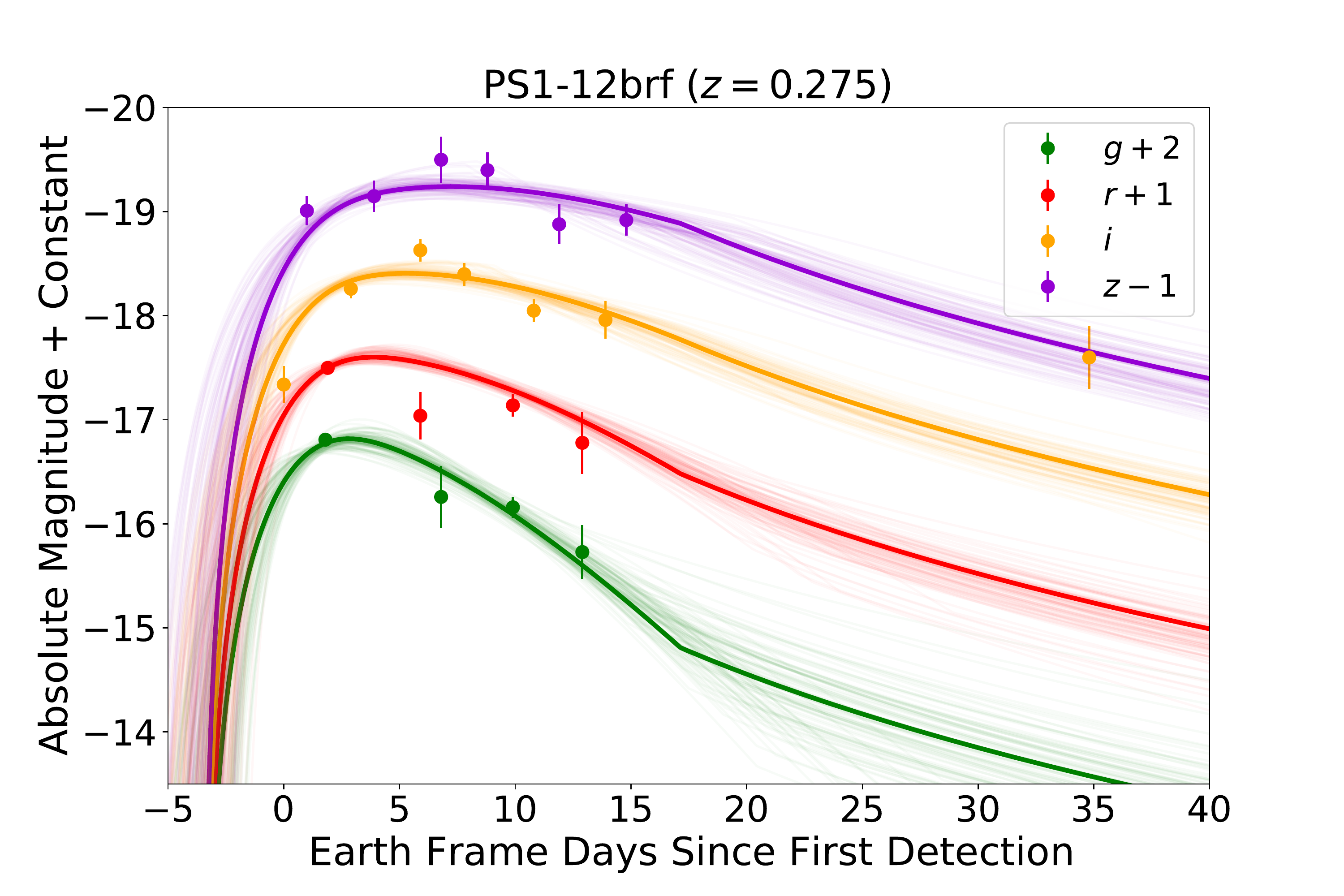}

\includegraphics[width=5.8cm]{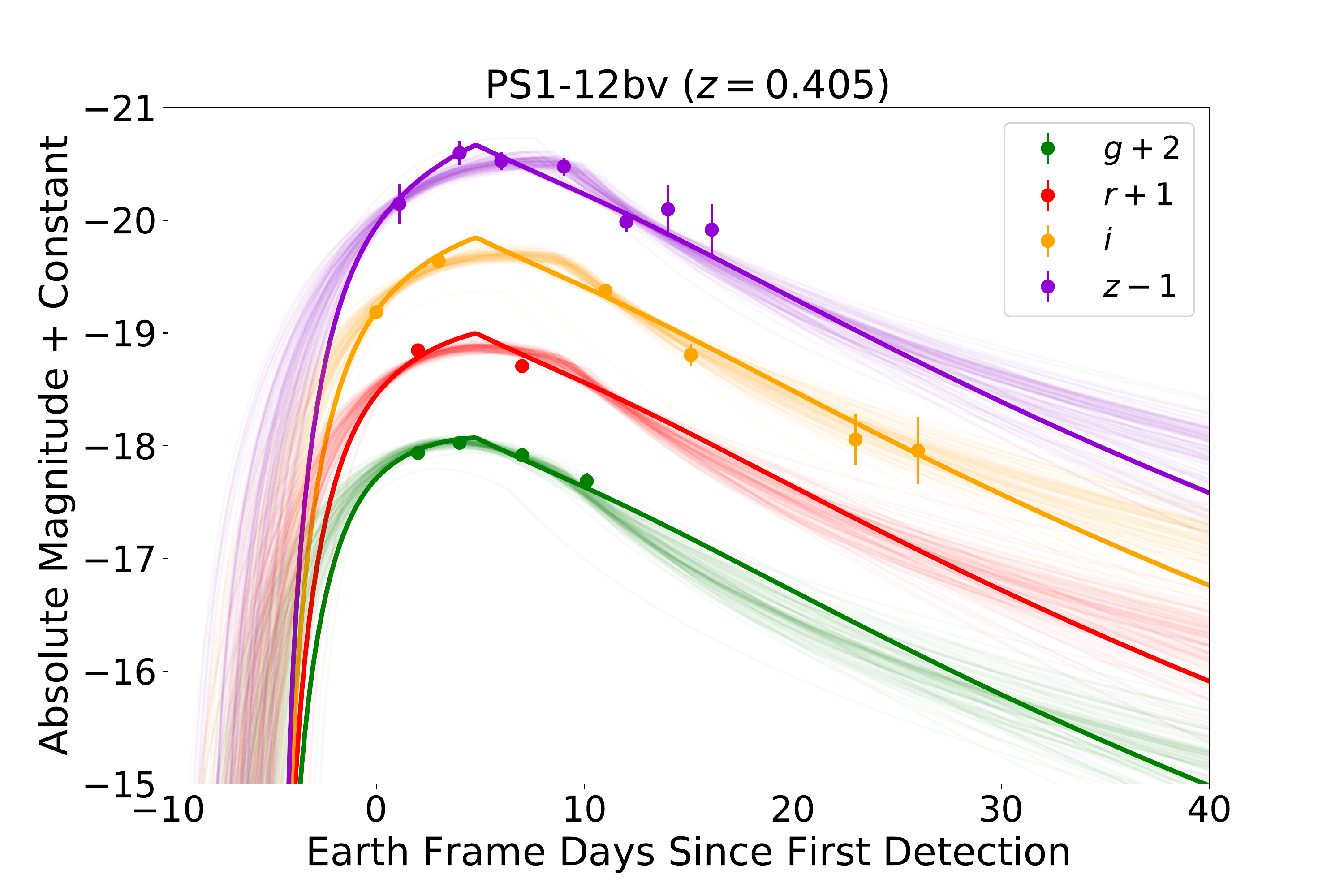}
\includegraphics[width=5.8cm]{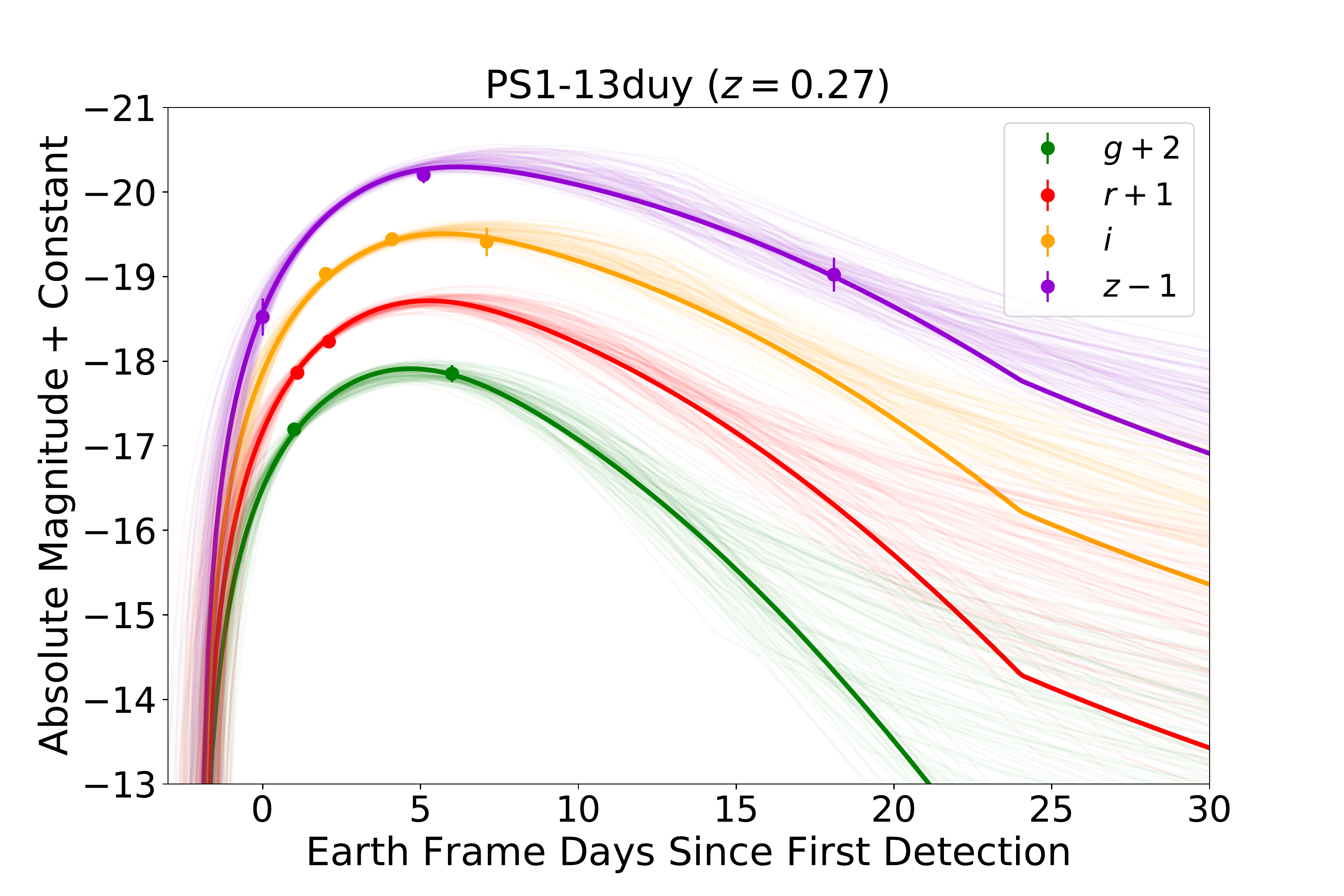}
\includegraphics[width=5.8cm]{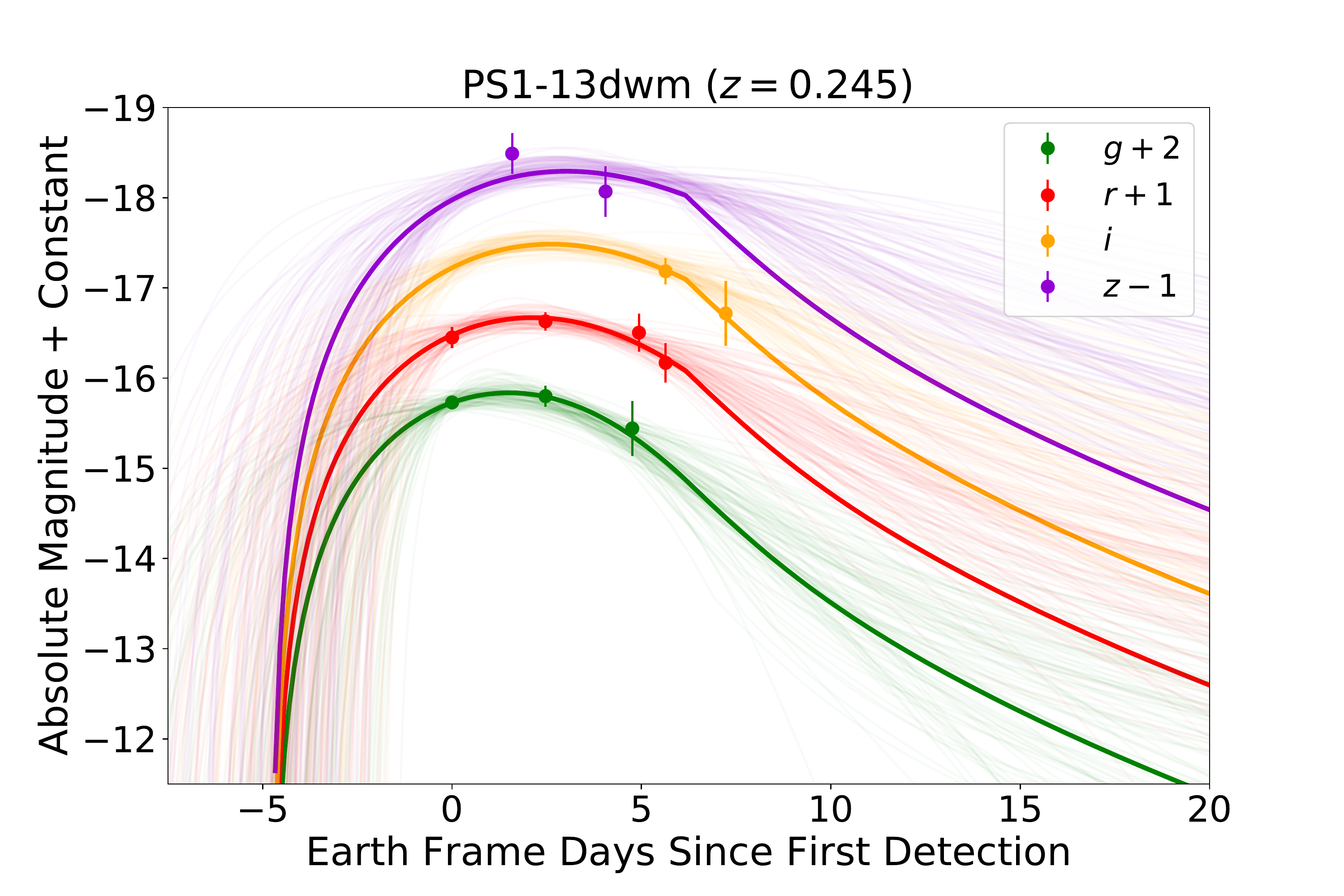}
\caption{(Continued.)}
\end{figure}

\addtocounter{figure}{-1}
\begin{figure}
\includegraphics[width=5.8cm]{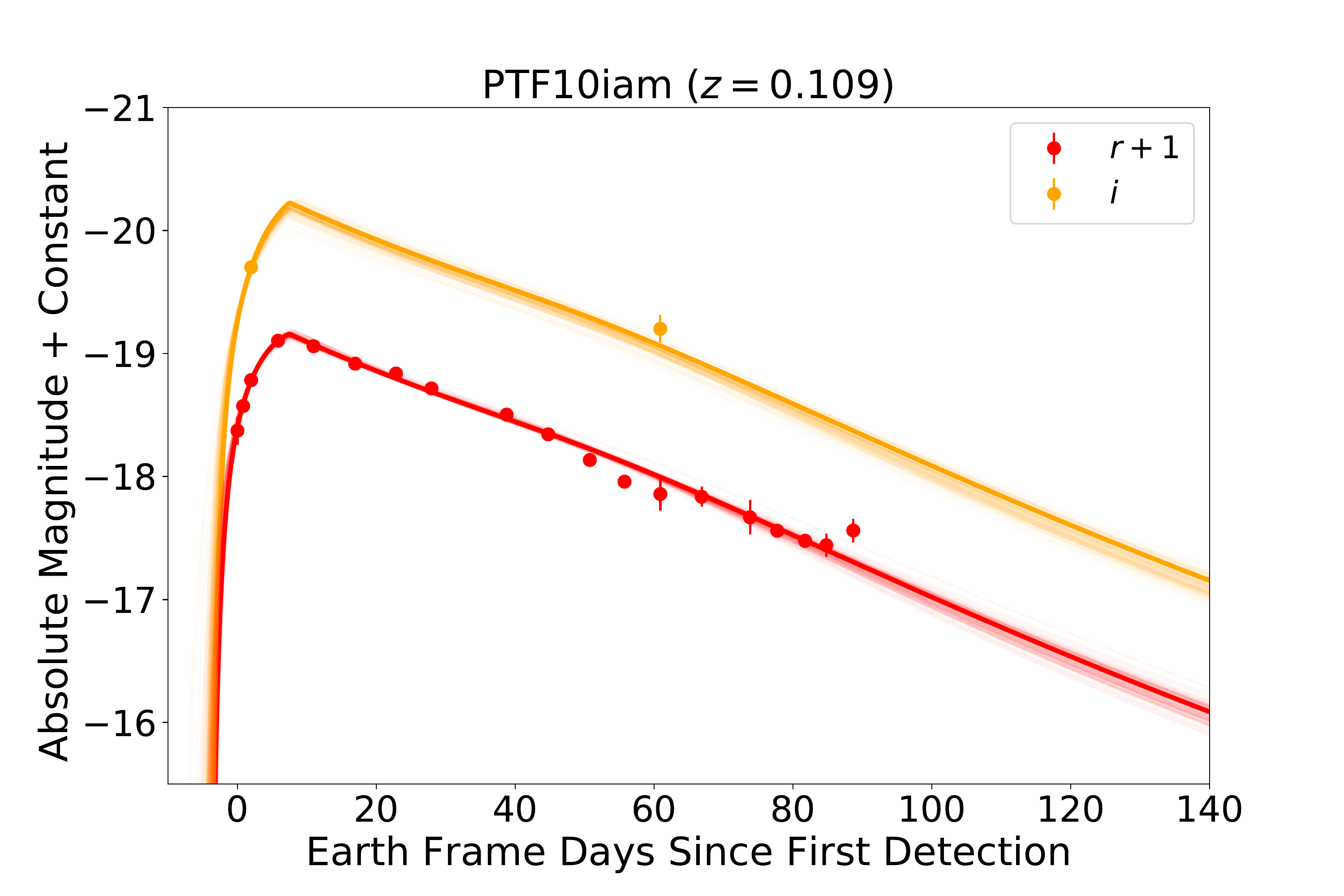}
\includegraphics[width=5.8cm]{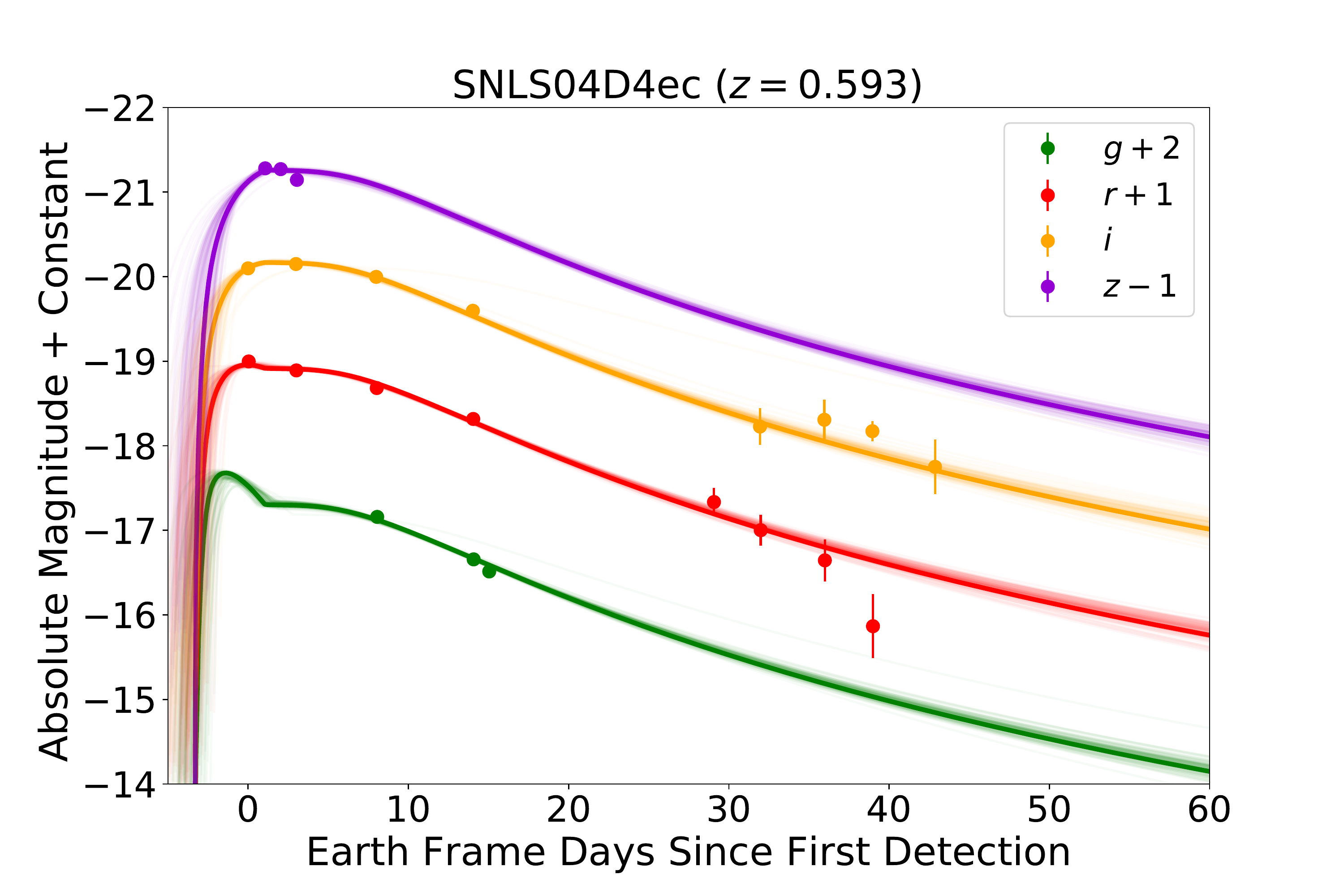}
\includegraphics[width=5.8cm]{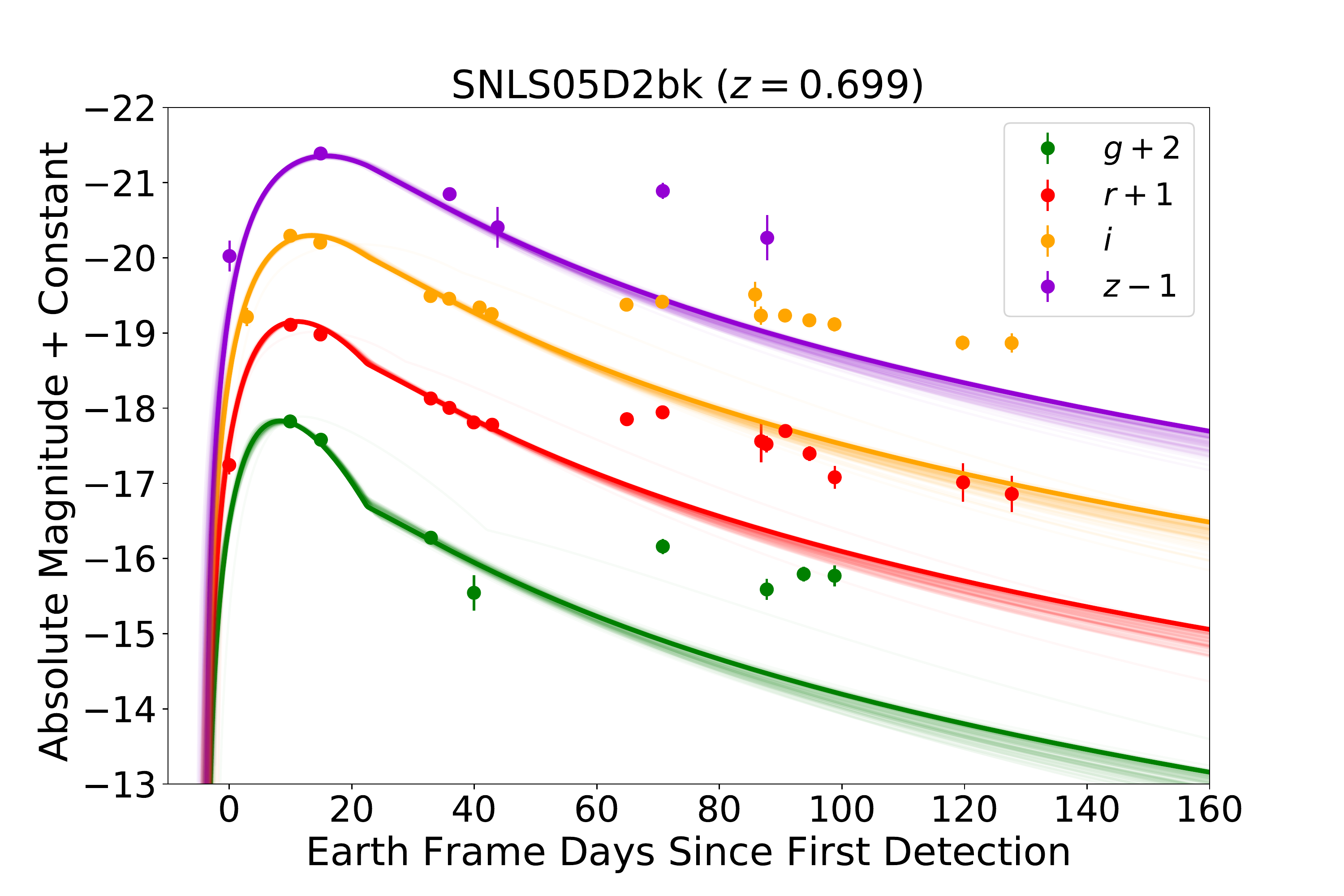}

\includegraphics[width=5.8cm]{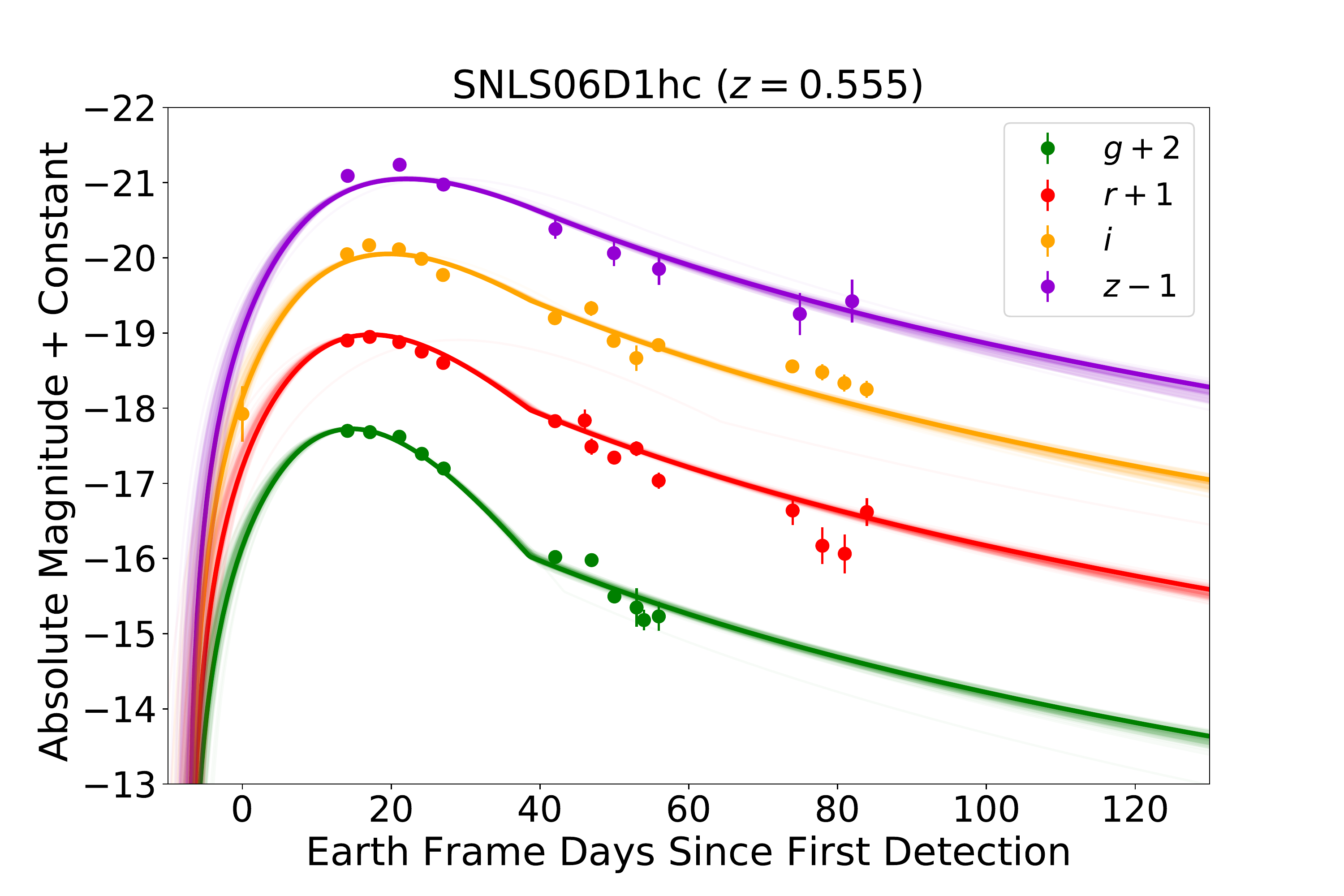}
\caption{(Continued.)}
\end{figure}

\begin{figure*}
    \centering
	\includegraphics[width = 1\linewidth]{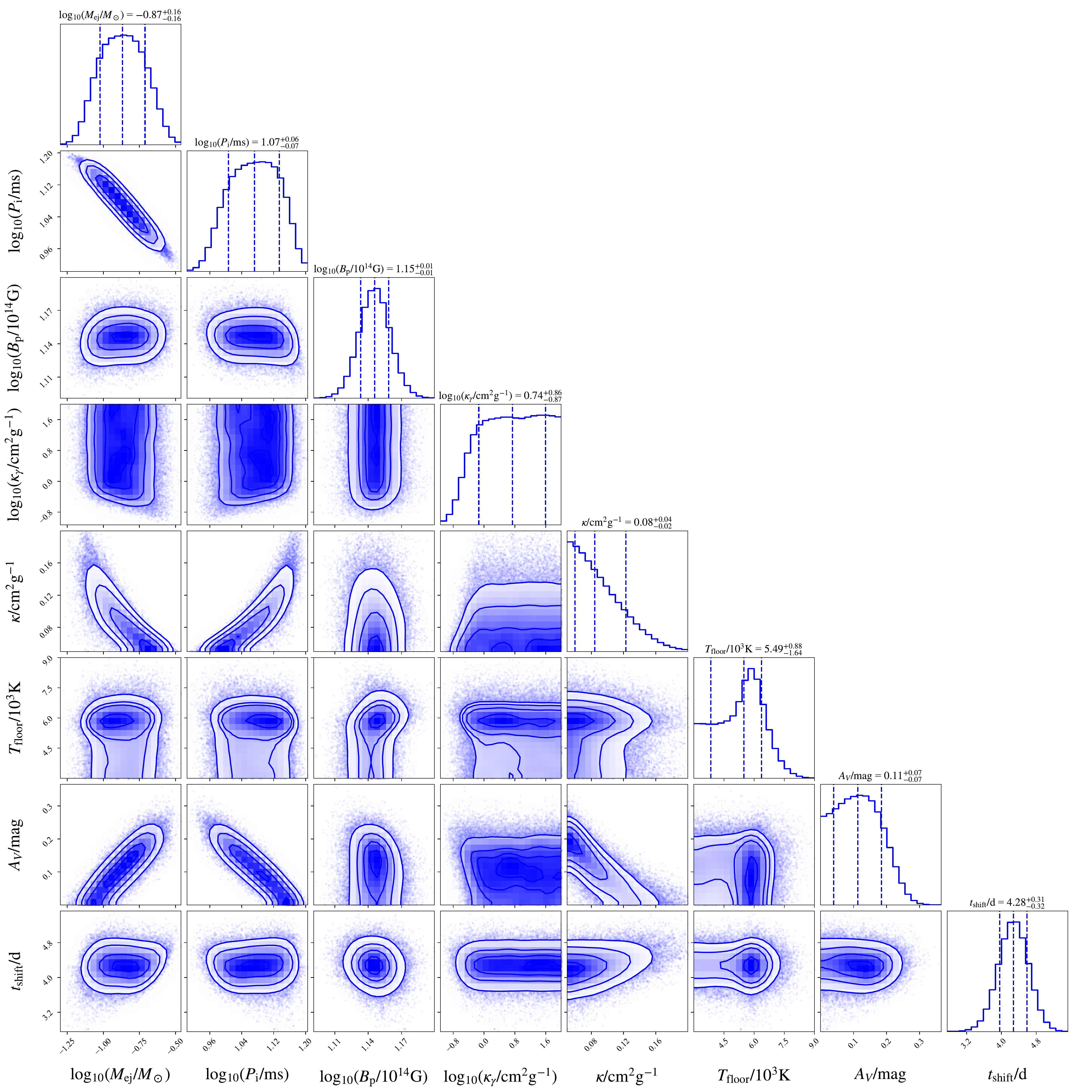}
    \caption{Posteriors for magnetar model fit to DES16C3gin. Medians and 1$\sigma$ ranges are labeled. }\label{fig:DES16C3gin_corner_fit}
\end{figure*}

%\begin{figure*}
%\centering\includegraphics[width = 0.50\linewidth, trim= 0 0 20 0,clip]{MT_corner_DES16C3gin.pdf}
%\caption{Corner plot of the posterior probability distributions of $M_{\rm peak}, t_{\rm rise}$, and $t_{\rm decl}$ of DES16C3gin. Medians and 1$\sigma$ ranges are labeled.}
%\label{fig:MT_corner_DES16C3gin_v2}
%\end{figure*}

\begin{deluxetable*}{l|ccc|rcrcccc}
\tablecaption{The lightcurve properties and the fitting results of the derived parameters for the collected FBOTs \label{tab:ParameterResults}}
\tablecolumns{8}
\tablewidth{0pt}
\tablehead{
\colhead{FBOT} &
\colhead{$M_{\rm peak}$} &
\colhead{$t_{\rm rise}$} &
\colhead{$t_{\rm decl}$} &
\colhead{$\log_{10}M_{\rm ej,\odot}$} &
\colhead{$\log_{10}P_{\rm i,-3}$} &
\colhead{$\log_{10}B_{\rm p,14}$ } &
\colhead{$\log_{10}\kappa_{\gamma}$} &
\colhead{$\kappa$}&
\colhead{$T_{\rm floor,3}$}&
\colhead{$A_{V}$}
}
\startdata
AT2018cow      &$-22.42_{-0.04}^{+0.03}$ & $0.80_{-0.00}^{+0.04}$ & $1.68_{-0.04}^{+0.04}$
               & $-1.78_{-0.01}^{+0.02}$ & $0.82_{-0.01}^{+0.00}$ & $0.84_{-0.01}^{+0.01}$  & $1.83_{-0.14}^{+0.12}$ & $0.19_{-0.01}^{+0.01}$ & $24.58_{-0.16}^{+0.27}$ & $0.50_{-0.00}^{+0.00}$ \\
DES13C3bcok    &$-19.53_{-0.13}^{+0.14}$ & $3.01_{-0.15}^{+0.30}$ & $4.96_{-0.30}^{+0.45}$
               & $-0.64_{-0.14}^{+0.13}$ & $1.06_{-0.07}^{+0.06}$ & $1.19_{-0.08}^{+0.07}$  & $-0.7_{-1.2}^{+1.8}$   & $0.06_{-0.01}^{+0.02}$ & $10.5_{-1.1}^{+2.0}$    & $0.03_{-0.02}^{+0.04}$ \\
DES13C3uig$^*$ &$-19.45_{-0.37}^{+0.33}$ & $1.65_{-0.45}^{+0.30}$ & $3.31_{-0.75}^{+0.60}$
               & $-1.46_{-0.26}^{+0.25}$ & $1.27_{-0.07}^{+0.05}$ & $1.19_{-0.06}^{+0.06}$  & $1.07_{-0.74}^{+0.64}$  & $0.10_{-0.04}^{+0.06}$ & $7.9_{-1.3}^{+1.2}$    & $0.15_{-0.11}^{+0.17}$ \\
DES13X1hav$^*$ &$-19.62_{-0.36}^{+0.41}$ & $2.0_{-0.7}^{+5.2}$    & $9.5_{-3.5}^{+5.3}$
               & $-1.69_{-0.33}^{+0.38}$ & $1.04_{-0.17}^{+0.07}$ & $0.56_{-0.50}^{+0.20}$  & $1.22_{-0.78}^{+0.54}$  & $0.13_{-0.05}^{+0.05}$ & $12.2_{-1.4}^{+1.7}$   & $0.29_{-0.18}^{+0.15}$ \\
DES13X3gms     &$-19.90_{-0.26}^{+0.23}$ & $4.36_{-0.45}^{+0.60}$ & $7.21_{-0.75}^{+0.90}$
               & $-0.49_{-0.16}^{+0.16}$ & $0.95_{-0.06}^{+0.05}$ & $0.87_{-0.03}^{+0.03}$  & $0.5_{-1.1}^{+1.0}$     & $0.08_{-0.02}^{+0.04}$ & $6.1_{-2.1}^{+1.9}$    & $0.14_{-0.09}^{+0.11}$ \\
DES13X3npb     &$-19.08_{-0.29}^{+0.26}$ & $4.21_{-0.45}^{+0.45}$ & $6.91_{-0.75}^{+0.90}$
               & $-0.63_{-0.20}^{+0.24}$ & $1.05_{-0.11}^{+0.09}$ & $1.17_{-0.12}^{+0.07}$  & $0.1_{-1.2}^{+1.3}$     & $0.12_{-0.05}^{+0.05}$ & $6.5_{-2.4}^{+3.1}$    & $0.23_{-0.15}^{+0.17}$ \\
DES13X3nyg$^*$ &$-21.11_{-0.34}^{+0.35}$ & $1.65_{-0.30}^{+0.45}$ & $3.01_{-0.45}^{+0.60}$
               & $-1.14_{-0.22}^{+0.25}$ & $0.93_{-0.09}^{+0.06}$ & $0.95_{-0.04}^{+0.04}$  & $0.82_{-0.86}^{+0.78}$  & $0.10_{-0.03}^{+0.06}$  & $10.7_{-4.0}^{+1.5}$  & $0.19_{-0.13}^{+0.14}$ \\
DES14C3tvw     &$-19.65_{-0.30}^{+0.27}$ & $3.46_{-0.75}^{+0.90}$ & $6.0_{-1.1}^{+1.5}$
               & $-0.77_{-0.24}^{+0.23}$ & $1.07_{-0.06}^{+0.05}$ & $0.95_{-0.06}^{+0.05}$  & $0.5_{-1.0}^{+1.0}$     & $0.08_{-0.02}^{+0.05}$  & $8.7_{-3.8}^{+2.4}$   & $0.10_{-0.07}^{+0.13}$ \\
DES14S2anq$^*$ &$-15.27_{-0.06}^{+0.04}$ & $5.71_{-0.15}^{+0.15}$ & $13.37_{-0.30}^{+0.45}$
               & $-0.16_{-0.11}^{+0.20}$ & $0.94_{-0.10}^{+0.07}$ & $2.03_{-0.02}^{+0.02}$  & $0.4_{-1.2}^{+1.1}$     & $0.15_{-0.05}^{+0.04}$  & $5.63_{-0.24}^{+0.29}$& $0.03_{-0.02}^{+0.05}$ \\
DES14S2plb     &$-15.94_{-0.54}^{+0.83}$ & $3.76_{-0.60}^{+0.60}$ & $7.8_{-2.1}^{+3.0}$
               & $-0.72_{-0.42}^{+0.68}$ & $1.39_{-0.74}^{+0.24}$ & $1.95_{-0.13}^{+0.28}$  & $0.2_{-1.2}^{+1.3}$     & $0.13_{-0.05}^{+0.05}$  & $5.7_{-1.3}^{+1.5}$   & $0.35_{-0.22}^{+0.12}$ \\
DES14X1bnh$^*$ &$-20.70_{-0.49}^{+0.44}$ & $2.7_{-0.8}^{+1.1}$    & $6.8_{-2.1}^{+5.4}$
               & $-0.99_{-0.30}^{+0.31}$ & $0.85_{-0.11}^{+0.06}$ & $0.61_{-0.32}^{+0.15}$  & $0.85_{-0.96}^{+0.81}$  & $0.11_{-0.04}^{+0.06}$  & $8.6_{-3.7}^{+5.5}$   & $0.25_{-0.17}^{+0.16}$ \\
DES15C3lpq     &$-20.08_{-0.14}^{+0.15}$ & $2.85_{-0.30}^{+0.45}$ & $4.96_{-0.60}^{+0.60}$
               & $-0.72_{-0.13}^{+0.12}$ & $1.02_{-0.03}^{+0.03}$ & $0.97_{-0.02}^{+0.02}$  & $0.66_{-0.83}^{+0.89}$  & $0.06_{-0.01}^{+0.01}$  & $7.66_{-0.76}^{+0.80}$& $0.03_{-0.02}^{+0.04}$ \\
DES15C3mgq$^*$ &$-18.20_{-0.36}^{+0.29}$ & $1.35_{-0.45}^{+0.75}$ & $2.55_{-0.60}^{+0.90}$
               & $-1.74_{-0.35}^{+0.35}$ & $1.53_{-0.10}^{+0.11}$ & $1.61_{-0.02}^{+0.02}$  & $1.07_{-0.75}^{+0.63}$  & $0.14_{-0.05}^{+0.04}$  & $5.23_{-0.70}^{+0.94}$& $0.39_{-0.17}^{+0.08}$ \\
DES15C3nat$^*$ &$-19.79_{-0.39}^{+0.35}$ & $2.25_{-0.60}^{+0.60}$ & $3.9_{-0.9}^{+1.1}$
               & $-0.90_{-0.32}^{+0.34}$ & $1.01_{-0.21}^{+0.14}$ & $1.29_{-0.16}^{+0.16}$  & $0.3_{-1.4}^{+1.2}$     & $0.09_{-0.03}^{+0.06}$  & $6.1_{-2.2}^{+2.6}$   & $0.12_{-0.09}^{+0.15}$ \\
DES15C3opk     &$-20.79_{-0.35}^{+0.25}$ & $3.01_{-0.45}^{+0.60}$ & $5.1_{-0.6}^{+1.2}$
               & $-0.80_{-0.17}^{+0.20}$ & $0.85_{-0.07}^{+0.06}$ & $0.79_{-0.09}^{+0.07}$  & $-0.4_{-0.5}^{+1.4}$    & $0.12_{-0.05}^{+0.05}$  & $17.5_{-1.3}^{+1.3}$  & $0.30_{-0.11}^{+0.10}$ \\
DES15C3opp$^*$ &$-18.36_{-0.39}^{+0.33}$ & $2.10_{-0.60}^{+0.75}$ & $3.8_{-1.1}^{+1.1}$
               & $-1.27_{-0.30}^{+0.30}$ & $1.41_{-0.09}^{+0.07}$ & $1.49_{-0.08}^{+0.08}$  & $0.4_{-1.1}^{+1.1}$     & $0.08_{-0.02}^{+0.05}$  & $7.1_{-2.7}^{+2.0}$   & $0.09_{-0.06}^{+0.13}$ \\
DES15E2nqh$^*$ &$-20.29_{-0.30}^{+0.28}$ & $2.70_{-0.75}^{+0.60}$ & $4.66_{-0.90}^{+0.90}$
               & $-0.88_{-0.26}^{+0.23}$ & $1.00_{-0.07}^{+0.06}$ & $0.95_{-0.05}^{+0.04}$  & $0.68_{-0.96}^{+0.91}$  & $0.08_{-0.02}^{+0.05}$  & $8.3_{-2.4}^{+1.2}$   & $0.09_{-0.07}^{+0.12}$ \\
DES15S1fli$^*$ &$-20.06_{-0.39}^{+0.29}$ & $2.25_{-0.60}^{+0.60}$ & $5.9_{-1.5}^{+2.4}$
               & $-1.23_{-0.27}^{+0.24}$ & $1.04_{-0.07}^{+0.04}$ & $0.78_{-0.15}^{+0.09}$  & $1.23_{-0.68}^{+0.53}$  & $0.12_{-0.05}^{+0.05}$  & $11.0_{-2.7}^{+2.3}$  & $0.17_{-0.12}^{+0.18}$ \\
DES15S1fll$^*$ &$-18.86_{-0.45}^{+0.37}$ & $2.9_{-0.9}^{+1.1}$    & $4.8_{-1.4}^{+1.8}$
               & $-1.01_{-0.28}^{+0.32}$ & $1.24_{-0.10}^{+0.08}$ & $1.29_{-0.07}^{+0.05}$  & $0.5_{-1.0}^{+1.1}$     & $0.10_{-0.03}^{+0.05}$  & $7.1_{-2.5}^{+2.7}$   & $0.09_{-0.06}^{+0.10}$ \\
DES15X3mxf     &$-20.61_{-0.18}^{+0.22}$ & $2.25_{-0.15}^{+0.30}$ & $4.06_{-0.45}^{+0.45}$
               & $-0.53_{-0.11}^{+0.09}$ & $0.74_{-0.04}^{+0.05}$ & $1.24_{-0.02}^{+0.02}$  & $0.69_{-0.93}^{+0.90}$  & $0.06_{-0.01}^{+0.03}$  & $11.29_{-0.66}^{+0.81}$& $0.03_{-0.02}^{+0.05}$ \\
DES16C1cbd     &$-19.96_{-0.32}^{+0.31}$ & $2.85_{-0.60}^{+0.60}$ & $5.6_{-1.1}^{+1.1}$
               & $-1.02_{-0.21}^{+0.20}$ & $1.04_{-0.06}^{+0.06}$ & $0.87_{-0.06}^{+0.06}$  & $0.70_{-0.93}^{+0.89}$  & $0.12_{-0.05}^{+0.05}$  & $7.2_{-2.9}^{+2.4}$    & $0.34_{-0.16}^{+0.11}$ \\
DES16C3gin     &$-19.81_{-0.25}^{+0.26}$ & $2.70_{-0.30}^{+0.15}$ & $4.36_{-0.30}^{+0.30}$
               & $-0.86_{-0.15}^{+0.16}$ & $1.07_{-0.06}^{+0.06}$ & $1.15_{-0.01}^{+0.01}$  & $0.73_{-0.86}^{+0.87}$  & $0.08_{-0.02}^{+0.04}$  & $5.5_{-1.7}^{+0.9}$    & $0.11_{-0.07}^{+0.07}$ \\
DES16E1bir$^*$ &$-22.47_{-0.29}^{+0.15}$ & $3.16_{-0.45}^{+0.45}$ & $5.86_{-0.75}^{+0.75}$
               & $-0.05_{-0.31}^{+0.26}$ & $0.16_{-0.17}^{+0.21}$ & $0.75_{-0.18}^{+0.11}$  & $0.4_{-1.1}^{+1.1}$     & $0.11_{-0.04}^{+0.05}$  & $9.8_{-4.8}^{+4.7}$    & $0.10_{-0.07}^{+0.14}$ \\
DES16E2pv$^*$  &$-20.27_{-0.42}^{+0.34}$ & $1.80_{-0.60}^{+0.63}$ & $4.4_{-1.2}^{+3.0}$
               & $-1.36_{-0.35}^{+0.40}$ & $1.02_{-0.13}^{+0.06}$ & $0.88_{-0.30}^{+0.19}$  & $0.7_{-1.1}^{+0.9}$     & $0.11_{-0.05}^{+0.06}$  & $16.7_{-5.6}^{+2.3}$   & $0.18_{-0.12}^{+0.17}$ \\
DES16X3cxn$^*$ &$-20.18_{-0.42}^{+0.34}$ & $2.25_{-0.45}^{+0.60}$ & $4.21_{-0.75}^{+0.90}$
               & $-1.09_{-0.20}^{+0.24}$ & $1.05_{-0.09}^{+0.06}$ & $0.96_{-0.03}^{+0.03}$  & $0.69_{-0.78}^{+0.88}$  & $0.10_{-0.04}^{+0.06}$  & $5.4_{-1.6}^{+2.3}$    & $0.24_{-0.15}^{+0.15}$ \\
DES16X3ega     &$-19.96_{-0.03}^{+0.04}$ & $3.76_{-0.15}^{+0.00}$ & $6.16_{-0.15}^{+0.00}$
               & $-0.38_{-0.03}^{+0.03}$ & $0.93_{-0.01}^{+0.01}$ & $1.03_{-0.01}^{+0.01}$  & $-1.20_{-0.05}^{+0.06}$ & $0.05_{-0.00}^{+0.00}$  & $3.45_{-0.30}^{+0.32}$ & $0.16_{-0.03}^{+0.03}$ \\
HSC17bhyl$^*$  &$-20.57_{-0.28}^{+0.21}$ & $0.90_{-0.15}^{+0.30}$ & $1.80_{-0.30}^{+0.45}$
               & $-1.89_{-0.16}^{+0.27}$ & $1.27_{-0.08}^{+0.04}$ & $1.30_{-0.02}^{+0.02}$  & $1.57_{-0.40}^{+0.30}$  & $0.14_{-0.04}^{+0.06}$  & $8.13_{-0.67}^{+0.60}$ & $0.08_{-0.06}^{+0.15}$ \\
HSC17btum$^*$  &$-18.43_{-0.27}^{+0.21}$ & $4.36_{-0.30}^{+0.30}$ & $8.11_{-0.75}^{+0.75}$
               & $-0.33_{-0.24}^{+0.22}$ & $0.91_{-0.14}^{+0.15}$ & $1.42_{-0.10}^{+0.08}$  & $-0.6_{-1.1}^{+1.7}$    & $0.10_{-0.04}^{+0.06}$  & $10.6_{-0.8}^{+1.0}$   & $0.10_{-0.06}^{+0.09}$ \\
Koala          &$-22.85_{-0.35}^{+0.32}$ & $0.90_{-0.15}^{+0.30}$ & $1.65_{-0.45}^{+0.30}$
               & $-1.10_{-0.25}^{+0.23}$ & $0.60_{-0.08}^{+0.09}$ & $1.10_{-0.07}^{+0.06}$  & $0.99_{-0.73}^{+0.69}$  & $0.09_{-0.03}^{+0.05}$  & $15.7_{-1.6}^{+1.7}$   & $0.04_{-0.03}^{+0.06}$ \\
PS1-10bjp      &$-20.30_{-0.04}^{+0.04}$ & $1.51_{-0.10}^{+0.00}$ & $2.51_{-0.10}^{+0.00}$
               & $-0.88_{-0.04}^{+0.03}$ & $0.95_{-0.01}^{+0.02}$ & $1.47_{-0.01}^{+0.01}$  & $1.03_{-0.65}^{+0.65}$  & $0.05_{-0.00}^{+0.00}$  & $5.44_{-0.21}^{+0.22}$  & $0.48_{-0.02}^{+0.01}$ \\
PS1-11qr       &$-20.24_{-0.32}^{+0.26}$ & $3.16_{-0.45}^{+0.60}$ & $5.56_{-0.60}^{+0.75}$
               & $-0.81_{-0.23}^{+0.19}$ & $0.81_{-0.10}^{+0.08}$ & $0.31_{-0.19}^{+0.20}$  & $-1.44_{-0.38}^{+0.51}$ & $0.13_{-0.05}^{+0.05}$  & $13.8_{-1.9}^{+2.9}$    & $0.19_{-0.11}^{+0.12}$ \\
PS1-12bb$^*$   &$-17.39_{-0.16}^{+0.19}$ & $0.64_{-0.60}^{+0.24}$ & $10.02_{-0.85}^{+0.85}$
               & $-2.71_{-0.16}^{+0.23}$ & $1.61_{-0.04}^{+0.05}$ & $1.11_{-0.04}^{+0.06}$  & $1.61_{-0.42}^{+0.28}$  & $0.13_{-0.05}^{+0.05}$  & $9.56_{-0.83}^{+0.59}$  & $0.35_{-0.19}^{+0.11}$ \\
PS1-12brf      &$-19.56_{-0.38}^{+0.33}$ & $1.80_{-0.45}^{+0.45}$ & $3.16_{-0.75}^{+0.60}$
               & $-1.27_{-0.28}^{+0.26}$ & $1.21_{-0.10}^{+0.08}$ & $1.31_{-0.04}^{+0.04}$  & $1.11_{-0.71}^{+0.62}$  & $0.11_{-0.04}^{+0.06}$   & $6.4_{-0.9}^{+1.2}$    & $0.11_{-0.07}^{+0.11}$ \\
PS1-12bv       &$-20.65_{-0.39}^{+0.32}$ & $2.85_{-0.60}^{+0.60}$ & $4.8_{-0.9}^{+1.2}$
               & $-0.76_{-0.25}^{+0.20}$ & $0.88_{-0.08}^{+0.08}$ & $0.91_{-0.10}^{+0.06}$  & $0.3_{-0.9}^{+1.2}$     & $0.09_{-0.03}^{+0.06}$   & $14.3_{-1.0}^{+1.4}$   & $0.18_{-0.10}^{+0.10}$ \\
PS1-13duy      &$-20.80_{-0.36}^{+0.42}$ & $1.95_{-0.30}^{+0.30}$ & $3.31_{-0.60}^{+0.90}$
               & $-0.81_{-0.33}^{+0.31}$ & $0.74_{-0.18}^{+0.16}$ & $1.24_{-0.34}^{+0.14}$  & $-0.6_{-0.9}^{+1.6}$    & $0.13_{-0.05}^{+0.05}$   & $6.4_{-2.3}^{+2.4}$    & $0.26_{-0.13}^{+0.12}$ \\
PS1-13dwm$^*$  &$-18.42_{-0.56}^{+0.44}$ & $1.80_{-0.60}^{+0.60}$ & $3.2_{-1.1}^{+1.1}$
               & $-1.17_{-0.40}^{+0.32}$ & $1.26_{-0.19}^{+0.15}$ & $1.73_{-0.16}^{+0.14}$  & $-0.1_{-1.4}^{+1.4}$    & $0.11_{-0.04}^{+0.06}$   & $6.7_{-2.5}^{+2.3}$    & $0.17_{-0.11}^{+0.16}$ \\
PTF10iam       &$-20.19_{-0.13}^{+0.14}$ & $2.1_{-0.4}^{+1.7}$    & $29.5_{-2.1}^{+2.6}$
               & $-1.54_{-0.09}^{+0.15}$ & $0.83_{-0.03}^{+0.03}$ & $0.10_{-0.03}^{+0.03}$  & $1.83_{-0.26}^{+0.13}$  & $0.13_{-0.05}^{+0.05}$   & $8.95_{-0.55}^{+0.71}$ & $0.40_{-0.13}^{+0.07}$ \\
SNLS04D4ec     &$-19.64_{-0.52}^{+0.45}$ & $4.0_{-2.4}^{+3.4}$    & $6.8_{-0.8}^{+4.4}$
               & $-1.92_{-0.33}^{+0.43}$ & $0.48_{-0.24}^{+0.21}$ & $-0.71_{-0.48}^{+0.43}$ & $1.40_{-0.44}^{+0.37}$  & $0.12_{-0.05}^{+0.05}$   & $8.25_{-0.13}^{+0.20}$ & $0.05_{-0.04}^{+0.08}$ \\
SNLS05D2bk     &$-19.48_{-0.10}^{+0.05}$ & $5.02_{-0.25}^{+0.25}$ & $12.54_{-0.50}^{+0.75}$
               & $-0.62_{-0.21}^{+0.19}$ & $0.45_{-0.09}^{+0.10}$ & $-0.71_{-0.19}^{+0.20}$ & $-0.65_{-0.22}^{+0.25}$ & $0.11_{-0.04}^{+0.06}$   & $7.35_{-0.19}^{+0.22}$ & $0.07_{-0.05}^{+0.10}$ \\
SNLS06D1hc     &$-19.23_{-0.07}^{+0.03}$ & $7.52_{-0.25}^{+0.25}$ & $14.54_{-0.25}^{+0.50}$
               & $-0.02_{-0.14}^{+0.12}$ & $0.44_{-0.07}^{+0.07}$ & $-0.68_{-0.13}^{+0.14}$ & $-1.77_{-0.15}^{+0.18}$ & $0.11_{-0.03}^{+0.04}$   & $6.48_{-0.09}^{+0.11}$ & $0.04_{-0.03}^{+0.06}$ \\
\hline
Median & $-19.9^{+1.3}_{-1.0}$          &$2.5^{+1.5}_{-1.2}$      & $5.0^{+3.6}_{-2.5}$     & $-0.95^{+0.48}_{-0.64}$    & $0.96^{+0.30}_{-0.28}$     & $1.03^{+0.42}_{-0.50}$ & $0.6^{+0.7}_{-1.0}$      & $0.103^{+0.030}_{-0.033}$      & $7.9^{+4.8}_{-2.7}$          & $0.15^{+0.18}_{-0.11}$ \\
\enddata
\tablecomments{Asterisk ($*$) marks the FBOT event that lacked the detection epoch before the peak. The values at the last line represent the median with $1\sigma$ deviation of each parameter for the collected FBOTs. The columns from left to right are (1) FBOT event; (2) peak absolute magnitude in units of mag; (3) rise time above half-maximum in units of day; (4) decline time above half-maximum in units of day; (5) ejecta mass in units of $M_\odot$; (6) initial spin period in units of ms; (7) polar magnetic field strength in units of $10^{14}\,{\rm G}$; (8) opacity to high-energy photons in units of ${\rm cm}^2{\rm g}^{-1}$; (9) opacity in units of ${\rm cm}^2{\rm g}^{-1}$; (10) floor temperature in units of $10^3\,{\rm K}$; (11) extinction in units of ${\rm mag}$.}
\end{deluxetable*}

\clearpage

\bibliography{FBOT}{}
\bibliographystyle{aasjournal}

\end{document}